\documentclass{SciPost}

\usepackage{braket}
\usepackage{float}
\usepackage{soul}
\usepackage{subcaption}
\usepackage[dvipsnames]{xcolor}
\usepackage{comment}
\usepackage[inkscapepath=svgdir/svgpath]{svg}
\usepackage{wrapfig}
\usepackage{tabularx}
\usepackage{diagbox,makecell}
\usepackage{multirow}

\graphicspath{{./sections/figures/}}

\hypersetup{
    colorlinks,
    linkcolor={red!50!black},
    citecolor={blue!50!black},
    urlcolor={blue!80!black}
}

\usepackage[bitstream-charter]{mathdesign}
\DeclareSymbolFont{usualmathcal}{OMS}{cmsy}{m}{n}
\DeclareSymbolFontAlphabet{\mathcal}{usualmathcal}

\fancypagestyle{SPstyle}{
\fancyhf{}
\lhead{\colorbox{scipostblue}{\bf \color{white} ~SciPost Physics }}
\rhead{{\bf \color{scipostdeepblue} ~Submission }}

\fancyfoot[C]{\textbf{\thepage}}
}

\begin{document}

\pagestyle{SPstyle}

\begin{center}{\Large \textbf{\color{scipostdeepblue}{
Quasi-solitons of the 1D Bose gas \\ at arbitrary-strength contact repulsion
}}}\end{center}

\begin{center}\textbf{
Anahita Sarvi\textsuperscript{1$\star$},
Cesare Vianello\textsuperscript{2$\dagger$} and
Jean-Sébastien Caux\textsuperscript{1$\ddagger$}
}\end{center}

\begin{center}
{\bf 1} Institute of Physics, University of Amsterdam, \\ Science Park 904, 1098 XH Amsterdam, The Netherlands
\\
{\bf 2} Dipartimento di Fisica e Astronomia ``Galileo Galilei'', Università di Padova, \\ and INFN Sezione di Padova, via Marzolo 8, 35131 Padova, Italy
\\[\baselineskip]
$\star$ \href{mailto:a.sarvi@uva.nl}{\small a.sarvi@uva.nl}\,,\quad
$\dagger$ \href{mailto:cesare.vianello@phd.unipd.it}{\small cesare.vianello@phd.unipd.it}\,,\quad
$\ddagger$ \href{mailto:prof@jscaux.org}{\small prof@jscaux.org}
\end{center}

\section*{\color{scipostdeepblue}{Abstract}}
\textbf{\boldmath{
%Abstract
Harnessing integrability at arbitrary interaction strength, we build long-lived and spatially localized soliton-like excitations in the 1D Bose gas with contact repulsion (Lieb-Liniger model). We show how these density profiles, built from selective superpositions of multi-hole eigenstates, display long lifetimes tunable through parameters like dispersion width and number of holes. We further illustrate a scaling protocol that in principle enables unbounded increase in lifetimes in the thermodynamic limit. Our results showcase classical-like excitations originating from unitary quantum dynamics, away from the semi-classical regime, and in regions where interactions cannot be treated perturbatively.
}}

\vspace{\baselineskip}

\noindent\textcolor{white!90!black}{%
\fbox{\parbox{0.975\linewidth}{%
\textcolor{white!40!black}{\begin{tabular}{lr}%
  \begin{minipage}{0.6\textwidth}%
    {\small Copyright attribution to authors. \newline
    This work is a submission to SciPost Physics. \newline
    License information to appear upon publication. \newline
    Publication information to appear upon publication.}
  \end{minipage} & \begin{minipage}{0.4\textwidth}
    {\small Received Date \newline Accepted Date \newline Published Date}%
  \end{minipage}
\end{tabular}}
}}
}

% \linenumbers

\vspace{10pt}
\noindent\rule{\textwidth}{1pt}
\tableofcontents
\noindent\rule{\textwidth}{1pt}

\section{Introduction} \label{sec:intro}
Classical solitons are localized, stable excitations that travel at fixed speed through their host system, without deformation or decay, and emerge undeformed after collision with other excitations of their kind\footnote{It is worth noting the distinction between a \textit{soliton} and a \textit{solitary wave}, the former being “a particular type of solitary wave, which is not destroyed when it collides with another wave of the same kind” \cite{Hirota_2004}. Non-integrable models, such as the $\phi^4$ equation, can admit solitary wave solutions which do \textit{not} display robustness under collision.}. Such entities arise in a variety of physical systems, including shallow water (Korteweg-de Vries equation \cite{KdV}), deep stratified fluids (Benjamin-Ono equation \cite{Benjamin, Ono}), and the plethora of systems modeled by the sine-Gordon equation, such as coupled torsion pendula and Josephson junctions. Models that admit soliton solutions generally possess a remarkable feature---that of \textit{integrability}---which endows the system with the capacity to display non-ergodic time evolution, first noticed by Fermi, Pasta, Ulam, and Tsingou (FPUT) \cite{FPUT} as an unusual reversal of thermalization in their numerical simulations of nonlinear strings. Zabusky and Kruskal \cite{Zabusky} (who eventually coined the term \textit{soliton}) thereafter put these findings on firmer footing with their discovery of soliton solutions to the Korteweg-de-Vries equation, which were shown to form the continuum limit of FPUT's string model with appropriate nonlinearity. Synthesis was achieved through the development of the classical inverse scattering transform \cite{Gardner} as a method to solve many instances of nonlinear partial differential equations. Not long thereafter, and inspired by the idea of factorized S-matrices \cite{1979_Zamolodchikov_AP_120} and commuting transfer matrices \cite{BaxterBOOK}, the quantum world was afforded the perks of inverse scattering, starting with \cite{Sklyanin,Takhtadzhyan,Faddeev} and the wider Leningrad school, resulting in the method known today as the Algebraic Bethe Ansatz (see \cite{Korepin1993} and references therein): a coherent algebraic framework to classify, obtain and utilize exact solutions to integrable many-body quantum systems.

One example of such an integrable many-body quantum system is a gas of delta-interacting bosons in 1+1 dimensions, first solved perturbatively for weak interactions by Bogoliubov \cite{Bogoliubov}, and in the strongly interacting regime by Tonks and Girardeau \cite{Tonks, Girardeau} via a mapping to free fermions. In 1963, Lieb and Liniger \cite{Lieb_Liniger} solved the model exactly at arbitrary interaction strength using the Bethe Ansatz \cite{1931_Bethe_ZP_71}, revealing a Pauli exclusion-based classification of eigenstates. In a subsequent paper, Lieb \cite{Lieb}  introduced a new branch of excitations complementing the ``Type I'' modes of Bogoliubov's theory, coined ``Type II'' or ``hole-like'' modes, corresponding to the removal of a previously occupied momentum from the Fermi sea. Since then, various aspects of solitonic excitations of the 1D Bose gas have been studied in a series of works, including analytical and numerical studies, non-equilibrium dynamics in driven systems, hydrodynamic approaches, and more \cite{Ballu_2026, Rabec_2025, Simmons_2023, Martin_2010, Girardeau_2003, Carr_2001, Tettamanti_2021, Calzavara_2020, Kopycinski_2022, Syrwid_2016}.
In this work, we focus on obtaining the quantum counterpart of the soliton as localized, long-lived excitations of the 1D Bose gas by using the Bethe Ansatz to probe non-perturbatively high energy states. Our strategy is simple: given a set of eigenstates $|\alpha\rangle$ with energy $\omega_\alpha$ and momentum $k_\alpha$, we consider wavefunctions
\begin{equation}
|\Psi (t) \rangle = \sum_\alpha \psi_\alpha e^{-i \omega_\alpha t} | \alpha \rangle
\end{equation}
built from states within a judiciously chosen subbasis, and with opportune amplitudes. Our object of interest is then the space- and time-dependent expectation value of the density,
\begin{equation}
\rho(x,t) = \sum_{\alpha', \alpha} \psi^*_{\alpha'} \psi_{\alpha} e^{i (\omega_{\alpha' \alpha} t - k_{\alpha' \alpha} x)} \langle \alpha' | \hat{\rho}(0) | \alpha \rangle
\end{equation}
in which $\omega_{\alpha' \alpha} \equiv \omega_{\alpha'} - \omega_{\alpha}$ (and similarly for $k_{\alpha' \alpha}$). All building blocks (states, energies and momenta, as well as density operator matrix elements) are provided by (Algebraic) Bethe Ansatz. The object of the game is to have the fully quantum-mechanically computed $\rho(x,t)$ yield depletion profiles with classical-like behavior---remaining simultaneously localized and long-lived--- by  making a suitable choice of the subbasis and amplitudes in the wavefunction.

Our search is encouraged by the exact correspondence between the dispersion relations of (classical) dark solitons of the Gross-Pitaevskii equation \cite{Gross, Pitaevskii} (also known as the nonlinear Schrödinger equation) and that of Lieb's Type II excitations in the limit of weak interaction \cite{Kulish_1976, Ishikawa, Karpiuk, Syrwid_2021, Cesare}, which provides a basis to use Type II modes as building blocks for the quantum soliton. However, the correspondence is at most only qualitative for moderate and strong interactions \cite{Kolomeisky, Karpiuk, Kopy, Syrwid_2021, Cesare}, which makes our task nontrivial. The main challenge already arises here: classically, a soliton retains a spatially localized profile and well-defined momentum. In  quantum mechanics, simultaneous localization in both momentum- and real-space violates the Heisenberg uncertainty principle. In a translationally invariant theory, the Hamiltonian and momentum operators commute: thus an eigenstate of the system---which has a well-defined energy and is  infinitely long-lived---also possesses a well-defined momentum and is thus spatially delocalized. A superposition of such eigenstates yields a localized wavepacket at the cost of introducing a finite lifetime due to nonlinear dispersion: the presence of various energy modes leads to the dephasing and spatial deformation of the wavepacket. Our endeavor therefore becomes a balancing act between spatial localization and stability under time evolution.

Current literature has mainly focused on the construction of localized wavepackets from a superposition of all available single-hole Type II states, weighted in various ways \cite{Sato2016, Shamailov, Golletz, Kaminishi_2020, Ishiguro, Kinjo2022, Kinjo2023} and at various interaction strengths. While summing over all available such eigenstates (and thus the widest distribution of momenta) produces a localized excitation, the large nonlinear energy spread between these states leads inevitably to a rapid dephasing and decay of the wavepacket.

Here, we introduce a different approach to address the challenge of balancing depletion depth, localization and lifetime. Using a selective subset of available eigenstates in the wavepacket, we decrease the nonlinear spread in energy levels, thus lowering the dephasing rate of the wavepacket and improving its lifetime. Second, we displace more particles by using multi-hole eigenstates as building blocks. Our construction, based on Bethe Ansatz, can be attempted at any interaction strength. While solitonic excitations have been well-studied in the weakly interacting limit of the 1D Bose gas, where semi-classical approximations are valid \cite{Gross, Pitaevskii, Zakharov, Tsuzuki, Jackson, Burger, Phillips, Carretero, Frantzeskakis, Pethick}, we show that for finite interactions our method yields novel, long-lived and localized wavepackets in the regime where the system is non-perturbatively interacting and inherently quantum mechanical.

Our work is organized as follows: we begin with a formulation of our method in the Tonks-Girardeau (TG) regime, where we obtain analytical results using coordinate Bethe Ansatz, and extend our work numerically using ABACUS \cite{2009_Caux_JMP_50} to obtain results at finite interaction strength. Our results for single-hole, full-width summation wavepackets with uniformly weighted coefficients are in agreement with those presented previously in the literature, both in the Gross-Pitaevskii regime and in the TG regime obtained via mapping to free fermions \cite{Sato2016}. We then introduce selective-width summations and present energetics-based arguments explaining the effect of summation width (and its placement in the Fermi sea) on the lifetime, velocity, and initial profile of the wavepacket.

Next, we present our results for multi-hole ($n=2,3,4$) uniform-amplitude wavepackets at arbitrary interaction strength, probing the effect of increasing $n$ on the spatial profile and dynamics of the wavepacket. We then explore an alternate choice of coefficients to maximize the depletion in a chosen spatial window via a variational method. We present the scaling behavior of spatial profiles and time evolution of both uniform-amplitude and variational wavepackets with system size, and introduce two methods to quantitatively measure lifetime: a time power series $\Delta \rho(t)$ tracking the central point of the depletion, and a shape-aware measure of lifetime $R(t)$. Using these, we quantify the effect of various factors such as choice of coefficients, interaction strength, number of holes, summation width, system-size scaling on lifetime.

\section{Preliminaries} \label{sec:prelim}
\subsection{The Lieb-Liniger model}\label{sec:LL_model}

The Lieb-Liniger model \cite{Lieb_Liniger, Lieb} describes $N$ identical bosons in 1D interacting via a delta function potential. In units where $\hbar = 2m=1$, the Hamiltonian is
\begin{equation}
    \hat H = -\sum_{i=1}^N \partial_{x_i}^2 + 2c\sum_{i<j}\delta(x_i-x_j)
\end{equation}
or, in second-quantized form,
\begin{equation}
    \hat H = \int_0^L dx\left[\partial_x \hat \Psi^\dag(x) \partial_x \hat \Psi(x) + c \hat \Psi^\dag(x)\hat \Psi^\dag(x)\hat \Psi(x)\hat \Psi(x)\right],
\end{equation}
where $L$ is the size of the system and $\hat \Psi^{(\dag)}(x)$ are canonical bosonic field operators. The total particle number $\hat N = \int_0^L dx\,\hat \Psi^\dag(x)\hat \Psi(x)$ and momentum $\hat P = -i \int_0^L dx\,\hat \Psi^\dag(x)\partial_x \hat \Psi(x)$ commute with the Hamiltonian and are therefore conserved charges (the first in an infinite hierarchy). Throughout this work we will consider $c>0$, which corresponds to repulsive interaction, and assume periodic boundary conditions. Furthermore, we take $N=L$, namely unit density, unless otherwise indicated.

As a consequence of the fact that the scattering phase shift between many particles can be reduced to a sequence of two-particle scattering phase shifts, the Lieb-Liniger model is exactly solvable via the Bethe Ansatz \cite{Lieb_Liniger, Lieb, Yang1967, Yang1968, Sutherland, YangYang, Gaudin1971a, Gaudin1971b, Dorlas, Korepin1993}. When $N$ is odd (even), for any set $\{I_j\}_{j=1,\dots,N}$ of non-coincident integer (half-integer) quantum numbers there exists a unique set of real rapidities $\{\lambda_j\}_{j=1,\dots,N}$ solving the Bethe equations
\begin{equation}\label{eq:Bethe_eq}
    \lambda_j + \frac{1}{L} \sum_{k=1}^N \phi(\lambda_j-\lambda_k) = \frac{2\pi}{L} I_j,
\end{equation}
where
\begin{equation}\label{eq:phaseshift}
    \phi(\lambda_j-\lambda_k) = 2 \arctan\left(\frac{\lambda_j-\lambda_k}{c}\right)
\end{equation}
is the two-particle scattering phase shift. These rapidities identify a simultaneous eigenstate $\ket{\{\lambda\}}$ of $\hat P$ and $\hat H$ with momentum
\begin{equation}\label{eq:P_Bethe}
    P = \sum_{j=1}^N \lambda_j = \frac{2\pi}{L}\sum_{j=1}^N I_j
\end{equation}
and energy 
\begin{equation}\label{eq:E_Bethe}
    E = \sum_{j=1}^N \lambda_j^2.
\end{equation}
In this way, the complete set of eigenstates of the Hamiltonian can be constructed. Each Bethe state is uniquely labeled either by its rapidities or its quantum numbers, which are in one-to-one correspondence and equally ordered, i.e. if $I_j>I_k$ then $\lambda_j>\lambda_k$. In an ordered sector of coordinate space, a Bethe state is represented by a superposition of plane waves over all permutation $\sigma$ of the rapidities, $\Psi(x_1,\dots,x_N) = \sum_\sigma A_\sigma \exp(i\sum_{j=1}^N\lambda_{\sigma(j)}x_j)$, with the amplitudes $A_\sigma$ determined by the two-particle scattering phase shifts.

The ground state is specified by a maximally compact set of quantum numbers centered around zero. Since quantum numbers cannot coincide, this is given by 
\begin{equation}
    \{I_j^0\}_N = \left\{-\frac{N+1}{2}+j\right\}_{j=1,\dots,N}.
\end{equation}
This set has two Fermi edges, in analogy with a 1D Fermi sea. 

\subsubsection{\emph{n}-hole states}

We call $n$-hole state an eigenstate constructed from the ground state by removing one occupied quantum number (creating a hole) and adding it to the first available position to the right of the Fermi edge (creating a particle), and repeating the process for $n$ distinct hole positions. The holes' quantum numbers are
\begin{equation}
    \{\tilde I_j^{\mathbf h}\}_n = \left\{\frac{N+1}{2}-h_i\right\}_{j=1,\dots,n},
\end{equation}
with $h_j \in \{1,\dots,N\}$ ($h_i \neq h_j$), and the quantum numbers of the resulting $n$ particles to the right of
the ground state interval are
\begin{equation}
    \{I_j^{\mathbf p}\}_n = \left\{\frac{N-1}{2}+j\right\}_{j=1,\dots,n}.
\end{equation}
The quantum numbers of a $n$-hole state are therefore
\begin{equation}\label{eq:qn_nholes}
    \{I_j\}_N = \left\{\{I_j^0\}_N \setminus \{\tilde I_j^{\mathbf h}\}_n\right\} \cup \{I_j^{\mathbf p}\}_n.
\end{equation}

\subsubsection{\emph{n}-hole wavepackets}

Being eigenstates of a translationally invariant Hamiltonian, $n$-hole states are completely delocalized in space. We can construct a spatially localized $n$-hole wavepacket $\ket{\psi_n(0)}$ as a linear superposition of normalized $n$-hole states $\ket{P_n^{\mathbf h}}$ with energy $E_n^{\mathbf h}$ as
\begin{equation}
    \ket{\psi_n(0)} = \frac{1}{\sqrt{\mathcal N}}\sum_{\mathbf h} C_n^{\mathbf h}\ket{P_n^{\mathbf h}},
\end{equation}
where $\mathcal N$ is the number of states in the summation. The coefficients $C_n^{\mathbf h}$ should be tuned in order to obtain the desired density profile at $t=0$. The wavepacket will then evolve unitarily in time according to
\begin{equation}
    \ket{\psi_n(t)} = \frac{1}{\sqrt{\mathcal N}}\sum_{\mathbf h} C_n^{\mathbf h} e^{-i E_n^{\mathbf h} t}\ket{P_n^{\mathbf h}}.
\end{equation}

The density profile is given by $\rho_n(x,t) = \bra{\psi_n(t)}{\hat\rho(x)}\ket{\psi_n(t)}$, where $\hat\rho(x) = \hat \Psi^\dag(x)\hat \Psi(x)$ is the density operator. Since by translational invariance $\hat\rho(x) = e^{-i\hat Px}\hat\rho(0)e^{i\hat Px}$, we have
\begin{equation}\label{eq:density_profile}
    \rho_n(x,t) = \frac{1}{\mathcal N} \sum_{\mathbf h, \mathbf h'}\left(C_n^{\mathbf h'}\right)^* C_n^{\mathbf h}\,e^{-i[(P_n^{\mathbf h'}-P_n^{\mathbf h})x-(E_n^{\mathbf h'}-E_n^{\mathbf h})t]}\bra{P_n^{\mathbf h'}}{\hat\rho(0)}\ket{P_n^{\mathbf h}}.
\end{equation}

\subsubsection{Density matrix elements}

The key ingredients to compute the density profile are the matrix elements of $\hat\rho(0)$ between two (unnormalized) Bethe states in the $N$-particle sector, call them $\ket{\{\mu\}}$ and $\ket{\{\lambda\}}$. These are clearly related to the matrix elements of the integrated density $\hat Q_1(x) = \int_0^x dx' \hat\rho(x')$ by
\begin{equation}\label{eq:me0}
        \bra{\{\mu\}}{\hat\rho(0)}\ket{\{\lambda\}} = e^{-i\sum_j(\lambda_j-\mu_j)x}\partial_x\bra{\{\mu\}}{\hat Q_1(x)}\ket{\{\lambda\}}.
\end{equation}
The matrix elements of $\hat Q_1(x)$ obtained from the Algebraic Bethe Ansatz are \cite{Izergin, Slavnov1989, Slavnov}
\begin{equation}\label{eq:meq}
        \bra{\{\mu\}} {\hat Q_1(x)}\ket{\{\lambda\}} = \left(e^{i\sum_j(\lambda_j-\mu_j)x}-1\right)\partial_\alpha\sigma_N^\alpha(\{\mu\},\{\lambda\})_{\alpha = 0},
\end{equation}
where
\begin{equation}
    \partial_\alpha\sigma_N^\alpha(\{\mu\},\{\lambda\})_{\alpha = 0} = \prod_j(V_j^+-V_j^-)\prod_{j,k}\left(\frac{\lambda_j-\lambda_k+ic}{\mu_j-\lambda_k}\right)\frac{\det_N(\delta_{jk} + U_{jk}(\lambda_p))}{V_p^+-V_p^-}.
\end{equation}
Here, $\lambda_p$ is an arbitrary parameter and the quantities $V_j^\pm$, $U_{jk}$ are defined by
\begin{gather}
    V_j^\pm = \prod_m \frac{\mu_m - \lambda_j \pm ic}{\lambda_m-\lambda_j \pm ic},\\[0.5ex]
    U_{jk}(\lambda_p) = \frac{i(\mu_j-\lambda_j)}{V_j^+-V_j^-}\prod_{m\neq j} \left(\frac{\mu_m-\lambda_j}{\lambda_m-\lambda_j}\right)(K_{jk} - K_{pk}),
\end{gather}
with
\begin{equation}
    K_{jk} = \frac{2c}{(\lambda_j-\lambda_k)^2+c^2}.
\end{equation}
Eqs. \eqref{eq:me0}-\eqref{eq:meq} thus give
\begin{equation}\label{eq:me_final}
       \bra{\{\mu\}}{\hat\rho(0)}\ket{\{\lambda\}}  = i \sum_j (\lambda_j-\mu_j)\partial_\alpha\sigma_N^\alpha(\{\mu\},\{\lambda\})_{\alpha = 0}.
\end{equation}
For proper normalization, the right-hand side should then be divided by $\Vert\{\mu\}\Vert \Vert\{\lambda\}\Vert$, where the norm of the Bethe states is given by the Gaudin-Korepin formula \cite{Gaudin1981, Korepin1982,Gaudin_2014}
\begin{equation}
    \|\{\lambda\}\|^2 = c^N \prod_{j>k}\frac{(\lambda_j-\lambda_k)^2+c^2}{(\lambda_j-\lambda_k)^2}\,\mathrm{det}_N\biggl[\delta_{jk}\biggl(L + \sum_l K_{jl}\biggr)-K_{jk}\biggr].
\end{equation}
 
\subsection{Solutions in analytically tractable limits}\label{sec:analytic_lims}
Below, as precursors to our finite $c$ results, we present general solutions in two limits, $c \rightarrow \infty$ and $c \rightarrow 0$, amenable to analytic treatment. We will use these expressions, adapted to various situations, in the following sections.
\subsubsection{$n$-hole density profiles in the $c \rightarrow \infty$ limit}\label{sec:nh_c_inf_lim}

In the Tonks-Girardeau regime, the scattering phase shift approaches zero, and the Bethe equations yield a simple map from quantum number to rapidities, $\lambda_j = 2\pi I_j/L$. With this mapping, we can easily calculate the momentum and energy of a $n$-hole state defined by the quantum numbers \eqref{eq:qn_nholes} according to Eqs. \eqref{eq:P_Bethe}-\eqref{eq:E_Bethe}:
\begin{align}
    P_n^{\mathbf h} &= \frac{2\pi}{L}\biggl[\frac{n(n-1)}{2}+\sum_{j=1}^n h_j\biggr],\\[0.5ex]
    E_n^{\mathbf h} &= E_{GS} + \left(\frac{2\pi}{L}\right)^2\sum_{j=1}^n \left[j(j-1) - h_j(h_j-1) + N(j+h_j-1)\right],
\end{align}
where $E_{GS} = \pi^2N(N^2-1)/3L^2$ is the ground state energy. 

As we recall in Appendix \ref{sec:app_dme}, in the Tonks-Girardeau regime the density matrix elements \eqref{eq:me_final} also simplify significantly. Matrix elements between Bethe states differing by more than one quantum number vanish, hence to have a nonzero matrix element the states $\ket{\{\mu\}}$ and $\ket{\{\lambda\}}$ must differ by exactly zero or one holes. In the first case, we have the diagonal matrix element $N/L$. In the second case, denoting by $\tilde I_h$ and $\tilde I_{h'}$ the quantum numbers of the single hole that is displaced between the states $\ket{\{\mu\}}$ and $\ket{\{\lambda\}}$, the matrix element is
\begin{equation}
    \bra{\{\mu\}}{\hat\rho(0)}\ket{\{\lambda\}} = \frac{(-1)^{M_{h,h'}}}{L} = (-1)^{\tilde M_{h,h'}+1}\Bigg(\frac{e^{i\pi(h-h')}}{L}\Bigg),
\end{equation}
where $M_{h,h'}$ and $\tilde M_{h,h'}$ are, respectively, the number of particles and the number of holes between $\tilde I_h$ and $\tilde I_{h'}$. 

Applying these results to Eq. \eqref{eq:density_profile} for the density profile, we obtain that the $n$-holes states $\ket{P_n^{\mathbf h}}$ and $\ket{P_n^{\mathbf h'}}$ must have $n-1$ holes $\bar h_1,\dots,\bar h_{n-1}$ in common, and one displaced hole, identified by $h$ in $\ket{P_n^{\mathbf h}}$ and by $h'$ in $\ket{P_n^{\mathbf h'}}$. The density profile is then
\begin{equation}\label{eq:rhon_TG}
    \rho^{(TG)}_n(x,t) = \frac{N}{L} + \frac{1}{\mathcal N L} \sum_{\substack{\bar h_1,\dots, \bar h_{n-1},\\{h,h'}}}\left(C_n^{\mathbf h'}\right)^* C_n^{\mathbf h}\,e^{-\frac{2\pi i}{L}(h'-h)[x+\frac{L}{2} -\frac{2\pi}{L}(N+1-h'-h)t]}(-1)^{\tilde M_{h,h'}+1},
\end{equation}
where the summation is over distinct hole positions. 

\subsubsection{Dark solitons in the $c \rightarrow 0$ limit}\label{sec:dark_sol_c_0_lim}

It is well-known that the one-dimensional Bose gas can support increasingly long-lived excitations in the $c \to 0$ limit. In fact, in the weakly interacting regime the dynamics of the system is well captured by the mean-field Gross-Pitaevskii (GP) equation \cite{Gross, Pitaevskii, Pethick}, 
\begin{equation}
    i\partial_t \psi(x,t) = \left[-\partial_x^2 + 2c|\psi(x,t)|^2\right]\psi(x,t).
\end{equation}
This admits the analytic dark-soliton solution 
\begin{equation}\label{gpe_soliton}
    \psi_s(x,t) = \sqrt{\rho_0}\left[i\frac{v}{u}+\sqrt{1-\frac{v^2}{u^2}}\tanh\left(\frac{x-x_0-vt}{\sqrt{2}\xi_v}\right)\right]e^{-i u^2t/2},
\end{equation}
where $u = 2\sqrt{c \rho_0}$ is the Bogoliubov sound velocity and $\xi_v = \xi_0(1-v^2/u^2)^{-1/2}$, with $v \le u$ the propagation velocity of the soliton and $\xi_0 = (2 c \rho_0)^{-1/2}$ the healing length \cite{Tsuzuki, Zakharov, Pethick}. In the small-amplitude limit, the dark solitons of the GP equation are perturbatively connected to those of the Korteweg-de Vries equation \cite{Gardner, Tsuzuki, Carretero, Frantzeskakis}, which has historically been shown to describe the asymptotic development of shallow-water waves \cite{KdV}. The density associated to $\psi_s(x,t)$ is
\begin{equation}\label{eq:depletion_density}
    \rho_s(x,t) \equiv |\psi_s(x,t)|^2 = \rho_0 - (\rho_0-\rho_\text{min})\,\text{sech}^2\left(\frac{x-x_0-vt}{\sqrt{2}\xi_v}\right),
\end{equation}
where $\rho_\text{min}=\rho_s(x=x_0+vt,t) = \rho_0v^2/u^2$ is the minimal value of the density, corresponding to the bottom of the depletion. Therefore the propagation velocity is related to the minimum density by $v=2\sqrt{c\rho_\text{min}}$, i.e. it is the bulk sound velocity evaluated at $\rho_\text{min}$. In particular, it is nonzero whenever that minimum is nonzero, and vice versa.
\subsection{Dispersion-based estimates of lifetime}\label{sec:scaling_arg}

\begin{figure}
    \centering
    \includegraphics[width=\linewidth]{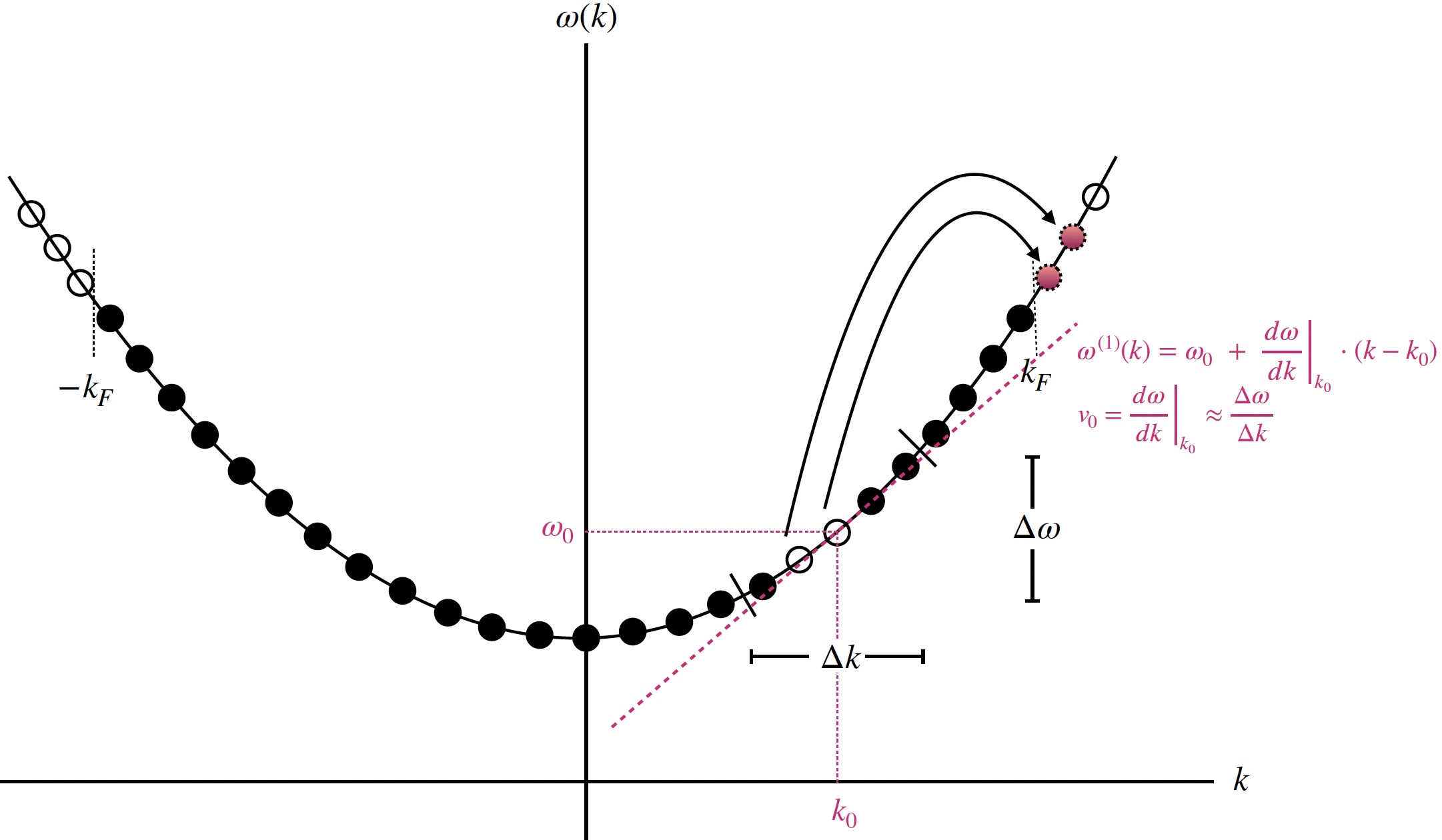}
    \caption{Schematic of an $n=2$ particle-hole excitation in the Fermi sea with the first order expansion of energies $\omega$ around $k_0$.}
    \label{fig:intro_omega_diag_half}
\end{figure}

Equipped with an overview of the model, energies, and states, we now outline our approach for building the Type II wavepackets, and present a scaling relation between the lifetime $\tau$ and momentum width $\Delta k$ using the dispersion $\omega (k)$.   
A schematic of energies $\omega$ as a function of momentum $k$ is depicted in Fig. \ref{fig:intro_omega_diag_half}, where each circle represents a momentum value admitted by the system. In the ground state of an $N$-particle system, $N$ spots are occupied (filled-in circles) consecutively and symmetrically about the center of the curve, and thus form the Fermi sea. A Type II (hole-like) excitation corresponds to removing one (or more) previously occupied momentum value(s), with excitations deep in the sea evidently costing more energy than those closer to the Fermi edge. For our purposes, we choose to always conserve the number of particles in the system by replacing each eliminated momentum at the first available spot on the right Fermi edge. To build an $n$-hole Type II wavepacket, we first select an interval of width $\Delta k$ within the Fermi sea (with corresponding window of energy $\Delta \omega$), in which the $n$ holes will be placed to form a single momentum eigenstate labeled by the choice of hole position. The wavepacket is then built from a superposition of all possible $n$-hole placements within $\Delta k$ with the desired choice of coefficients. We call this a \textit{selective width} wavepacket. In the example given in Fig. \ref{fig:intro_omega_diag_half}, an $n=2$-hole eigenstate is depicted in a hole window of width $\Delta w = \Delta k \cdot \frac{L}{2\pi} =5$, thus representing one of $_5C_2=10$ possible such eigenstates. The placement of the interval $\Delta k$ within the right half of the Fermi sea gives a group velocity $v_g = \frac{d\omega}{dk}|_{k_0}$ (i.e., the linear term of the expansion) to the wavepacket, with $k_0$ marking the center of the hole window. To obtain an estimate for the lifetime of the wavepacket, we consider the deviation of the linear expansion from the energy curve, as depicted in Fig. \ref{fig:intro_omega_diag_inset}. It is useful to define some guidelines for what we would like to achieve: we may demand that the wavepacket travel a certain distance (given an initial velocity) while retaining a set percentage of its initial depth. For instance, given a group velocity of $\mathcal{O}(1)$ (i.e. scaling as $N^0$), we may aim for an excitation that traverses the system (of length $L=N$) once while retaining at least half of its initial depth. While this condition is purely illustrative, it can serve as a useful reference point. Second, we are interested in wavepackets that, to a large extent, evolve \textit{rigidly} and \textit{symmetrically} about the center of the depletion. Rigid evaporation is such that the spatial profile does not display a large \textit{qualitative} change over the course of traversing the system. As we will see, some wavepackets display these qualities to a larger extent than others, depending on factors such as choice of coefficients and interaction strength.

\subsubsection{Scaling of lifetime with $N$, $\Delta k$}

Let us take a system with $N=L$ in dimensionless units, sufficiently large such that we have ample choice for the momentum hole window $\Delta k = \frac{2\pi}{L}w$ (where $w$ is the width in units of quantum numbers) and subsequent placement of the $n$ holes: $1 <<  w << N$. Let us then take $w \propto  N^{\alpha}$ for $0<\alpha <1$. We then have $w/L \sim N^{\alpha}/L \sim L^{\alpha-1}$, with the physical width of the depletion $D_w$ scaling as the reciprocal of $\Delta k$, thus $D_w \propto L/w \sim L^{1-\alpha}$. For near-parabolic dispersions,\footnote{The parabolic estimate becomes exact in Tonks-Girardeau regime.} we have $\Delta \omega \propto (\Delta k)^2 \sim L^{2(\alpha - 1)}$, which gives a first (linear) scaling estimate of lifetime $\tau$ as $\tau \sim 1/\Delta \omega \sim L^{2(1-\alpha)}$. The distance traversed by the wavepacket traveling with a velocity $v \sim \mathcal{O}(1)$ during its lifetime thus also scales as $d \equiv\tau \cdot v \sim L^{2(1-\alpha)}$. Setting the desired distance traversed within the lifetime to be the length of the system, we find $\alpha = 1/2$. Therefore, a choice of hole-window $w \sim \sqrt{N}$ is constructive to a wavepacket with the desired criteria stated above. We emphasize that the arguments herein presented are based upon large interaction values; at smaller $c$, we expect the general scaling behavior to continue to hold.

\begin{figure}
    \centering
    \includegraphics[width=\linewidth]
    {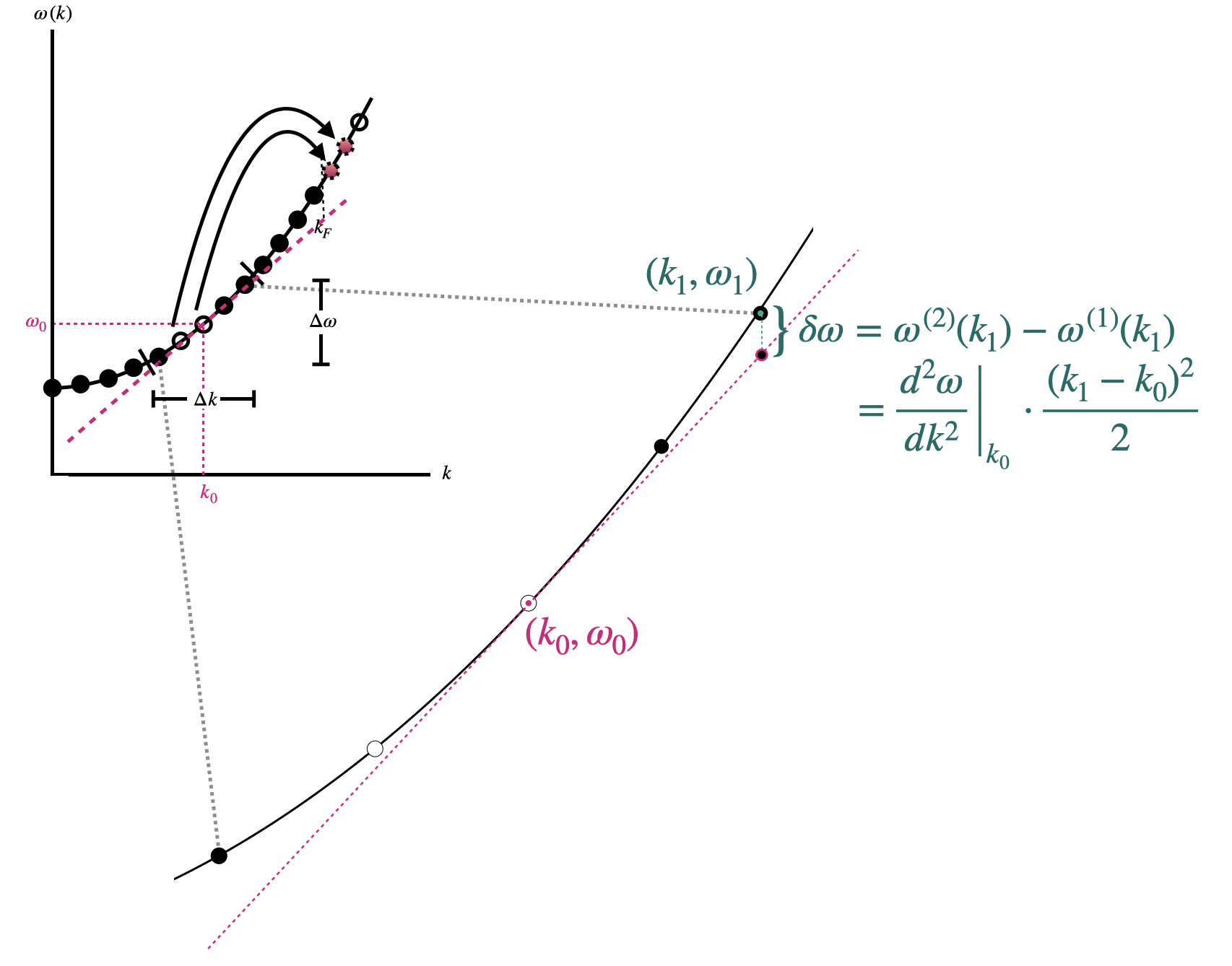}
    \caption{Schematic of an $n=2$ particle-hole excitation in the (right half of the) Fermi sea, with the second-order expansion of energies $\omega$ around $k_0$.}
    \label{fig:intro_omega_diag_inset}
\end{figure}

Next, since deviations from linear dispersion are responsible for the dephasing of the wavepacket and its ultimate demise, let us turn to the second-order expansion of $\omega(k)$ for an improved estimate of lifetime. As illustrated in Fig. \ref{fig:intro_omega_diag_inset}, the deviation from linearity $\delta \omega$ of the second order expansion $\omega^{(2)}(k)$ becomes 
\begin{equation}\label{eq:delta_omega}
    \delta \omega = \omega^{(2)}(k_1) - \omega^{(1)}(k_1) = \frac{d^2 \omega}{dk^2}\Bigg|_{k_0} \cdot \frac{(k_1-k_0)^2}{2},
\end{equation}
thereby yielding 
\begin{equation}\label{eq:tau_scaling}
  \tau \propto \frac{1}{(\Delta k)^2~ \omega''(k_0)}. 
\end{equation}
This directly implies that at large $c$, where the dispersion is predominantly quadratic and the curvature $\omega''$ is everywhere constant, lifetime scales simply as $(\Delta k)^{-2}$, without any effect from the position of the interval in the Fermi sea. As we will see, it is this scaling of lifetime with $(\Delta k)^{-2}$ that yields the strongest effect on lifetime. We shall confirm these observations in the coming sections (Sec. \ref{sec:lifetime}) using quantitative measures for lifetime.

\section{Uniform-amplitude wavepackets}

Let us now begin our discussion of specific instances of quasi-solitons. We begin with the most basic choice for the amplitudes of the wavefunction components, namely
\begin{equation}\label{eq:unif_amp_coef}
    C_n^{\mathbf h} = e^{-i P_n^{\mathbf h}x_0},
\end{equation}
which yields an $n$-hole wavepacket displaying a depletion centered at $L/2-x_0$. We call this a \textit{uniform amplitude} wavepacket. In the special case where we choose $x_0=L/2$ (and thus a depletion spatially centered at $x=0$), we refer to the result as a \textit{staggered-amplitude} (or simply \textit{staggered}) wavepacket due to the amplitudes alternating in sign at each hole step.

Throughout this work, we take $N/L \equiv \rho_0 = 1$ without loss of generality since the thermodynamic behavior of the system is simply dependent on the Lieb-Liniger parameter $\gamma$ defined as $cL/N = c/\rho_0 \equiv \gamma$. Thus the system at a higher (lower) density is equivalent to one at a weaker (stronger) interaction strength.

\subsection{Single-hole wavepackets}

\subsubsection{Wavepackets in the Tonks-Girardeau regime}\label{sec:c_inf_wavepackets}

Let us start with the simplest case for which we can obtain an analytic expression: a 1-hole wavepacket in the Tonks-Girardeau regime. With the choice \eqref{eq:unif_amp_coef} for the coefficients and $n = 1$, Eq. \eqref{eq:rhon_TG} becomes
\begin{equation}\label{eq:rho1_TG_uniform-amplitude}
    \rho_1^{(TG)}(x,t) = \frac{N}{L} - \frac{1}{\mathcal N L}\sum_{h\neq h'} e^{-\frac{2\pi i}{L}(h'-h)[x-x_0+\frac{L}{2} -\frac{2\pi}{L}(N+1-h'-h)t]}.
\end{equation}
Consider the wavepacket constructed as a superposition of all possible 1-hole states; in this case $h$, $h'$ take on all values in $\{1,\dots,N\}$ and $\mathcal N=N$. We call this a \emph{full-width summation}, and the corresponding wavepacket is a discrete Fourier transform of 1-hole type II eigenstates. Taking for instance $x_0 = L/2$, at time $t=0$ we find
\begin{equation}
    \rho_1^{(TG)}(x,0) = \frac{N+1}{L} - \frac{1}{NL}\sum_{h, h'} e^{-\frac{2\pi i}{L}(h-h')x} = \frac{N+1}{L} - \frac{1}{NL}\frac{\sin^2(N \pi x /L)}{\sin^2(\pi x/L)},
\end{equation}
which coincides with the result obtained in Ref. \cite{Sato2016} through a mapping to free fermions. 

\subsubsection*{Drift speed}

Concerning the dynamical evolution, we first note that such wavepacket is produced at rest, and therefore the density profile remains symmetric with respect to $x=x_0$ for any $t \ge 0$. This can be easily understood by rewriting Eq. \eqref{eq:rho1_TG_uniform-amplitude} in the form 
\begin{equation}
    \rho_1^{(TG)}(x,t)=\frac{N}{L}-\frac{2}{\mathcal NL}\sum_{h'>h} \cos\left[\frac{2\pi}{L}\left(x-x_0+\frac{L}{2}-v_{hh'}t\right)\right],
\end{equation}
where $v_{hh'}=2\pi(N+1-h-h')/L$ takes values in the symmetric interval around zero $[-v_c, v_c]$, with $v_c=2\pi(N-2)/L$. Our density is thus a superposition of left- and right-traveling cosine waves. Since all waves have the same amplitude and relative phase, we have pairs of counter-propagating waves with opposite velocities, which results in a stationary density depletion. We see that this is a very special situation, as a consequence of the peculiar structure of the density matrix elements in the TG regime. At finite $c$, even if $v_{hh'}$ assumed values in a symmetric interval around zero, the more complicated dependence of the matrix elements on $h$ and $h'$ due to the asymmetric distribution of particles to the right of the Fermi sea would introduce a relative phase between cosine waves propagating with opposite velocities, leading to a non-stationary density depletion.
We observe that under time evolution, the initial depletion at $x=x_0$ is redistributed throughout the whole system. At finite size, however, the disappearance of the initial depletion is not definitive, because the time evolution shows a recurrence (quantum revival). The revival time $t_{rev}$ can be identified as the smallest value of $t>0$ such that $\exp(-iE_1^ht)=1$ for any $h \in\{1,\dots,N\}$, that is $t_{rev}=L^2/4\pi$ for $N$ even and $L^2/2\pi$ for $N$ odd. Since $t_{rev}$ scales with $L^2$, there is no revival in a thermodynamically large system. Furthermore, exact revival occurs only in the Tonks-Girardeau regime, due to the commensurability of energy values. 

\subsubsection*{Thermodynamic limit}
Taking the thermodynamic limit with $N, L\to\infty$ and $N/L=\rho_0=\text{const.}$, we obtain
\begin{equation}
    \rho_1^{(TG)}(x,0) \asymp \rho_0 - \frac{\sin^2(\pi \rho_0 x)}{\pi^2\rho_0 x^2}.
\end{equation}
More generally, in the thermodynamic limit the quantity $z \equiv h/L$ becomes continuous and takes values in the interval $[0, \rho_0]$. Taking again $x_0=L/2$, Eq. \eqref{eq:rho1_TG_uniform-amplitude} becomes
\begin{align}
    \rho_1^{(TG)}(x,t) &\asymp \rho_0- \frac{1}{\rho_0}\int_0^{\rho_0} dz \int_0^{\rho_0} dz'\,e^{-2\pi i(z-z')[x-2\pi(\rho_0-z-z')t]}\nonumber\\[0.5ex]
    &= \rho_0 - \frac{1}{16\pi \rho_0 t}\left|\text{Erf}\left[\frac{(-1)^{1/4}(x-2\pi t)}{2\sqrt t}\right] - \text{Erf}\left[\frac{(-1)^{1/4}(x+2\pi t)}{2\sqrt t}\right]\right|^2.
\end{align}

As one would expect, this wavepacket describes a single localized density depression, of width (measured at the base) equal to $2/\rho_0$. The density takes values between $1/L$ and $(N+1)/L$, and thus between zero and $\rho_0$ in the thermodynamic limit. This is physically clear considering that we remove a particle within a range of momenta (hence a depletion  localized in space), and add it back with a fixed momentum, hence delocalized in space. In analogy with a liquid, a localized depletion causes the overall level of the liquid to rise. The density profile shows long-range oscillations of wavelength $\pi/\lambda_F=\rho_0^{-1}$, whose amplitude decays with the inverse square of $x-x_0$ (Friedel oscillations). This is a signature of the fermionization of the 1D Bose gas in the Tonks-Girardeau regime.

We have benchmarked these analytical results in the $c\to\infty$ limit against the numerical results obtained with ABACUS for $c\gg 1$. At $t=0$, the density profiles show agreement up to differences of order $\mathcal O(c^{-1})$. As expected, these discrepancies grow under time evolution. These results are shown in Appendix \ref{app:exact_vs_num}. Our results (corresponding to ones obtained earlier in \cite{Sato2012}) already show good correspondence with the thermodynamic limit starting with system size $N=250$.

\subsubsection*{Wavepackets at finite $c$}\label{sec:wavepackets_finite_c}
ABACUS enables construction of wavepackets at any interaction strength $c>0$ according to Eq. \eqref{eq:density_profile}. For all $c$, the density profile of the one-hole wavepacket exhibits a single localized depletion, but varying $c$ produces important qualitative changes. First, as anticipated the density profile is no longer perfectly stationary, but acquires a small velocity (due to our protocol) with the resulting displacement becoming visible at long times. Second, with all other parameters fixed, smaller $c$ yields a more long-lived depletion than its large-$c$ counterpart, as demonstrated in Fig. \ref{fig:c_1_1024_n_1_time_evol}, and as we demonstrate quantitatively in Sec. \ref{sec:lifetime}. This gain in lifetime at finite $c$ is clearly reflected in the curvature of Type II excitations: noting that the lifetime of the wavepacket is inversely proportional to the dispersion curvature $\omega''$, i.e. $\tau \sim \frac{1}{\omega_0''(\Delta k)^2}$ at a fixed momentum window $\Delta k$ centered at $(k_0,\omega_0)$ (as shown in Fig. \ref{fig:intro_omega_diag_inset}), the vanishing curvature at small $c$ leads to a multiple-order-of-magnitude gain in lifetime, as shown in Fig. \ref{fig:typeII_inv_curv}. Third, reducing $c$ progressively suppresses Friedel oscillations (Fig. \ref{fig:c_1_1024_n_1_t0}) as the system returns to the truly bosonic regime. On this note, let us turn to the small-$c$ limit, where we can make further analytic progress. 

\subsubsection{Wavepackets in the $c\to 0$ limit}

As discussed in Sec. \ref{sec:analytic_lims}, the system dynamics in the $c \rightarrow 0$ limit is well captured by the Gross-Pitaevskii equation, which admits the analytic dark soliton solutions \eqref{gpe_soliton}. The weight (area) of the depletion is, according to Eq. \eqref{eq:depletion_density},
\begin{equation}
    A=\int dx\,\left(\rho_0-\rho_s\right)=2\sqrt{\frac{\rho_0}{c}\left(1-\frac{v^2}{u^2}\right)}.
\end{equation}
Therefore in the weakly interacting regime, a single hole displaces up to $2\sqrt{\rho_0/c}$ particles, a number which diverges as $c^{-1/2}$ for $c\to 0$. Physically, we must have $A \le N$, hence if we take $\rho_0=1$ and $v=0$ (corresponding to a stationary black soliton), the condition $c>4/N^2$ must be satisfied in order for the mean-field description to be applicable to a finite system. This fact may be interpreted by saying that if the system is finite and $c$ is below this critical value, we have effectively non-interacting bosons, which cannot support nontrivial solitonic excitations.

We may introduce a parameter $\theta \in (0, \pi)$ such that $v=u\cos\theta$ and $\sqrt{1-v^2/u^2}=\sin\theta$. The depth $\rho_0-\rho_\text{min}=\rho_0 \sin^2\theta$, width $\ell=\sqrt 2 \xi_v = (\sqrt{c\rho_0}\sin\theta)^{-1}$, area $A = 2\sqrt{\rho_0/c}\sin\theta$, and propagation velocity $v=u\cos\theta$ are then all fixed by the single parameter $\theta$. In terms of such parameter, the momentum and energy of the soliton read \cite{Kulish_1976, Pethick, Cesare, Syrwid_2021}
\begin{equation}
    k(\theta) = \rho_0 (2\theta - \sin 2\theta),\qquad \omega(\theta) = \frac{4}{3} \rho_0 u \sin^3\theta,
\end{equation}
which give $d\omega/dk=v(\theta)$ and $|\omega''(k)|= u / (4 \rho_0 \sin\theta)$, matching the Lieb-Liniger Type II dispersion in the limit $c \to 0$ \cite{Kulish_1976, Ishikawa, Cesare, Syrwid_2021}. This motivates considering a Type II eigenstate in the small-$c$ limit as a GP soliton completely delocalized on the ring. It is then natural that, in order to recover a localized GP-like profile, we should superpose many Type II eigenstates.

We thus construct a wavepacket by superposing Type II eigenstates, each carrying 1 hole placed within a momentum window $\Delta k = 2\pi w/L$, $w \in \{1,2,\dots,N\}$, centered on some momentum $k_0$ within the filled Fermi interval. For small enough $\Delta k$, the value of $k_0$ fixes the group velocity of the wavepacket via $v = \frac{d\omega}{dk}\big|_{k_0}$. At finite $L$, and away from the Tonks-Girardeau regime, the observed propagation velocity differs from $v$ by an $\mathcal O(L^{-1})$ effect due to the asymmetric distribution of particles to the right of the Fermi sea. Up to this residual, the value of $k_0$ fixes the propagation velocity and thus all the properties of the target GP soliton, in particular $A$ and $\ell$.

Superposing eigenstates within $\Delta k$ localizes the position of the depletion within a spatial range of width $\Delta x \sim 2\pi/\Delta k$. If we consider a full-width summation, namely the widest possible $\Delta k = 2\pi \rho_0$ obtained for $w=N$, the group velocity cannot be treated as constant, because $v(k) = v(k_0)$ $+ v'(k_0)(k-k_0)+\frac{1}{2}v''(k_0)(k-k_0)^2+\cdots$. Since the GP soliton family is parameterized by $k$, this translates into a $k$-dependence of $\ell$ and $A$. Such a superposition thus amounts to averaging over neighboring members of the family of the GP soliton at $k_0$:
\begin{equation}
    \rho(x, t=0) \approx \rho_0-\frac{\rho_0}{\Delta k}\int_{k_0-\frac{\Delta k}{2}}^{k_0+\frac{\Delta k}{2}} dk \left(1-\frac{v^2(k)}{u^2}\right) \text{sech}^2\!\left(\frac{x-x_0}{\ell(k)}\right).
\end{equation}
While such an initial profile may still agree with the profile of a GP dark soliton of some ``effective'' width and area \cite{Sato2016}, the full-width summation carries the largest available $\Delta k$ and therefore the shortest possible dephasing time for all other parameters fixed. 

This leads us to apply the idea of selective summations introduced in \ref{sec:scaling_arg}, restricting the summation in Eq. \eqref{eq:density_profile} over $h$, $h'$ to an interval $I \subset \{1,\dots,N\}$ of width $w$, in order to reduce $\Delta k$ and thus shed neighboring solitons with $k$ significantly different from $k_0$. Indeed, reducing $\Delta k$ improves the agreement with the target GP soliton until we reach the optimal value
\begin{equation}
    \Delta x^* \sim \ell(k_0) \qquad \Longleftrightarrow\qquad \Delta k^*\sim \frac{2\pi}{\ell(k_0)}\qquad \Longleftrightarrow\qquad w^* \sim\frac{L}{\ell(k_0)} = N\sqrt\gamma \sin\theta_0.
\end{equation}
If we keep decreasing $\Delta k$ beyond this value, we are convoluting the soliton of width $\ell(k_0)$ with a kernel of width $\sim 2\pi/\Delta k$. Since the area is conserved, this produces a depletion of depth $\sim A(k_0) \Delta k/2\pi \propto \Delta k$, namely a wide and shallow profile.

Given $\Delta k$, the dephasing time of our wavepacket is
\begin{equation}
    \tau \sim \frac{1}{|\omega''(k_0)|\,\Delta k^2} = \frac{2 \sin \theta_0}{\sqrt\gamma\,\Delta k^2},
\end{equation}
and in this time it travels through a fraction of system size given by
\begin{equation}
    \frac{d}{L} \sim \frac{v\,\tau}{L} = \frac{2 \rho_0 \sin( 2\theta_0)}{\Delta k^2\,L}.
\end{equation}
If we take the optimal value $\Delta k = \Delta k^*$, we find
\begin{equation}\label{eq:deph_time}
    \tau^* \sim \frac{1}{2\pi^2\rho_0^2 \gamma^{3/2} \sin\theta_0}\qquad \text{intensive}
\end{equation}
and
\begin{equation}
    \frac{d^*}{L} \sim \frac{\cot\theta_0}{\pi^2 \rho_0\gamma\,L} \propto \frac{1}{L}.
\end{equation}
Therefore, a genuine GP-like quasi-soliton cannot traverse a large system: its dephasing time is independent of system size and therefore $d/L \propto 1/L \to 0$ for $L \to \infty$. Nevertheless, $d^*/\Delta x^* \sim \cos\theta_0/\pi^2\sqrt\gamma$; therefore in the limit $\gamma \to 0$, the GP-like quasi-soliton traverses many of its own widths before dephasing.

\begin{figure}
    \centering
    \includegraphics[width=\linewidth]{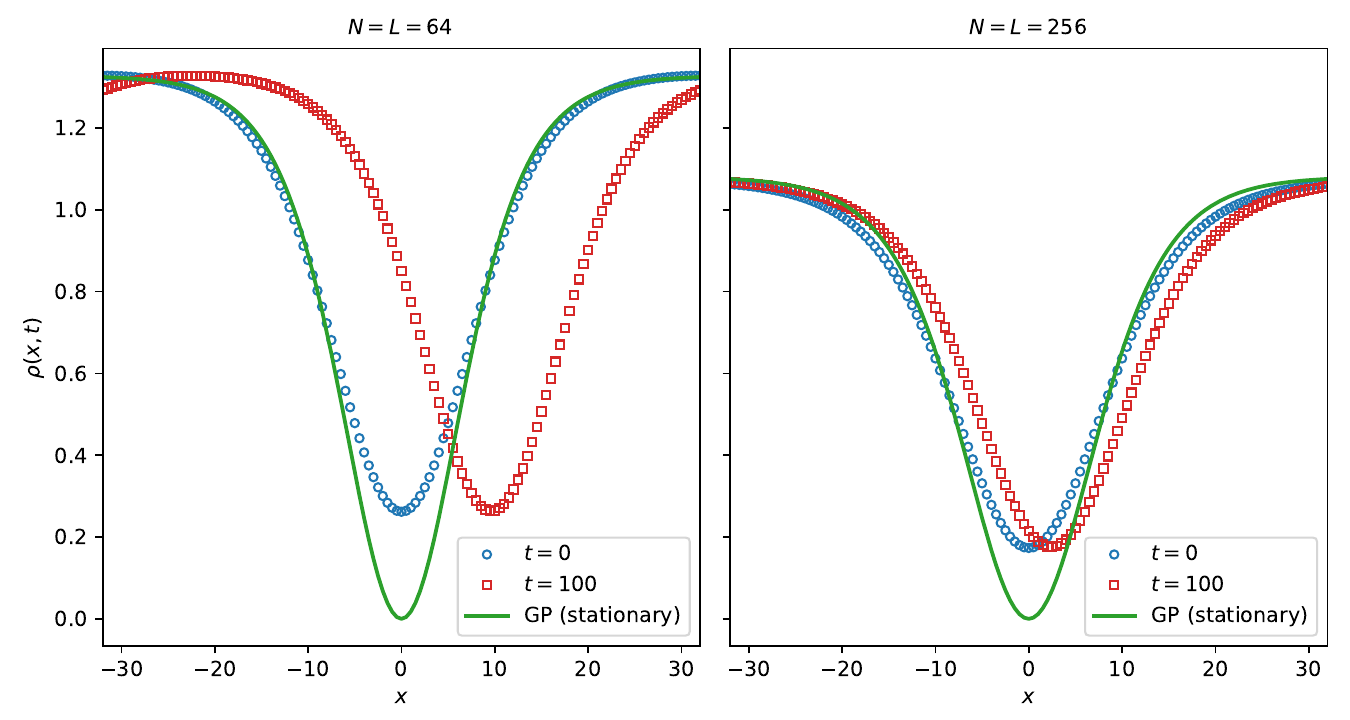}
    \caption{Comparison between selective 1-hole staggered wavepackets and analytical dark solitons of the GP equation, obtained for $c=0.01$ and two different system sizes, following the prescription given in the main text with target $v=0$.}
    \label{fig:compare_GP}
\end{figure}

According to the above discussion, we can set up a zero-parameter comparison between the density profiles of our wavepackets and that of GP dark solitons. Given $N=L$, $\gamma$, and target $\theta_0$, we should place the hole in an interval of width $w = N \sqrt \gamma \sin\theta_0$ and offset (with respect to the left Fermi edge) $o = h_0-\frac{w}{2}=\frac{k_0L}{2\pi} - \frac{w}{2} = \frac{N}{2\pi}[2\theta_0-\sin 2\theta_0]-\frac{N}{2}\sqrt\gamma\sin\theta_0$. In the GP expressions we must moreover replace $\rho_0$ everywhere by the observed background density $\rho_\text{bg}$, which accounts for the finite size of the system; since the construction conserves $N$, the depleted mass raises the plateau to $\rho_\text{bg} = \rho_0(1+A/N)$. For instance, with $N=L=64$, $\gamma=c=0.01$ and $\theta_0=\frac{\pi}{2}$ (namely $v=0$), we should take $w=6$ and $o=29$, and predict $\tau^* \sim 50$. Two examples of such comparison are shown in Fig. \ref{fig:compare_GP}. We notice that the stated prescription indeed produces a localized depletion having the correct width and area (as order of magnitude), and a lifetime which exceeds the predicted dephasing time \eqref{eq:deph_time} (for the quantitative definitions of possible lifetime measures, we refer the reader to Sec. \ref{sec:lifetime}). Moreover, as anticipated the residual propagation velocity due to the asymmetric distribution of particles to the right of the Fermi edge decreases with increasing size. 

The initial profiles in Fig. \ref{fig:compare_GP} do not show perfect agreement. Part of the discrepancy is due to the choice of uniform amplitudes; it has been shown previously that an exact match with the initial GP profile can be obtained using Gaussian amplitudes \cite{Kaminishi}. This, however, is not essential to our discussion. While previous literature has mainly focused on engineering specific initial profiles, matching for instance the GP solitons at $t=0$, less attention has been placed on their behavior under time evolution. Throughout this work, our focus will be on the structural factors that lead to a prolonged lifetime, and as will be discussed in detail in the following, the effect of the specific choice of normalized amplitudes on the lifetime is subleading with respect to other parameters, such as the width and offset of the hole window in the Fermi interval. In the present case, the spatial width of the GP soliton fixes the optimal value of $\Delta k$ in our zero-parameter set-up, which in turn yields an intensive lifetime $\tau$, which is independent of the chosen amplitudes.

\subsubsection{Selective-width wavepackets at arbitrary $c$}\label{sec:selective_width_summations}

\begin{figure}[p]
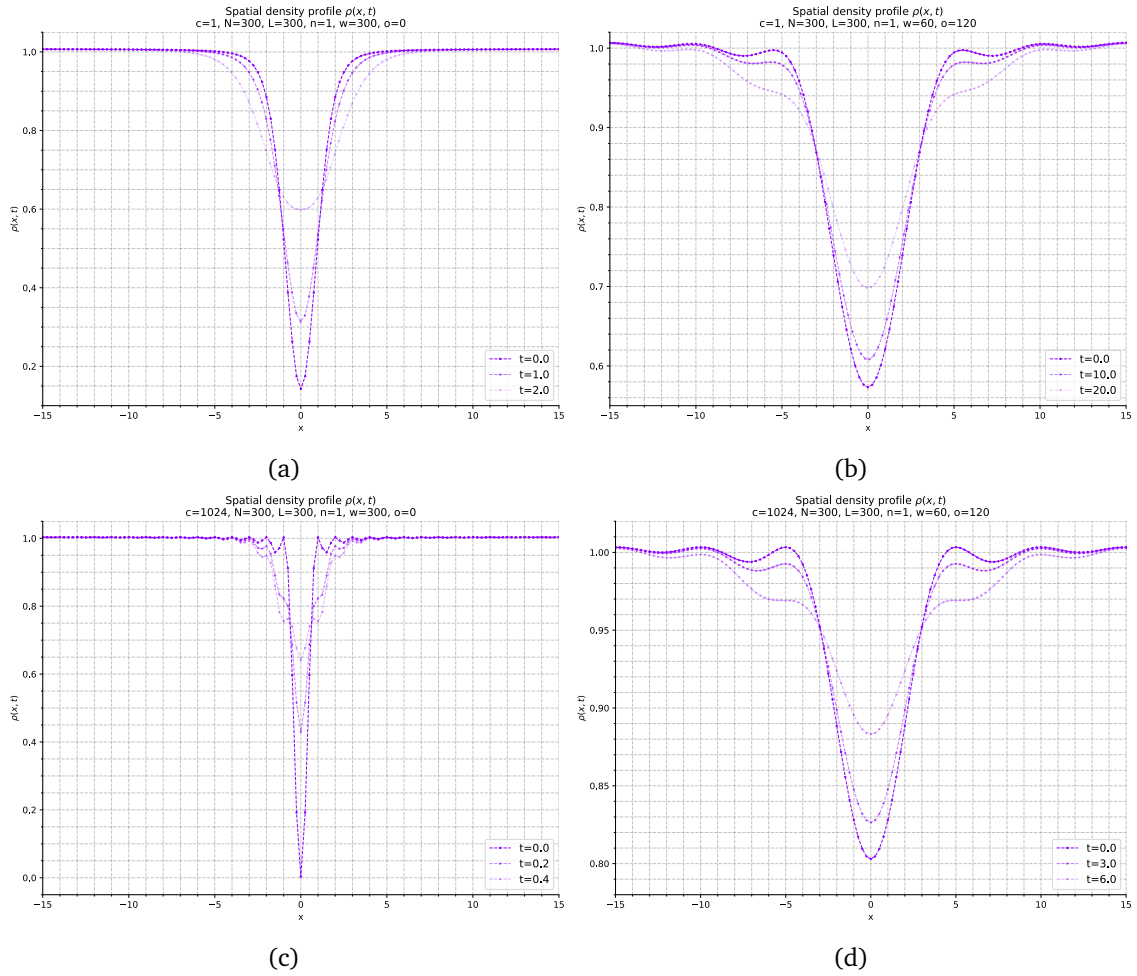

\centering
    \subfloat[]{\includesvg[width=0.5\textwidth]{sections/figures/wavepacket_profiles/uniform_amplitudes/n_1/c_1_N_300_L_300_nholes_1_width_300_offset_0_staggered_Nx_1200_Nt_1200.svg}\label{fig:c_1_N_300_n_1_w_300_o_0_staggered}}
    \hfill
    \subfloat[]{\includesvg[width=0.5\textwidth]{sections/figures/wavepacket_profiles/uniform_amplitudes/n_1/c_1_N_300_L_300_nholes_1_width_60_offset_120_staggered_Nx_1200_Nt_1200.svg}\label{fig:c_1_N_300_n_1_w_60_o_120_staggered}}
    \hfill
    \subfloat[]{\includesvg[width=0.5\textwidth]{sections/figures/wavepacket_profiles/uniform_amplitudes/n_1/c_1024_N_300_L_300_nholes_1_width_300_offset_0_staggered_Nx_1200_Nt_1200.svg}\label{fig:c_1024_N_300_n_1_w_300_o_0_staggered}}
   \subfloat[]{\includesvg[width=0.5\textwidth]{sections/figures/wavepacket_profiles/uniform_amplitudes/n_1/c_1024_N_300_L_300_nholes_1_width_60_offset_120_staggered_Nx_1200_Nt_1200.svg}\label{fig:c_1024_N_300_n_1_w_60_o_120_staggered}}
    \caption[width=0.5\textwidth]{Time evolution of the full-width (left column) and selective-width (right column) $n=1$-hole staggered wavepackets at $c=1$ (top row) and $c=1024$ (bottom row), plotted over 1/10th of system size $L=300$. Note the differences in the scales of the vertical axis and plotted times.}\label{fig:c_1_1024_n_1_time_evol}
\end{figure}

Let us now apply our selective summation approach to wavepackets at arbitrary interaction strength. As discussed in Sec. \ref{sec:scaling_arg}, given a nonlinear dispersion, full-width summations result in the largest spread among velocities of the constituent eigenmodes, thereby significantly limiting the wavepacket's lifetime. Decreasing $w$ leads to an increased lifetime, as shown in Fig. \ref{fig:c_1_1024_n_1_time_evol} and demonstrated more quantitatively in Sec. \ref{sec:lifetime}, where we also demonstrate scaling of lifetime with $w^2$ for the 1-hole wavepacket at finite $c$.

\begin{figure}[p]
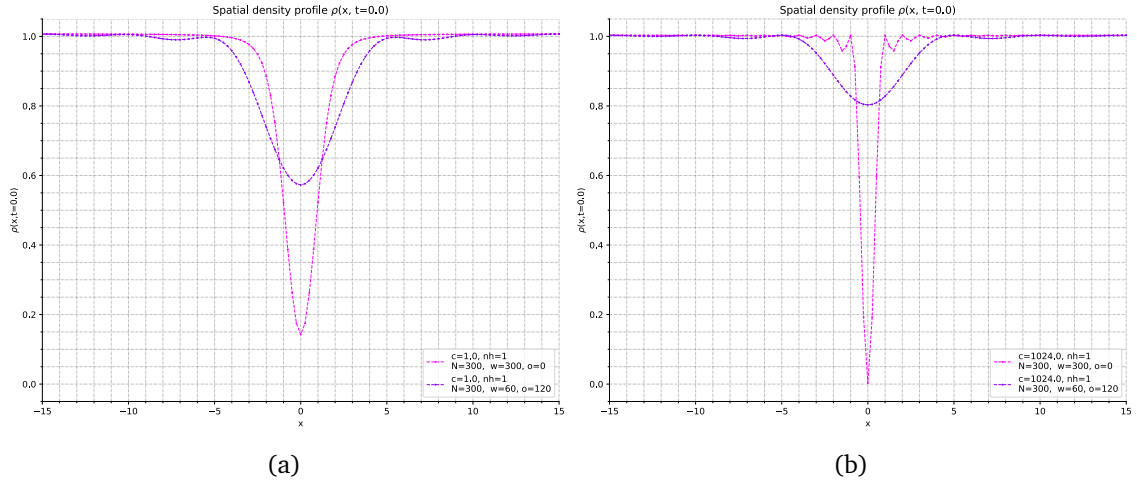

\centering
    \subfloat[]{\includesvg[width=0.5\textwidth]{sections/figures/wavepacket_profiles/uniform_amplitudes/n_1/tStar_0p0_c_1_N_300_L_300_nholes_1_width_300_60_offset_0_120_staggered_Nx_1200_Nt_1200.svg}\label{fig:c_1_N_300_n_1_w_300_60_staggered_tStar}}
    \hfill
    \subfloat[]{\includesvg[width=0.5\textwidth]{sections/figures/wavepacket_profiles/uniform_amplitudes/n_1/tStar_0p0_c_1024_N_300_L_300_nholes_1_width_300_60_offset_0_120_staggered_Nx_1200_Nt_1200.svg}\label{fig:c_1024_N_300_n_1_w_300_60_staggered_tStar}}
    \caption{Initial profiles of the full-width and selective-width one-hole staggered wavepackets at $c=1$ (\ref{fig:c_1_N_300_n_1_w_300_60_staggered_tStar}) and $c=1024$ (\ref{fig:c_1024_N_300_n_1_w_300_60_staggered_tStar}), plotted over 1/10th of system size $L=300$.}\label{fig:c_1_1024_n_1_t0}
\end{figure}

Let us begin with the initial profile of the one-hole staggered wavepacket for both $c >> 1$ and $c \sim \mathcal{O}(1)$\footnote{As for drift speed at finite-$c$ discussed in \ref{sec:c_inf_wavepackets}, we note that the time scales displayed in the subsequent plots are too small to show any appreciable displacement.} (Fig. \ref{fig:c_1_1024_n_1_t0}): As foreseen, the selective-width wavepacket has a shallower profile than its full-width counterpart, but proves to be advantageous in lifetime: at $c=1$ the selective wavepacket (Fig. \ref{fig:c_1_N_300_n_1_w_60_o_120_staggered}) loses only about 30\% of its initial depth by $t=20$, whereas the full-width wavepacket (Fig. \ref{fig:c_1_N_300_n_1_w_300_o_0_staggered})
loses over 50\% of its initial depth in an order-of-magnitude smaller period of time. Similar results hold at $c\sim\mathcal{O}(10^3)$ (Figs. \ref{fig:c_1024_N_300_n_1_w_300_o_0_staggered} and \ref{fig:c_1024_N_300_n_1_w_60_o_120_staggered}), with the overall time scales decreased by an order of magnitude due to the effect of interaction on lifetime: as discussed above and shown in Fig. \ref{fig:typeII_inv_curv}, wavepackets at small $c$ enjoy longer lifetimes, in accordance with semi-classical expectations.

As discussed in Sec. \ref{sec:scaling_arg}, the reduced summation width $w$ puts a new parameter at our disposal: the placement of the summation window, centered at a point $k_0$ within the original Fermi sea. We discuss the effects of this parameter below. 

\subsubsection*{Placement of summation interval: effect on lifetime, velocity, and initial profile}
The placement of the summation interval $I$ affects the lifetime, propagation speed, and initial profile of the wavepacket. As we outline below, these effects become (non-)negligible depending on interaction strength.

\begin{figure}
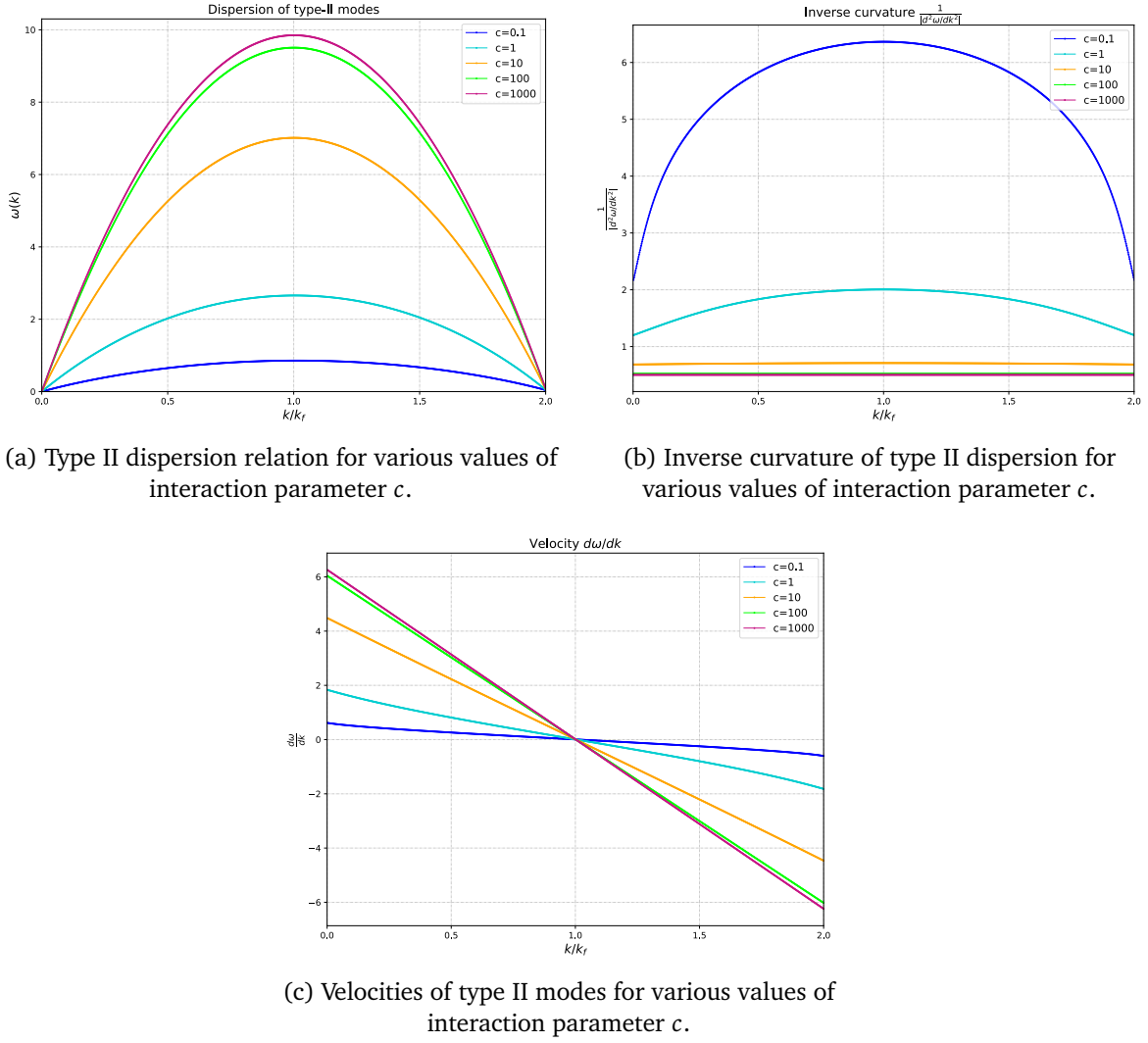

    \begin{subfigure}{0.5\textwidth}
        \includesvg[width=\linewidth]{sections/figures/intro_figs/disp_plt.svg}
        \centering
        \subcaption{\centering Type II dispersion relation for various values of interaction parameter $c$.} 
        \label{fig:typeII_disp}
    \end{subfigure}
    \hspace{0.3cm}
    \begin{subfigure}{0.5\textwidth}
        \includesvg[width=\linewidth]{sections/figures/intro_figs/inv_curv_plt.svg}
        \centering
        \subcaption{\centering Inverse curvature of type II dispersion for various values of interaction parameter $c$.} 
        \label{fig:typeII_inv_curv}
    \end{subfigure}
    \begin{center}
        \begin{subfigure}{0.5\textwidth}
        \includesvg[width=\linewidth]{sections/figures/intro_figs/velos_plt.svg}
        \subcaption{\centering Velocities of type II modes for various values of interaction parameter $c$.} 
        \label{fig:typeII_velo}
    \end{subfigure} 
    \end{center}
    \caption{Properties of the type II dispersion relation.}
\end{figure}

At finite-$c$, the lifetime of the wavepacket (at fixed $w$) improves by positioning $I$ symmetrically with respect to the center of the Fermi sea. This is again clearly reflect in the curvature of type-II modes $\omega''$, which we recall are related to the lifetime $\tau$ of the wavepacket with momentum width $\Delta k$ as $\tau \sim \frac{1}{\omega_0''(\Delta k)^2}$, where $\omega_0''$ corresponds to the curvature at the center of the interval $\Delta k$ (see Fig. \ref{fig:intro_omega_diag_inset}). As shown in Fig. \ref{fig:typeII_inv_curv}, $c \sim \mathcal{O}(1)$ and below demonstrate a variation in curvature as a function of momentum $k$, with a peak in inverse curvature at the center of the Fermi sea, corresponding to the optimal placement of $I$ for maximal lifetime. This advantage shrinks as we approach the Tonks-Girardeau regime, where the curvature becomes nearly constant, and the lifetime of the wavepacket becomes independent of the position of $I$ within the Fermi sea. We show this effect quantitatively for both $n=1$- and $n=2$-hole wavepackets---using two independent measures of lifetime---in Sec. \ref{sec:lifetime}.

We note that in the Tonks-Girardeau regime, if $I$ is symmetric with respect to the center of the Fermi sea (i.e., symmetric with respect to $(N+n)/2$), the resulting wavepacket is still produced at rest. Introducing an offset from this symmetric position (e.g. displacing $I$ from middle of Fermi sea to the edge) imparts a drastic boost to the wavepacket, while same displacement in $I$ has a much milder effect on the velocity at sub-$\mathcal{O}(1)$ interaction strengths. This is easily understood by looking at the slopes of the type-II dispersion ($\frac{d\omega}{dk}$) as depicted in Fig. \ref{fig:typeII_velo}: velocities span a wide range in the limit $c \rightarrow \infty$ but nearly vanish at smaller c, where the dispersion curve flattens out significantly. Fig. \ref{fig:typeII_velo} also demonstrates the ballistic nature of propagation velocities in the Tonks-Girardeau regime, which softly deviate from linearity in the $c\rightarrow 0$ limit. 
We further note that at very small interactions ($c \sim \mathcal{O}(0.1)$), the minimum density of the wavepacket (at fixed summation width) depends on the placement of the summation interval: with $I$ placed at the center of the Fermi sea (and thus a smaller velocity) produces a deeper minimum than placing $I$ at the edge of the sea, which produces a shallower, faster-propagating wavepacket. This is once again in-line with mean-field calculations: as shown in Sec. \ref{sec:wavepackets_finite_c}, dark solitons of the Gross-Pitaevskii equation demonstrate the same relation between depth and velocity.

\subsection{Multiple-hole wavepackets}\label{sec:multi_holes}

Going beyond single-hole wavepackets, another parameter at our disposal is the number of holes in a given type-II eigenstate. We wish to study the effect of increasing the number of holes $n$ on the spatial profile and lifetime of the wavepacket. Using Eqs. \eqref{eq:rhon_TG} and \eqref{eq:unif_amp_coef}, we obtain an expression for the density profile of uniform-amplitude $n$-hole wavepackets in the Tonks-Girardeau regime:
\begin{equation}
    \rho^{(TG)}_n(x,t) = \frac{N}{L} + \frac{1}{\mathcal N L} \sum_{\substack{\bar h_1,\dots, \bar h_{n-1},\\{h,h'}}}e^{-\frac{2\pi i}{L}(h'-h)[x-x_0+\frac{L}{2} -\frac{2\pi}{L}(N+1-h'-h)t]}(-1)^{\tilde M_{h,h'}+1}.
\end{equation}
As shown in the plots below, increasing the number of holes (within the uniform-amplitude formulation)  yields a density profile with $n$ localized depletions. As we will see in Secs. \ref{sec:time_tables} and \ref{sec:R_t_plots}, this configuration also leads to an improvement in lifetime. In the Tonks-Girardeau regime, both full and selective summations with $I$ centered in the middle of the Fermi sea produce a stationary density profile, as they do in the single-hole case. Throughout the evaporation process, uniform-amplitude wavepackets in the $c \rightarrow\infty$ limit remain symmetric with respect to the center of mass of the depletion. At finite $c$, wavepackets can exhibit asymmetric evaporation as a function of the (a)symmetry of the summation interval and the distribution of particles in momentum space. As with the single-hole case, we have compared our analytical results with numerical data obtained using ABACUS for $n=\{2,~3, ~4\}$ and $c\gg 1$, and we find that they agree to within $\mathcal O(c^{-1})$ (App. \ref{app:exact_vs_num}) throughout the wavepacket's lifetime.

\subsubsection*{n=2-hole wavepackets}
\begin{figure}[H]
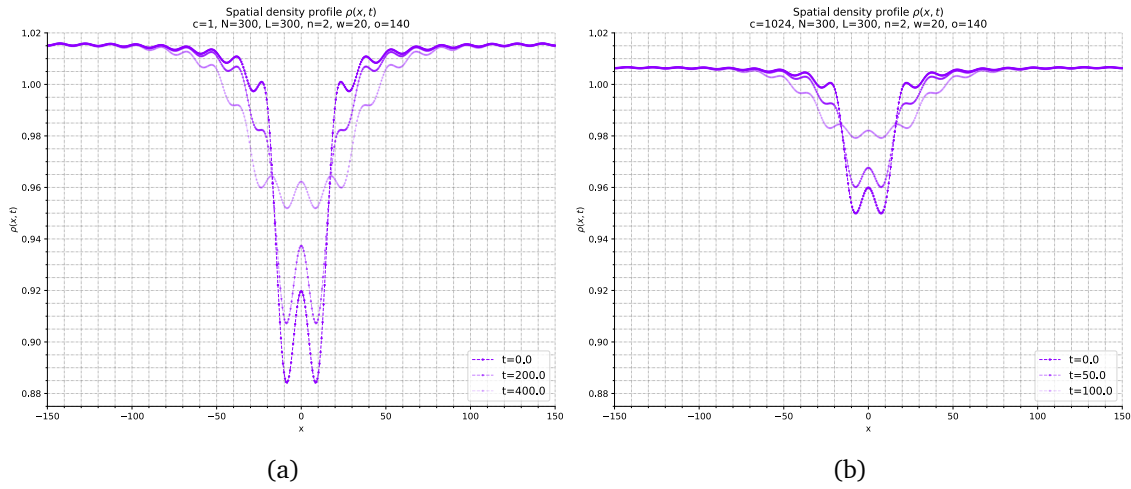

\centering
    \subfloat[]{\includesvg[width=0.5\textwidth]{sections/figures/wavepacket_profiles/uniform_amplitudes/n_2/c_1_N_300_L_300_nholes_2_width_20_offset_140_staggered_Nx_500_Nt_500_tmax_400.svg}\label{fig:c_1_N_300_n_2_w_20_o_140_stag}}
    \subfloat[]{\includesvg[width=0.5\textwidth]{sections/figures/wavepacket_profiles/uniform_amplitudes/n_2/c_1024_N_300_L_300_nholes_2_width_20_offset_140_staggered_Nx_500_Nt_500_tmax_100.svg}\label{fig:c_1024_N_300_n_2_w_20_o_140_stag}}
     \caption{Time evolution of $n=2$-hole staggered wavepackets at $c=1$ (\ref{fig:c_1_N_300_n_2_w_20_o_140_stag}) and $c=1024$ (\ref{fig:c_1024_N_300_n_2_w_20_o_140_stag}), $N=L=300$, width = 20, offset = 140.}
     \label{fig:c_1_1024_N_300_n_2_w_20_o_140_stag}
\end{figure}
As shown in Fig. \ref{fig:c_1_1024_N_300_n_2_w_20_o_140_stag}, summation over two-hole eigenstates using the uniform-amplitude formulation produces a double-dip initial profile at both small and large values of interaction. However, the same set of parameters at $c=1$ produces a deeper, longer lived wavepacket than its large-$c$ counterpart (note that the final plotted time in Fig. \ref{fig:c_1_N_300_n_2_w_20_o_140_stag} is four times larger than that of Fig. \ref{fig:c_1024_N_300_n_2_w_20_o_140_stag}). These characteristics further extend to higher-$n$ wavepackets. 

\subsubsection*{n=3-, 4-hole wavepackets}
\begin{figure}[H]
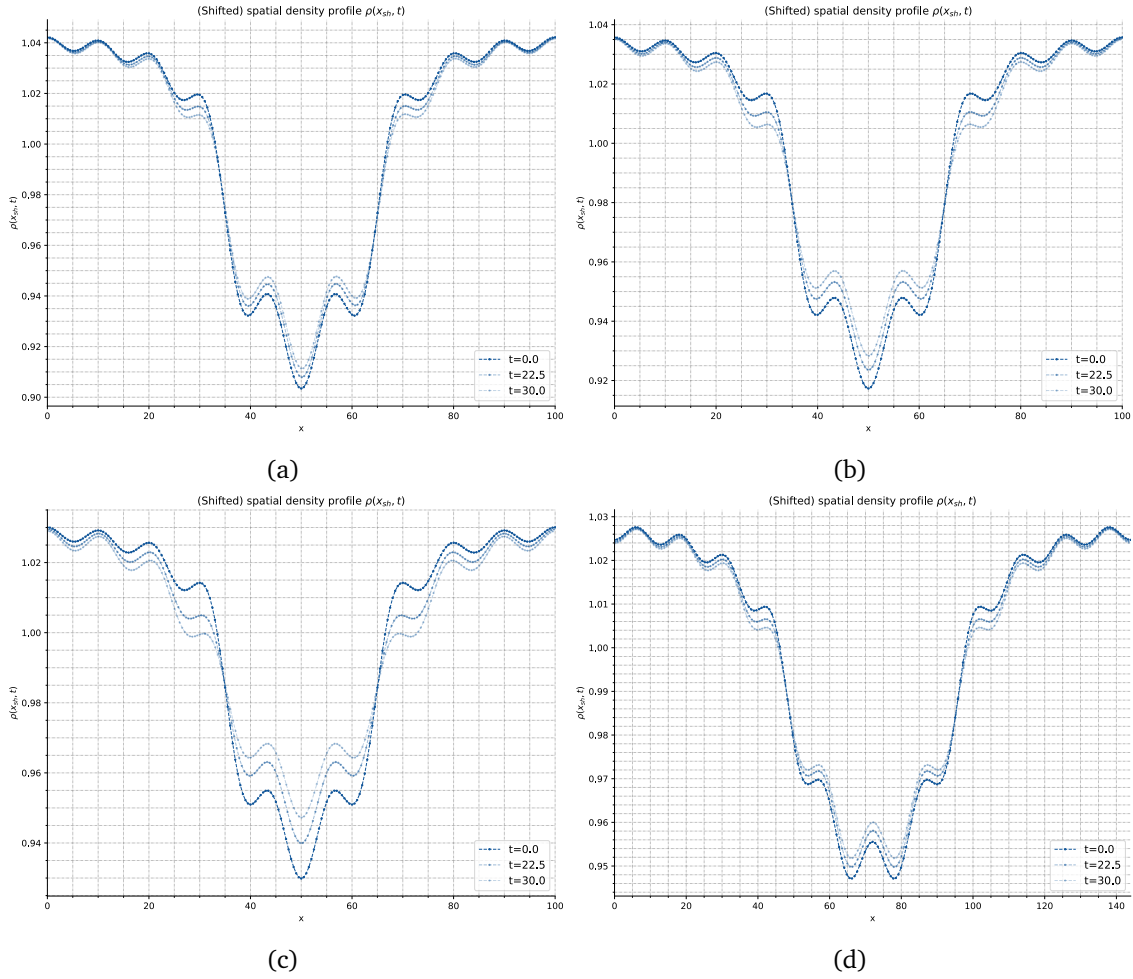

\centering
    \subfloat[]{\includesvg[width=0.5\textwidth]{sections/figures/wavepacket_profiles/uniform_amplitudes/n_3/c_4_N_100_L_100_nholes_3_width_10_Nx_200_Nt_200_tmax_30_shifted.svg}\label{fig:uni_amp_c_4_n_3_tStar}}
    \hfill
    \subfloat[]{\includesvg[width=0.5\textwidth]{sections/figures/wavepacket_profiles/uniform_amplitudes/n_3/c_10_N_100_L_100_nholes_3_width_10_Nx_200_Nt_200_tmax_30_shifted.svg}\label{fig:uni_amp_c_10_n_3_tStar}}
    \hfill
    \subfloat[]{\includesvg[width=0.5\textwidth]{sections/figures/wavepacket_profiles/uniform_amplitudes/n_3/c_1024_N_100_L_100_nholes_3_width_10_Nx_200_Nt_200_tmax_30_shifted.svg}\label{fig:uni_amp_c_1024_n_3_tStar}}
   \subfloat[]{\includesvg[width=0.5\textwidth]{sections/figures/wavepacket_profiles/uniform_amplitudes/n_4/c_10240_N_144_L_144_nholes_4_width_12_Nx_200_Nt_200_tmax_30_shifted.svg}\label{fig:uni_amp_c_10240_n_4_tStar}}
    \caption[width=0.5\textwidth]{Time evolution of density profile of $n=3$-hole uniform-amplitude wavepacket with $c=\{4, 10, 1024\}$ (\ref{fig:uni_amp_c_4_n_3_tStar}, \ref{fig:uni_amp_c_10_n_3_tStar}, \ref{fig:uni_amp_c_1024_n_3_tStar}) at $~N=L=100$, width = 10, offset = 0. \ref{fig:uni_amp_c_10240_n_4_tStar}: Shifted density profile of $n=4$-hole uniform-amplitude wavepacket at $t=\{0,22.5,30\}$, $c=10240, ~N=L=144$, width = 12, offset = 0.}
\end{figure}

Here, we note again the expected improvement in lifetime in the weak interaction regime, in line with our observation of increasingly long-lived $n=1$-hole uniform amplitude wavepackets at small $c$. As shown in Figs. \ref{fig:uni_amp_c_4_n_3_tStar}, \ref{fig:uni_amp_c_10_n_3_tStar}, and \ref{fig:uni_amp_c_1024_n_3_tStar}, decreasing $c$ results in a slightly deeper depletion and markedly slower evaporation in the $n=3$-hole wavepacket, extending the previously obtained results to higher particle-hole wavepackets. 

We also note that the effect of offset on initial profile is a function of interaction strength: at very weak interactions ($c\sim \mathcal{O}(10^{-1})$), the same summation interval can produce wavepackets with a different minimum, depending on the placement of the interval within the Fermi sea: the summation interval centered in the middle of the sea produces a deeper minimum (and a smaller velocity, as expected) than an interval at the edge of the sea, which produces a shallower, faster-propagation wavepacket.

This is once again in line with expected behavior of Gross-Pitaevskii in the mean-field regime, where the minimum value of the depletion is proportional to its propagation velocity (see Sec. \ref{sec:wavepackets_finite_c}.) Here, we see an extension of this effect to multiple-hole wavepackets.
Let us now compare the evolution of single and multiple-hole wavepackets over time.

\subsection{Comparison of multi-hole wavepacket profiles}
\begin{figure}[H]
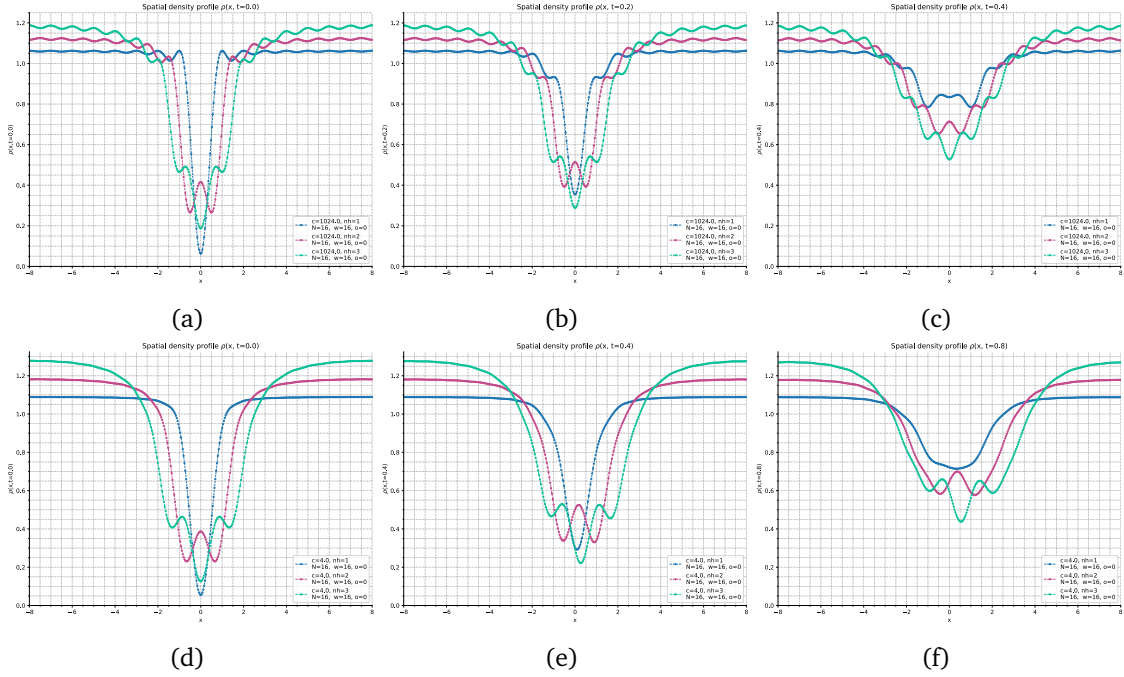

 \begin{flushleft}
    \subfloat[]{\includesvg[width=0.33\linewidth]{sections/figures/wavepacket_profiles/profile_comparisons/nholes_comparisons/c_1024/tStar_0p0_c_1024_N_16_L_16_nholes_1_2_3_width_16_offset_0_staggered_Nx_400_Nt_100_tmax_2.svg}}
    \subfloat[]{\includesvg[width=0.33\linewidth]{sections/figures/wavepacket_profiles/profile_comparisons/nholes_comparisons/c_1024/tStar_0p2_c_1024_N_16_L_16_nholes_1_2_3_width_16_offset_0_staggered_Nx_400_Nt_100_tmax_2.svg}}
    \subfloat[]{\includesvg[width=0.33\linewidth]{sections/figures/wavepacket_profiles/profile_comparisons/nholes_comparisons/c_1024/tStar_0p4_c_1024_N_16_L_16_nholes_1_2_3_width_16_offset_0_staggered_Nx_400_Nt_100_tmax_2.svg}}\\
    \subfloat[]{\includesvg[width=0.33\linewidth]{sections/figures/wavepacket_profiles/profile_comparisons/nholes_comparisons/c_4/tStar_0p0_c_4_N_16_L_16_nholes_1_2_3_width_16_offset_0_staggered_Nx_400_Nt_100_tmax_2.svg}}
    \subfloat[]{\includesvg[width=0.33\linewidth]{sections/figures/wavepacket_profiles/profile_comparisons/nholes_comparisons/c_4/tStar_0p4_c_4_N_16_L_16_nholes_1_2_3_width_16_offset_0_staggered_Nx_400_Nt_100_tmax_2.svg}}
    \subfloat[]{\includesvg[width=0.33\linewidth]{sections/figures/wavepacket_profiles/profile_comparisons/nholes_comparisons/c_4/tStar_0p8_c_4_N_16_L_16_nholes_1_2_3_width_16_offset_0_staggered_Nx_400_Nt_100_tmax_2.svg}}
\end{flushleft}
\caption{Time evolution of $n=1,2,3$-hole full-width staggered wavepackets at $c=1024$ (top) and $c=4$ (bottom).}
\label{fig:c_4_1024_n_1_2_3_stag}
\end{figure}
As shown in Fig. \ref{fig:c_4_1024_n_1_2_3_stag}, the $n=3$-hole wavepacket keeps its shape longer than its lower-$n$ counterparts at both $c=4$ and $c=1024$, displaying a marked improvement in lifetime as compared with the $n=1$ case. In Secs. \ref{sec:time_tables} and \ref{sec:R_t_plots}, we show this improvement of lifetime with $n$ quantitatively.
We also note that many of the $n=1$ wavepacket characteristics extend to higher $n$:\footnote{
The introduction of higher particle-hole pairs leads to some interesting observations, including the occurrence of ``fixed points'' in wavepacket profiles---points at which the density remains constant---under certain combinations of parameters. Given that the change in density is expected to propagate throughout the system, we find it worthwhile to take note of such cases, and give an analytic derivation of such points of fixed density for single-hole wavepackets in the Tonks-Girardeau regime in Appendix \ref{app:fixed_pts}.} for instance, we continue to observe Friedel-like oscillations at large $c$ while the smaller $c$ wavepackets yield slightly deeper, smoother profiles.

Let us note here that while increasing $n$ improves lifetime, the resulting multiple-dip profile becomes increasingly wide in real space: above $n \sim 5$ holes, the wavepacket (including the variational type, discussed next) becomes too wide with respect to system size to be considered a classical-like (quasi-)soliton. Therefore, throughout the remainder of the work, we focus on $n \leq 3$.

Seeing as increasing $n$ introduces a wider profile with multiple localized depletions, a natural question arises next: can we build multiple-hole wavepackets which retain the narrow, single-depletion profile of the $n=1$ case, using a different set of coefficients than uniform amplitudes? In the next section, we address this question using a variational approach.
\section{Variational amplitudes}
As we have seen so far, uniform-amplitude coefficients can yield long-lived wavepackets with a lifetime that increases as a function of $n$ and $1/w^2$. Evaporation of these wavepackets is for the most part rigid and symmetric about the center of depletion. However, $n>1$ produces wider spatial profiles with multiple localized depletions. It is therefore natural to ask: to what extent can we control the shape of the initial profile, especially in the case of $n>1$ wavepackets? To this end, we introduce a variational approach to obtaining the amplitudes.

First, let us define the quantity to be optimized. Among a plethora of possibilities, we consider a simple and easily evaluated one; namely, the density integrated within a specified spatial interval $\Delta x > 0$ at $t=0$. If we represent the $t=0$ wavefunction as $|\Psi \rangle$, we are thus looking to minimize the quantity
\begin{equation}
I_\Psi (\Delta x) \equiv \frac{1}{\Delta x} \int_{-\Delta x/2}^{\Delta x/2} dx \langle \Psi | \hat{\rho}(x) | \Psi \rangle.
\end{equation}
Let the state $|\Psi \rangle$ be expressed as a linear combination of eigenstates (in any chosen sub-basis) as
\begin{equation*}
~| \Psi \rangle = \sum_\alpha \psi_\alpha | \alpha \rangle.
\end{equation*}
Since
\begin{equation*}
\rho(x, t=0) = \sum_{\alpha_1, \alpha_2} \psi^*_{\alpha_1} \psi_{\alpha_2} e^{-i k_{\alpha_1 \alpha_2} x} \langle \alpha_1 | \hat{\rho}(0) | \alpha_2 \rangle,
\end{equation*}
in which $k_{\alpha_1 \alpha_2} \equiv K_{\alpha_1} - K_{\alpha_2}$, we get
\begin{equation*}
I_\Psi (\Delta x) = \rho_0 + \sum_{\alpha_1, \alpha_2} \psi^*_{\alpha_1} M_{\alpha_1 \alpha_2} \psi_{\alpha_2},
\end{equation*}
where $\rho_0 = N/L$ is the background density, ${\boldsymbol \psi}$ is the vector containing the amplitudes and ${\boldsymbol M}$ is the real symmetric matrix (with zeros on the diagonal):
\[
{\boldsymbol M}_{\alpha_1 \alpha_2} (\Delta x) \equiv (1 - \delta_{\alpha_1 \alpha_2}) ~\langle \alpha_1 | \hat{\rho}(0) | \alpha_2 \rangle ~\mbox{sinc} (k_{\alpha_1 \alpha_2} \Delta x/2),
\]
and $\mbox{sinc}(x) \equiv \frac{\sin(x)}{x}$. 
The optimal value of ${\boldsymbol \psi}$ is then simply given by the eigenvector of ${\boldsymbol M}$ with the lowest eigenvalue. In this set-up, $\Delta x$ remains a free parameter: we may choose it individually for each wavepacket to yield the desired profile.

\subsection{Initial profile}
\begin{figure}[H]
\centering
\includesvg[width=\linewidth]{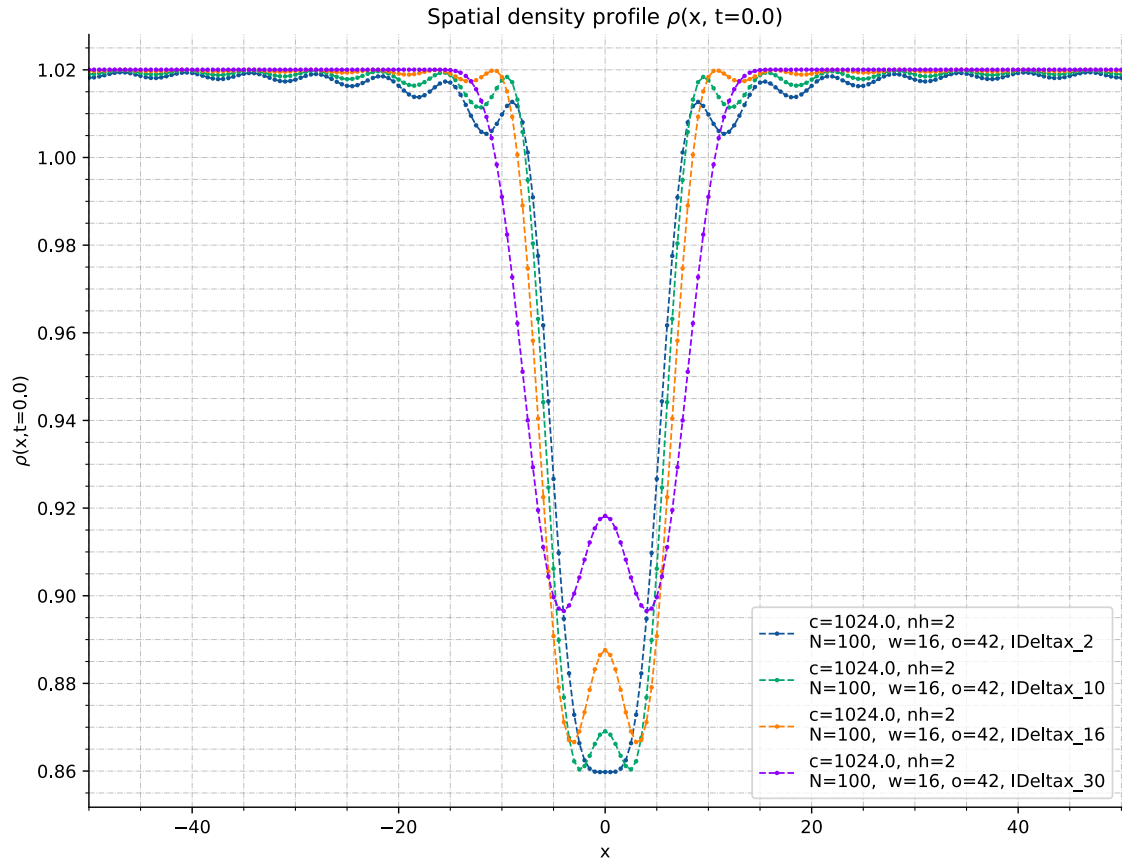}
 \caption{Effect of $\Delta x$ on the initial profile of the $n=2$-hole variational wavepacket at $c=1024, ~N=L=100$, $w=16$, $o=42$.}
\label{fig:c_1024_n_2_IDeltax_comparison}
\end{figure}

Using variational coefficients, a suitable choice of the parameter $\Delta x$ allows us to construct a multi-hole wavepacket with a single-depletion initial profile: in other words, an $n-$hole wavepacket resembling a one-hole wavepacket at $t=0$. Suitable choice(s) of $\Delta x$ vary with system parameters such as $c$ and $n$. Fig. \ref{fig:c_1024_n_2_IDeltax_comparison} shows the effect of $\Delta x$ for a fixed set of parameters. Figs. \ref{fig:c_1024_n_2_init_stag_var} and \ref{fig:c_1024_n_3_init_stag_var} show two examples of variational wavepackets in comparison with their staggered counterparts. 

\begin{figure}[H]
    \centering
    \begin{subfigure}{0.49\textwidth}
    \includesvg[width=\linewidth]{sections/figures/wavepacket_profiles/profile_comparisons/tStar_0p0_c_1024_N_100_L_100_nholes_2_width_16_offset_0_staggered_IDeltax_2.svg}
    \subcaption[width=0.2\textwidth]{Initial profile  of $n=2$-hole staggered (blue) and variational (purple) wavepackets with $c=1024$, $N=L=100$, $w=16$, $o=0$, $\Delta x=2$ for variational wavepacket.}
    \label{fig:c_1024_n_2_init_stag_var}
    \end{subfigure}
    \hspace{0.005\textwidth}
    \begin{subfigure}{0.49\textwidth}
    \includesvg[width=\linewidth]{sections/figures/wavepacket_profiles/profile_comparisons/tStar_0p0_c_1024_N_100_L_100_nholes_3_width_16_offset_42_staggered_Deltax_2.svg}
    \subcaption[width=0.2\textwidth]{Initial profile  of $n=3$-hole staggered (blue) and variational (green) wavepackets with $c=1024$, $N=L=100$, $w=16$, $o=42$, $\Delta x=2$ for variational wavepacket.}
    \label{fig:c_1024_n_3_init_stag_var}
    \end{subfigure}
    \caption{}
\end{figure}

\subsection{Time evolution}
\begin{figure}[H]
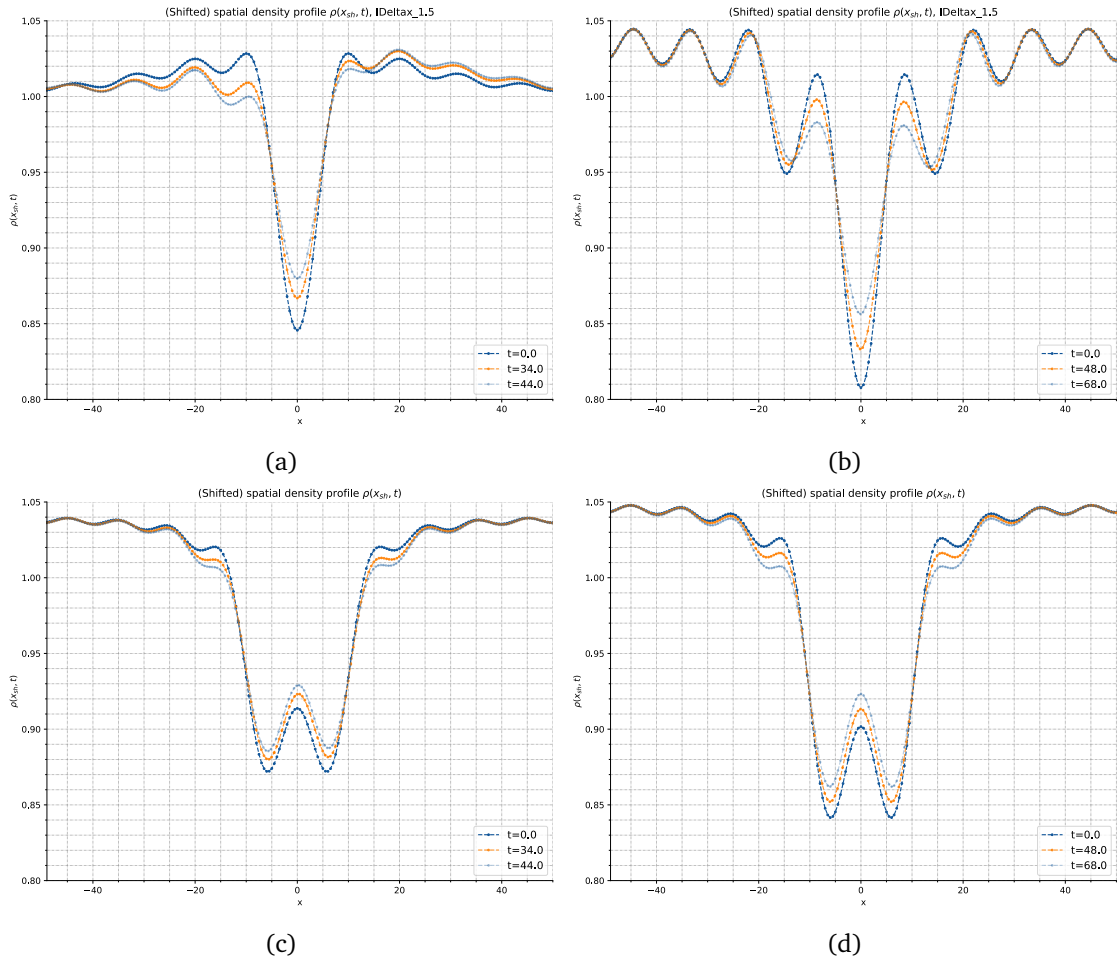

\centering
\begin{subfigure}{0.49\textwidth}
        \includesvg[width=\linewidth]{sections/figures/wavepacket_profiles/variational_amplitudes/n_2/c_1_N_100_L_100_nholes_2_width_10_offset_0_IDeltax_1p5_Nx_200_Nt_200_tmax_200_lf_v3.svg}
            \caption{}
        \label{fig:c_1_n_2_o_0_var}
\end{subfigure}
\begin{subfigure}{0.49\textwidth}
        \includesvg[width=\linewidth]{sections/figures/wavepacket_profiles/variational_amplitudes/n_2/c_1_N_100_L_100_nholes_2_width_10_offset_45_IDeltax_1p5_Nx_200_Nt_200_tmax_200_lf_v2.svg}
        \caption{}
        \label{fig:c_1_n_2_o_45_var}
\end{subfigure}

\centering
\begin{subfigure}{0.49\textwidth}
        \includesvg[width=\linewidth]{sections/figures/wavepacket_profiles/uniform_amplitudes/n_2/c_1_N_100_L_100_nholes_2_width_10_offset_0_staggered_Nx_200_Nt_200_tmax_200_lf_v2}
        \caption{}
        \label{fig:c_1_n_2_o_0_stag}
\end{subfigure}
\begin{subfigure}{0.49\textwidth}
        \includesvg[width=\linewidth]{sections/figures/wavepacket_profiles/uniform_amplitudes/n_2/c_1_N_100_L_100_nholes_2_width_10_offset_45_staggered_Nx_200_Nt_200_tmax_200_lf_v2}
        \caption{}
        \label{fig:c_1_n_2_o_45_stag}
\end{subfigure}
  \caption{Profile of $n=2$-hole variational (top) and staggered (bottom) wavepackets at $c=1, ~N=L=100$, $w=10$, $o=0$ (left column) and $o=45$ (right column) $\Delta x=1.5$ for variational wavepacket.}
    \label{fig:c_1_n_2_o_0_o_45_stag_var}
\end{figure}

The next natural question to ask is how such single-depletion profiles evolve in time, and specifically, how their time evolution compares with that of uniform-amplitude wavepackets.
\begin{figure}[H]
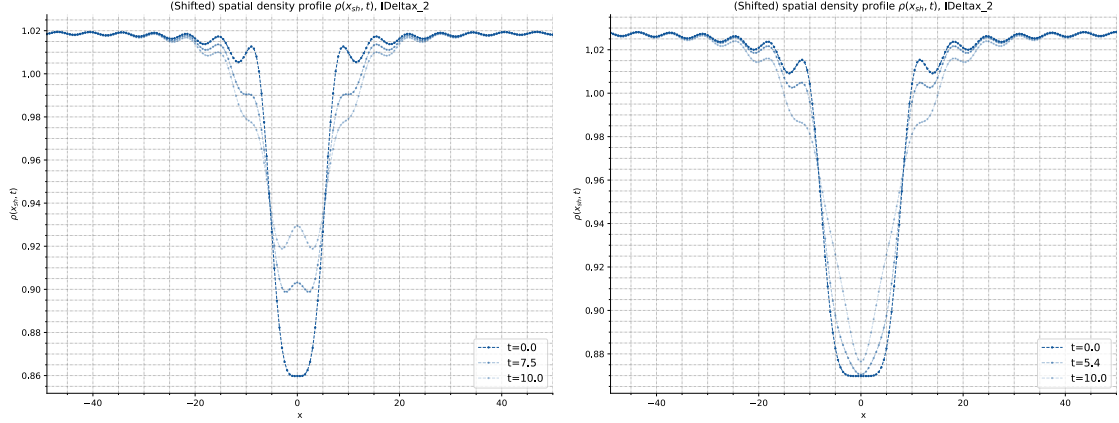

\centering
\begin{subfigure}{0.49\textwidth}
        \includesvg[width=\linewidth]{sections/figures/wavepacket_profiles/variational_amplitudes/n_2/c_1024_N_100_L_100_nholes_2_width_16_offset_42_IDeltax_2_Nx_200_Nt_200_tmax_10_shifted.svg}
\end{subfigure}
\begin{subfigure}{0.49\textwidth}
        \includesvg[width=\linewidth]{sections/figures/wavepacket_profiles/variational_amplitudes/n_3/c_1024_N_100_L_100_nholes_3_width_16_offset_42_IDeltax_2_Nx_200_Nt_200_tmax_30_shifted.svg}
\end{subfigure}
    \caption{Time evolution of $n=2$ (left) and $n=3$-hole (right) variational wavepackets at $c=1024, ~N=L=100$, $w=16$, $o=42$, $\Delta x=2$.}\label{fig:c_1024_n_2_3_var_tStar}
\end{figure}
As shown in Fig. \ref{fig:c_1024_n_2_3_var_tStar}, the variational method produces a flatter, wider depletion (at fixed $\Delta x$) for the $n=3$-hole wavepacket than its $n=2$ counterpart. This may in part be due to the fact that $\Delta x < n$ for the former, though this is not always the case (see e.g. Fig. \ref{fig:c_1_n_2_o_0_o_45_stag_var} with $\Delta x=1.5$, $n=2$, which produces a narrow profile). In both cases, however, the initial profile quickly loses its single-depletion character: in the $n=2$ case, we see a splitting of the depletion with time (as we observe also with the $n=1$ staggered wavepackets at large $c$; see e.g. Figs. \ref{fig:c_4_1024_n_1_2_3_stag} and
\ref{fig:app:TG_n_1_analytics_numerics}). In the $n=3$ case, the single depletion is maintained, but the shape of the wavepacket changes qualitatively from a wide, flat depletion to a narrow, pointed one.

It is informative to compare directly the time evolution of the staggered and variational wavepackets. Below, we present two categories of such examples at $c \sim \mathcal{O}(1)$ for $n=2$. Sec. \ref{sec:scaling_plots}, where we introduce an $N$-scaling protocol, includes further comparisons of staggered and variational $n=2,~3$-hole wavepackets at both $c=4$ and $c=1024$.

\begin{figure}[H]
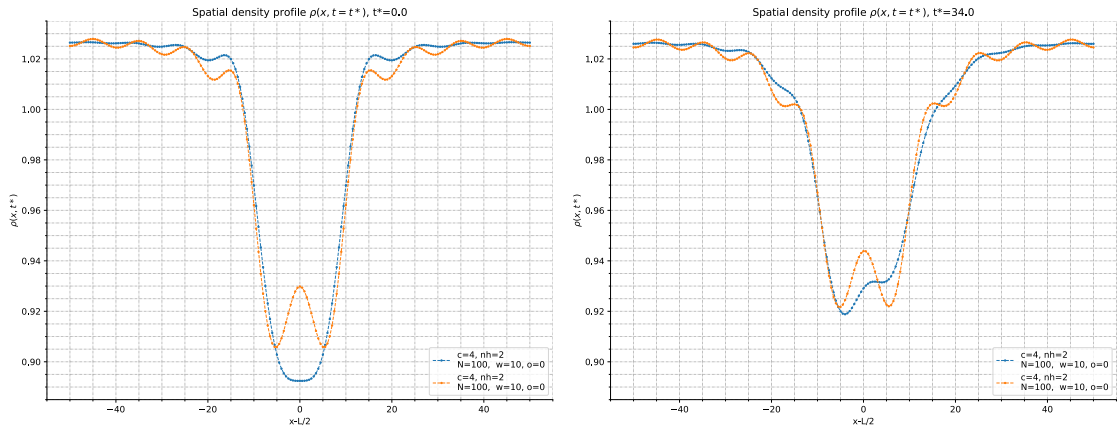
 
\centering
\begin{subfigure}{0.49\textwidth}
        \includesvg[width=\linewidth]{sections/figures/wavepacket_profiles/profile_comparisons/tStar_0p0_c_4_N_100_L_100_nholes_2_width_10_offset_0_IDeltax_4_staggered.svg}
\end{subfigure}
\begin{subfigure}{0.49\textwidth}
        \includesvg[width=\linewidth]{sections/figures/wavepacket_profiles/profile_comparisons/tStar_34p0_c_4_N_100_L_100_nholes_2_width_10_offset_0_IDeltax_4_staggered.svg}
\end{subfigure}    
\caption{Profiles of staggered (orange) and variational (blue) n=2-hole wavepackets at $t=\{0, 34\}$, $c=4$, $N=L=100$, $w=10$, $o=0$, $\Delta x = 4$ for variational wavepacket.}
\label{fig:c_4_n_2_stag_var_tStar}
\end{figure}
We note that Fig. \ref{fig:c_4_n_2_stag_var_tStar} depicts clearly the advantages of the variational wavepacket's initial profile as compared to that of the staggered: a deeper, smoother single-depletion profile, with wing-oscillations softened out. This advantage, however, does not survive time evolution: the depletion quickly evaporates, changing shape and losing symmetry in the process. While the staggered wavepacket also evaporates, it does so to a lesser extent, and in a manner that preserves its initial form (i.e., rigid and symmetric evaporation). In Sec. \ref{sec:time_tables}, we quantitatively compare the lifetimes of staggered and variational wavepackets.

Here, we note that the observed asymmetry is the result of a combination of factors: such spatial asymmetry is most visible at small $c$, becoming most pronounced for wavepackets with an asymmetric offset (i.e., with the summation interval \textit{not} centered in the middle of the Fermi sea). In spite of all of these factors, the uniform-amplitude wavepacket still evolves symmetrically to a large extent: as seen in Fig. \ref{fig:c_1_n_2_o_0_stag}, the $n=2$-hole staggered wavepacket at $c=1$ with offset 0 (i.e. summation window at the edge of the Fermi sea) produces a very small---almost imperceptible---amount of asymmetry by $t=44$, whereas the variational wavepacket (with the same remaining parameters) shows a drastic asymmetry early on (Fig. \ref{fig:c_1_n_2_o_0_var}). We note that introducing a symmetric offset in this case virtually eliminates any visible asymmetry in the staggered case (Fig. \ref{fig:c_1_n_2_o_45_stag}), and reduces it by a great degree in the variational wavepacket (Fig. \ref{fig:c_1_n_2_o_45_var}).

\subsection{Amplitude plots}
\begin{figure}
    \begin{subfigure}{0.5\textwidth}
      \centering
      \includegraphics[width=\linewidth]{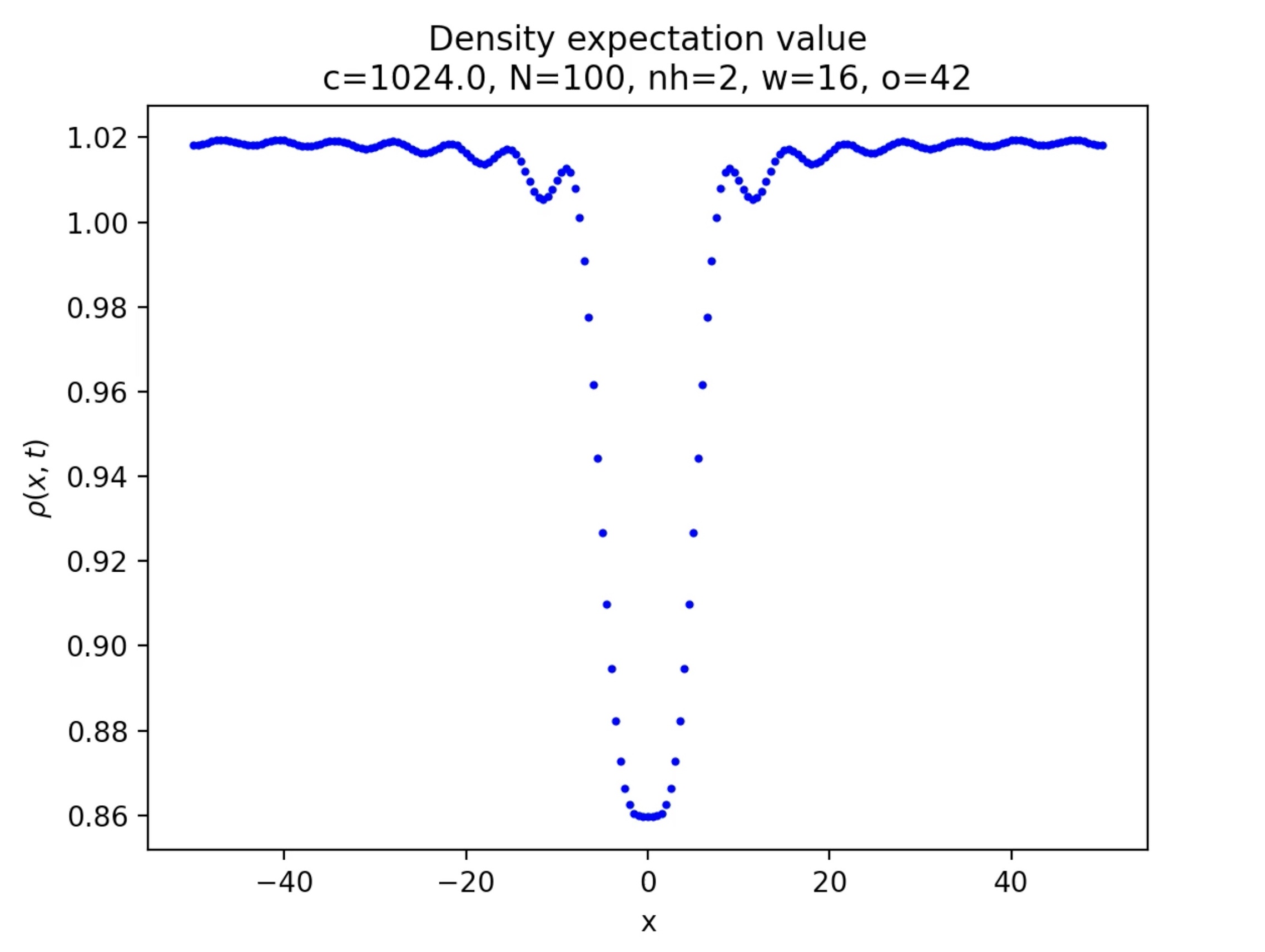}
     \caption{}
     \label{fig:heatmap_1}
     \end{subfigure}
    \begin{subfigure}{0.5\textwidth}
      \centering
      \includegraphics[width=0.85\linewidth]{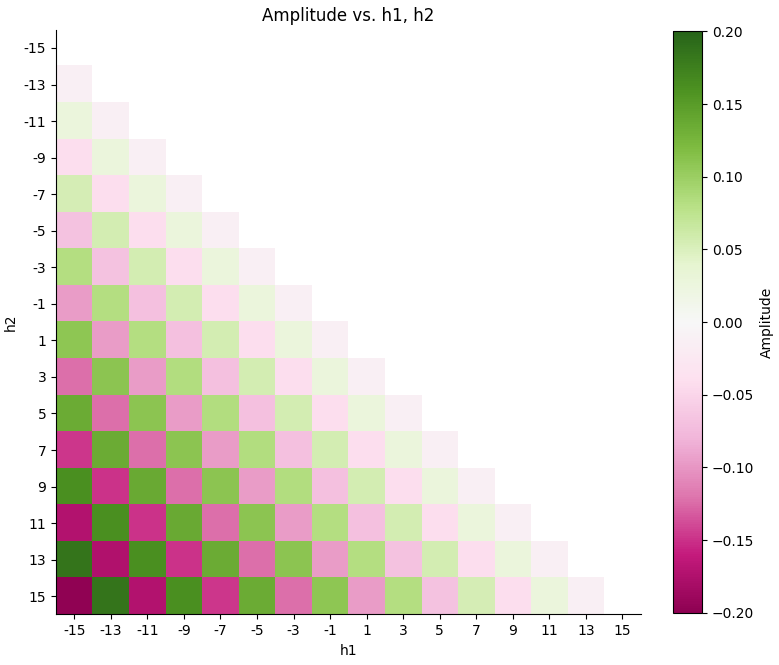}
      \caption{}
      \label{fig:heatmap_2}
    \end{subfigure}
    \newline
    \begin{subfigure}{0.5\textwidth}
      \centering
      \includegraphics[width=\linewidth]{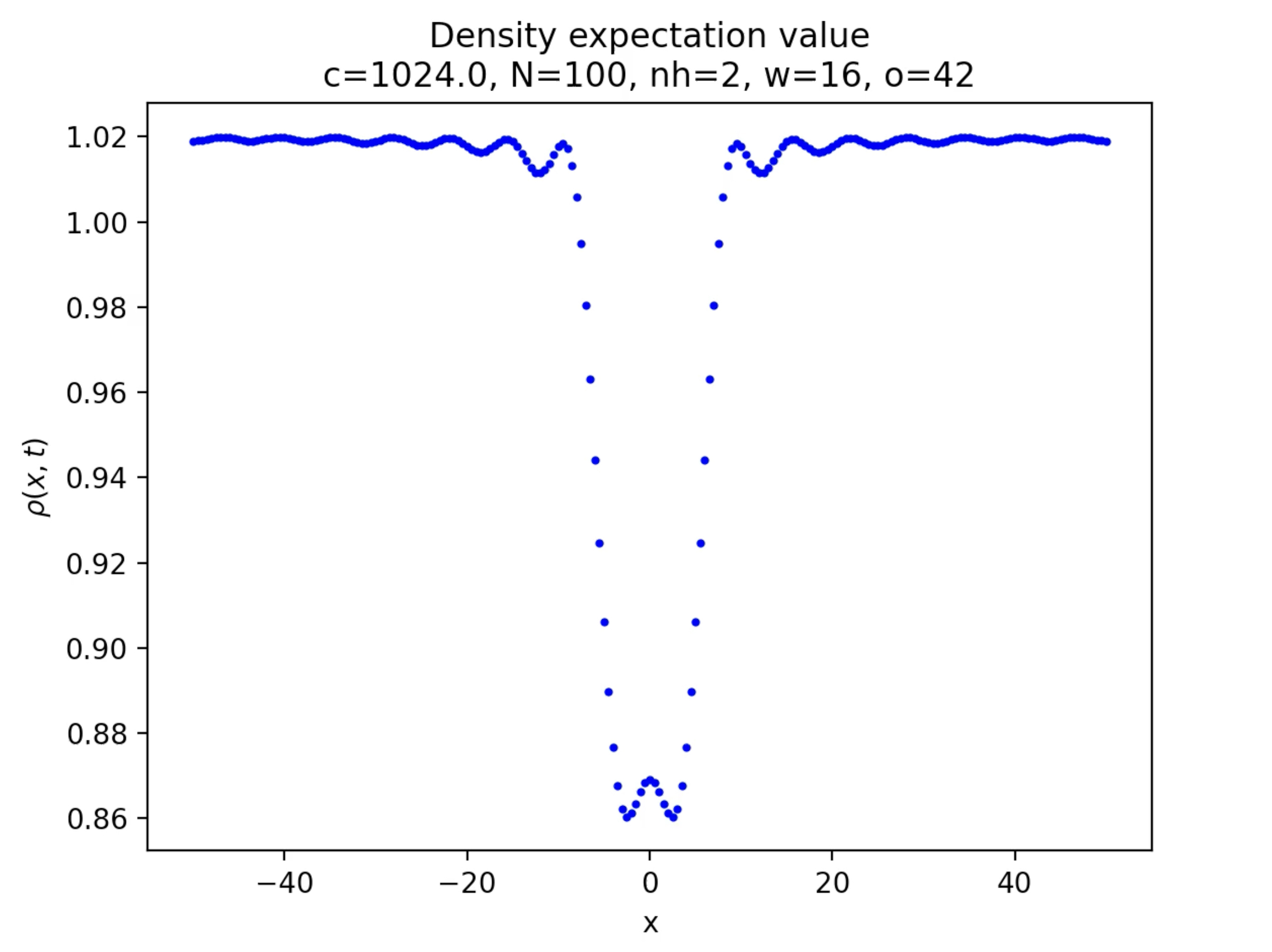}
     \caption{}
     \label{fig:heatmap_3}
     \end{subfigure}
    \begin{subfigure}{0.5\textwidth}
      \centering
      \includegraphics[width=.85\linewidth]{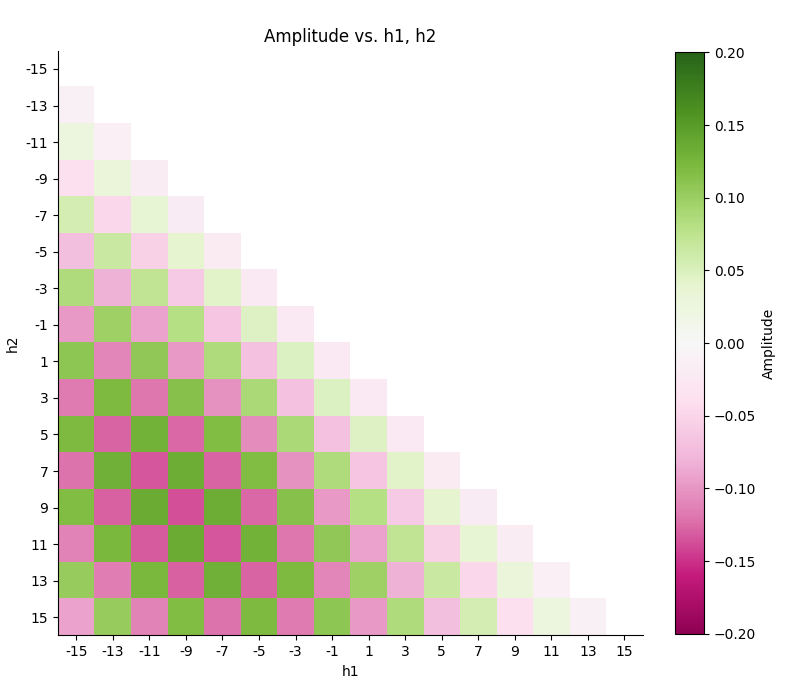}
      \caption{}
      \label{fig:heatmap_4}
    \end{subfigure}
    \newline
    \begin{subfigure}{0.5\textwidth}
      \centering
      \includegraphics[width=\linewidth]{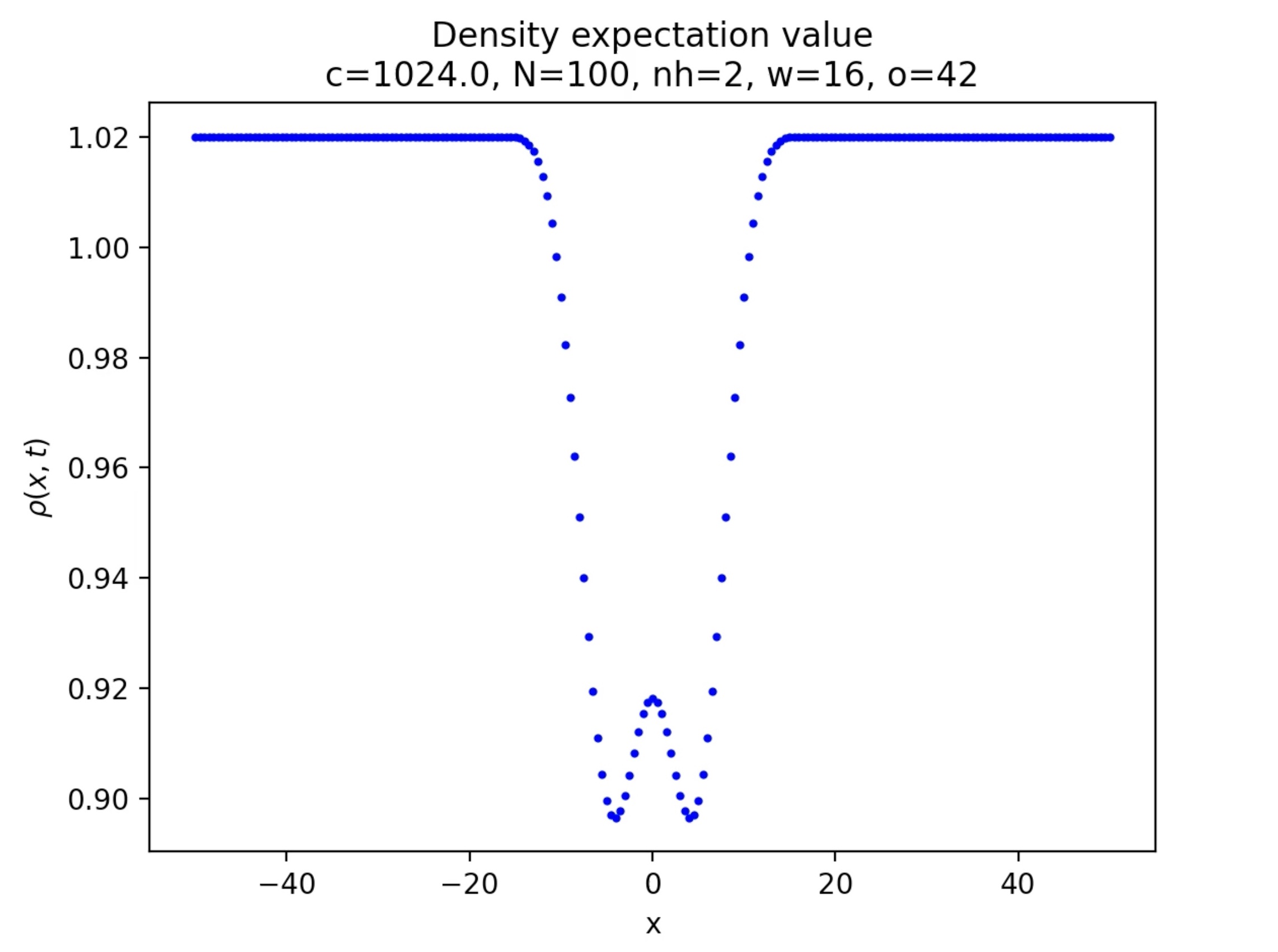}
     \caption{}
     \label{fig:heatmap_5}
     \end{subfigure}
    \begin{subfigure}{0.5\textwidth}
      \centering
      \includegraphics[width=.85\linewidth]{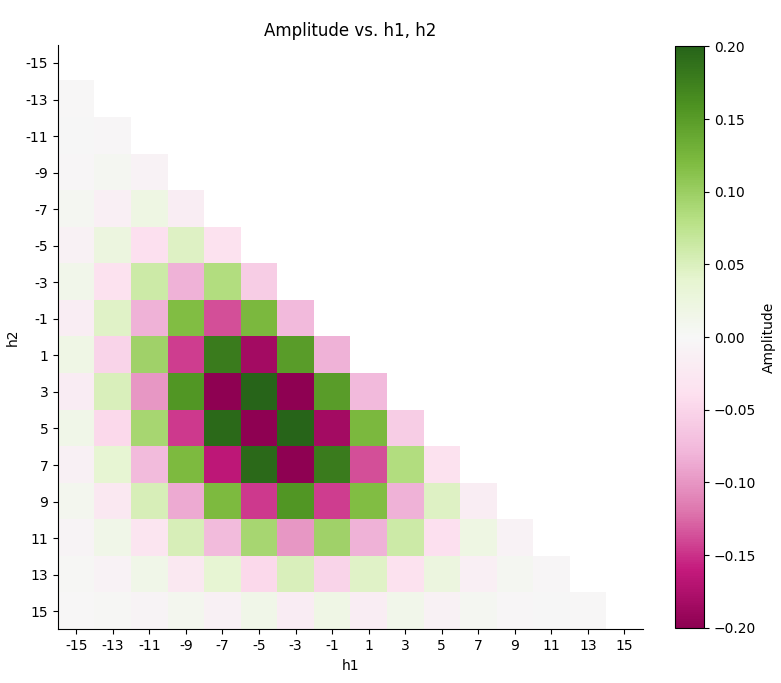}
      \caption{}
      \label{fig:heatmap_6}
    \end{subfigure}
    \caption{Amplitudes of variational wavepacket as a function of doubled quantum numbers of hole position for $c=1024$, $N=L=100$, $n=2$, $w=16$, $o=42$, for $\Delta x = \{2,~10,~30\}$ (fig. \ref{fig:heatmap_2},  \ref{fig:heatmap_4},  \ref{fig:heatmap_6}, respectively). Corresponding density profiles at $t=0$ are shown on the left.}
    \label{fig:ampPlot_n2}
\end{figure}

\begin{figure}
    \begin{subfigure}{0.5\textwidth}
      \centering
      \includegraphics[width=.85\linewidth]{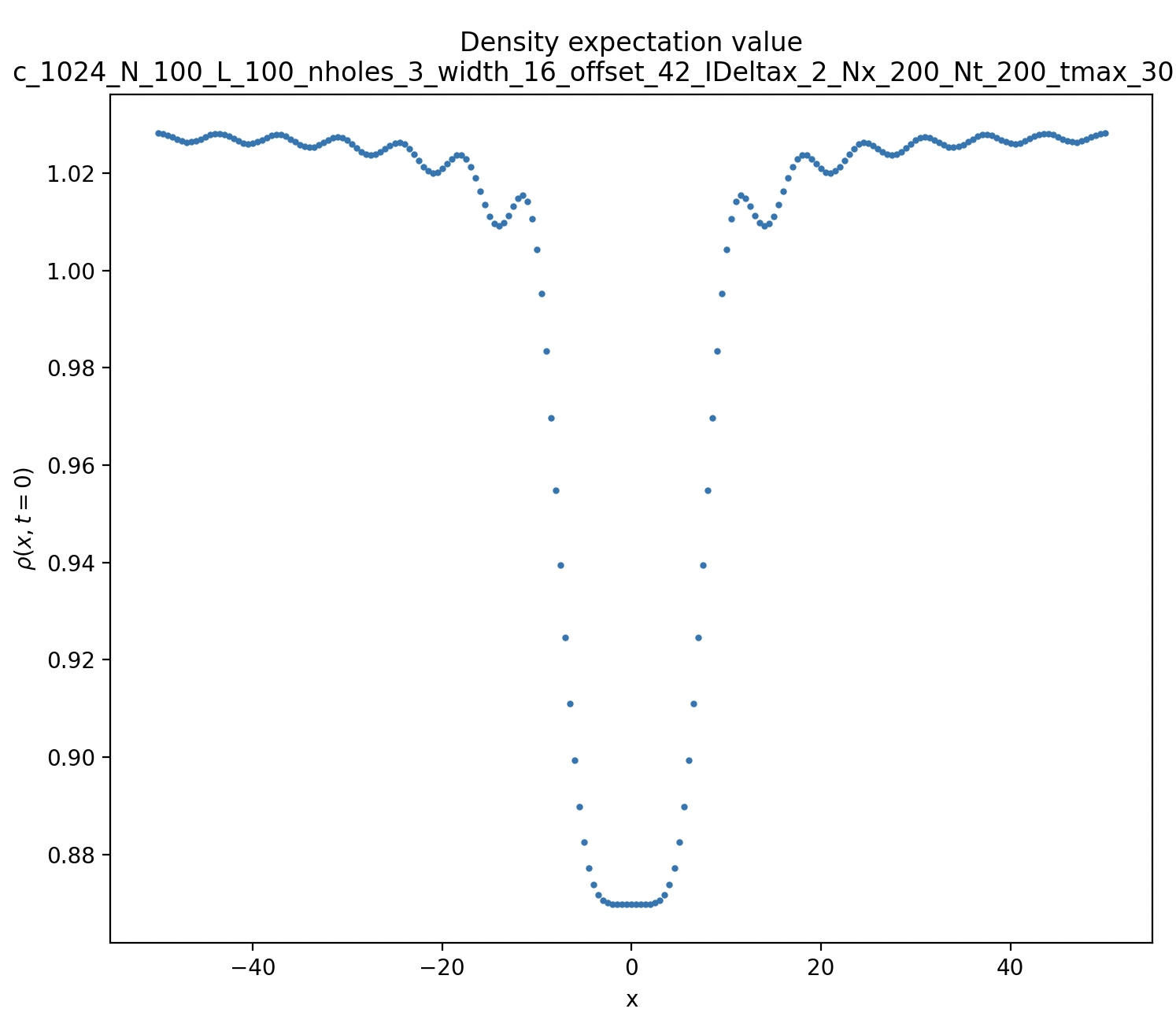}
      \caption{}
      \label{fig:heatmap_n3_t0}
    \end{subfigure}
    \begin{subfigure}{0.5\textwidth}
      \centering
      \includegraphics[width=\linewidth]{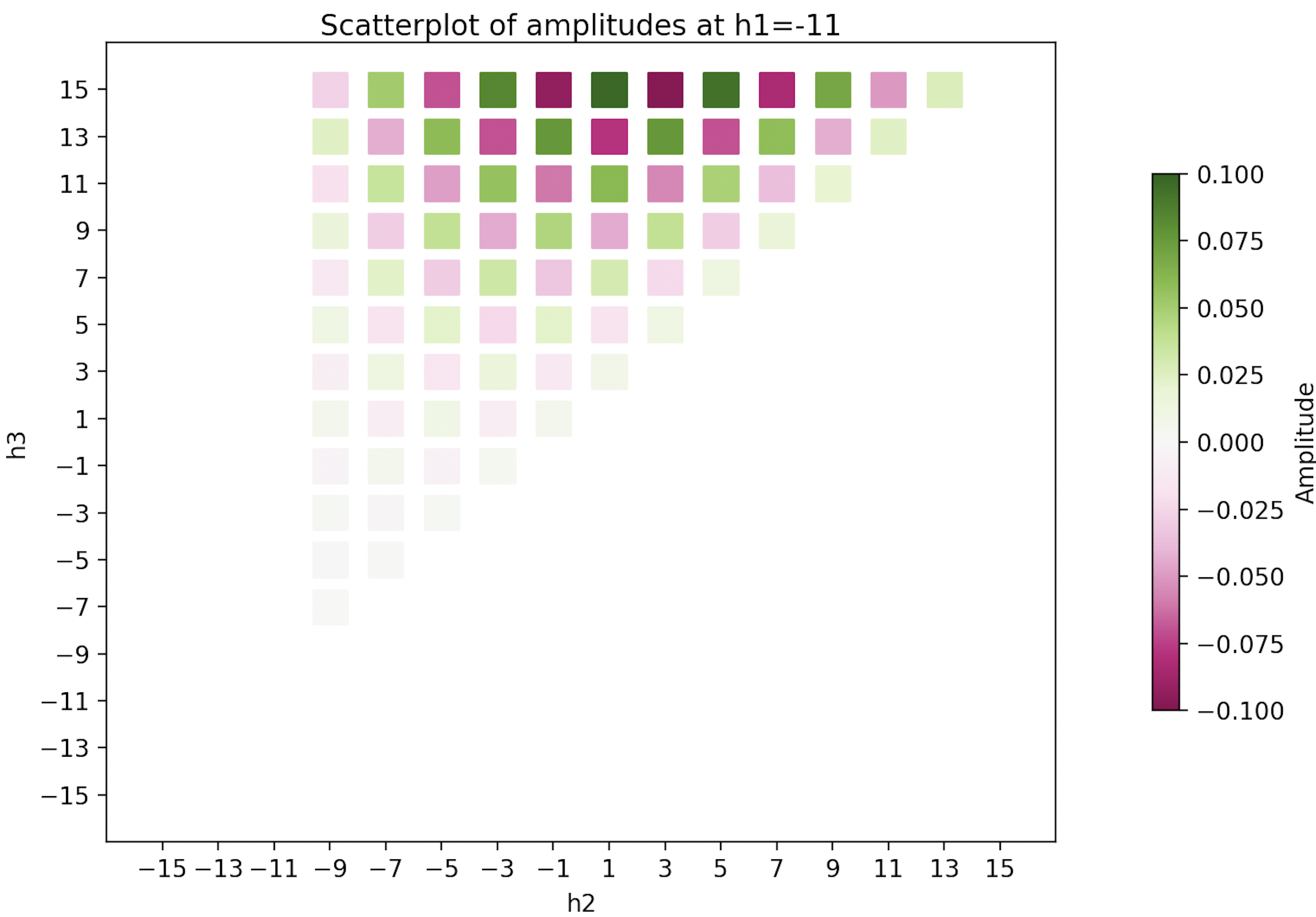}
     \caption{}
     \label{fig:heatmap_n3_h1}
     \end{subfigure}
       \newline
    \begin{subfigure}{0.5\textwidth}
      \centering
      \includegraphics[width=0.88\linewidth]{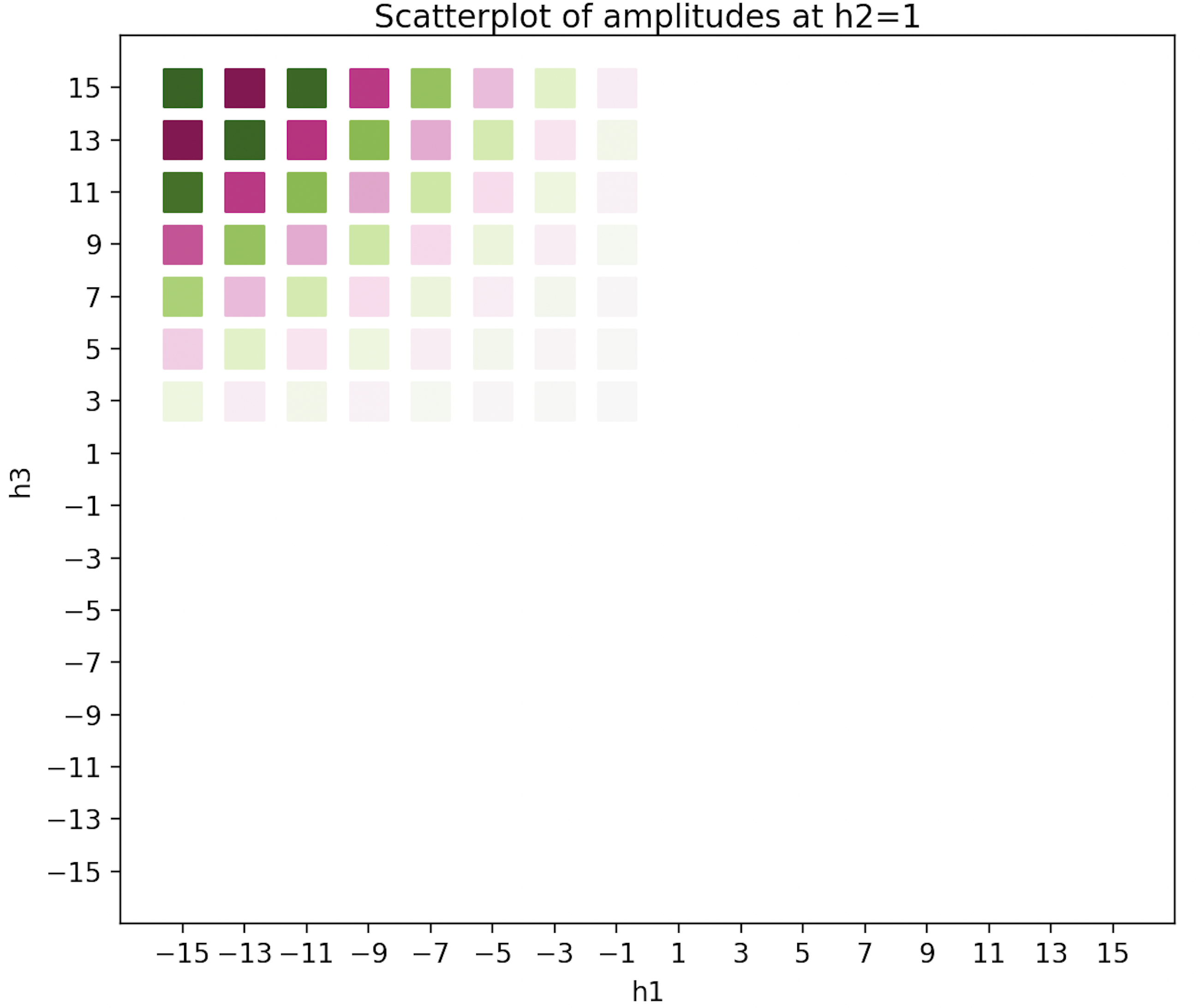}
     \caption{}
     \label{fig:heatmap_n3_h2}
     \end{subfigure}
    \begin{subfigure}{0.5\textwidth}
      \centering
      \includegraphics[width=\linewidth]{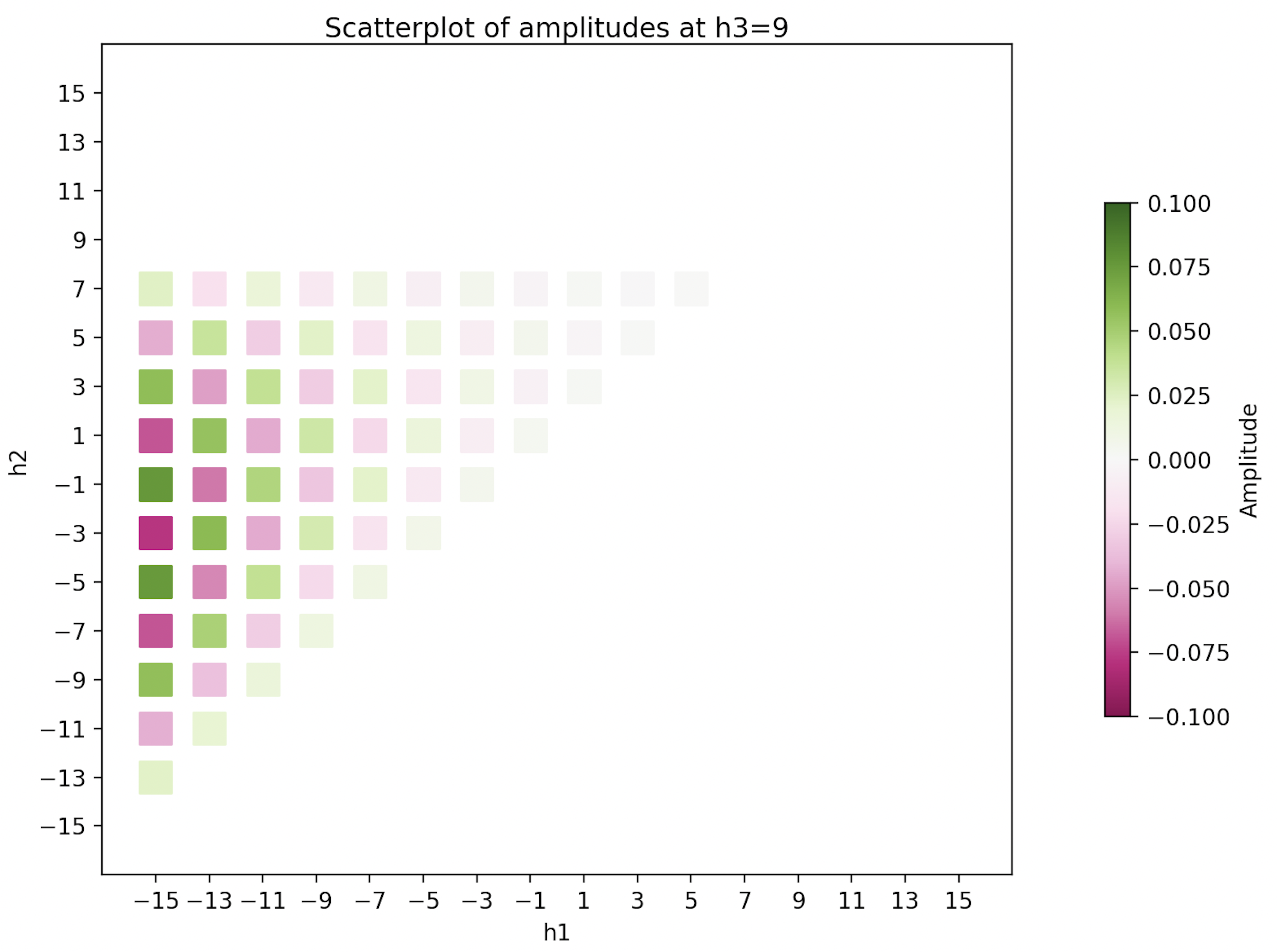}
     \caption{}
     \label{fig:heatmap_n3_h3}
     \end{subfigure}
    \caption{Amplitudes of variational wavepacket as a function of doubled quantum numbers of hole-positions for $c=1024$, $N=L=100$, $n=3$, $w=16$, $o=42$, $\Delta x=2$ at slices along fixed $h1$ (\ref{fig:heatmap_n3_h1}), $h2$ (\ref{fig:heatmap_n3_h2}), and $h3$ (\ref{fig:heatmap_n3_h3}), and corresponding density profile at $t=0$ (\ref{fig:heatmap_n3_t0}).}
    \label{fig:ampPlot_n3}
\end{figure}

For further insight into the distribution of eigenstates within a variational wavepacket, let us take a look at the amplitudes resulting from the optimization process. Below, we show the distribution of amplitudes as a function of hole position for three different values of $\Delta x$ (Fig. \ref{fig:ampPlot_n2}) for the $n=2$-hole wavepacket. Corresponding density profiles at $t=0$ are shown on the left. Fig. \ref{fig:ampPlot_n3} shows the distribution of amplitudes along three slices of the octant defined by the hole position 3-tuple $(h1,~h2,~h3)$. Note that the axes are labeled with doubled quantum numbers for the convenience of working with integer values.

Our observations are as follows: in all cases, the variational method yields amplitudes that alternate in sign with hole position (to localize the depletion around $x=0$, similarly to the one-hole case). Not surprisingly, amplitudes are more localized in momentum space for the wider, double depletion profiles (Figs. \ref{fig:heatmap_5} and \ref{fig:heatmap_6}). We note that a ``mild'' degree of amplitude localization in k-space with maximally separated hole positions (i.e., at the corner of the plotted axes) yields a narrow profile resembling the single-depletion of the $n=1$-hole uniform-amplitude wavepacket (Figs. \ref{fig:heatmap_1} and \ref{fig:heatmap_2}).

We conclude this section by noting that in many cases, the variational method can produce localized, single-depletion density profiles for $n>1$---some with lifetimes comparable to the corresponding uniform-amplitude wavepacket---for appropriately chosen values of $\Delta x$. In contrast, the initial profile of the uniform-amplitude $n$-hole wavepacket displays $n$ depletions that are each localized yet separated in space. While the narrow, localized initial profile of the variational wavepacket can decay rather quickly---and at times in an asymmetric or non-rigid manner---its lesser-localized uniform-amplitude counterpart demonstrates a slower decay, maintaining the qualitative shape of its initial profile for longer. We present a quantitative comparison of wavepacket lifetimes as a function of coefficient choice in Sec. \ref{sec:lifetime}. Our observations here are in line with the expected trade-off rooted in the uncertainty principle, as outlined in Sec. \ref{sec:intro}: given a nonlinear dispersion, increased localization of the initial profile results in reduced lifetime due to earlier dephasing of wavepacket components. Despite this inevitable constraint, there exists a silver lining to which we turn next: scaling of lifetime with system size.
\section{Scaling with system size}\label{sec:scaling_plots}
Based on the arguments presented in Sec. \ref{sec:scaling_arg}, we introduce the following scaling protocol: at each step, we double the system size and increase the summation width by a factor of 1.5, starting with $(N,w) = (32,8)$. This yields a computationally feasible integer-preserving protocol over a number of steps. Below, 
we show the results of the scaling protocol for $n=2$ at $c=4$ (Fig. \ref{fig:c_4_n_2_N_scaling_stag_var}) and $c=1024$ (Fig. \ref{fig:c_1024_n_2_N_scaling_stag_var}) for both staggered (top tow) and variational (bottom row) wavepackets. For the latter, the parameter $\Delta x$ is chosen at each step to yield a suitably localized, single-depletion wavepacket. Each column depicts a single snapshot in time at $t=\{0,~t_f/2, ~t_f\}$. We note a significant improvement in lifetime with system size: in all cases, the wavepacket at the largest step of the scaling remains virtually unchanged while others deform appreciably or evaporate altogether. In Secs. \ref{sec:time_tables} and \ref{sec:R_t_plots}, we measure this improvement in lifetime quantitatively.

\subsection*{$c=1024$:}
\begin{figure}[H]
    \centering
    %staggered
    \subfloat[]{\includesvg[width=0.34\linewidth]{sections/figures/wavepacket_profiles/N_scaling_plots/stag_scaling_plots/c_1024/tStar_0p0_c_1024_n_2_N_scaling_stag.svg}}
    \subfloat[]{\includesvg[width=0.34\linewidth]{sections/figures/wavepacket_profiles/N_scaling_plots/stag_scaling_plots/c_1024/tStar_4p0_c_1024_n_2_N_scaling_stag.svg}}
    \subfloat[]{\includesvg[width=0.34\linewidth]{sections/figures/wavepacket_profiles/N_scaling_plots/stag_scaling_plots/c_1024/tStar_8p0_c_1024_n_2_N_scaling_stag.svg}}\\
    %variational
    \subfloat[]{\includesvg[width=0.34\textwidth]{sections/figures/wavepacket_profiles/N_scaling_plots/var_scaling_plots/c_1024/tStar_0p0_c_1024_n_2_N_scaling_var.svg}}
    \subfloat[]{\includesvg[width=0.34\linewidth]{sections/figures/wavepacket_profiles/N_scaling_plots/var_scaling_plots/c_1024/tStar_4p0_c_1024_n_2_N_scaling_var}}
    \subfloat[]{\includesvg[width=0.34\linewidth]{sections/figures/wavepacket_profiles/N_scaling_plots/var_scaling_plots/c_1024/tStar_8p0_c_1024_n_2_N_scaling_var}}
    \caption{Scaling plots for $n=2$-hole staggered (top) and variational (bottom) wavepackets at $c=1024$.} 
\label{fig:c_1024_n_2_N_scaling_stag_var}
\end{figure}

\begin{figure}[H]
    \centering
    \includegraphics[width=\linewidth]{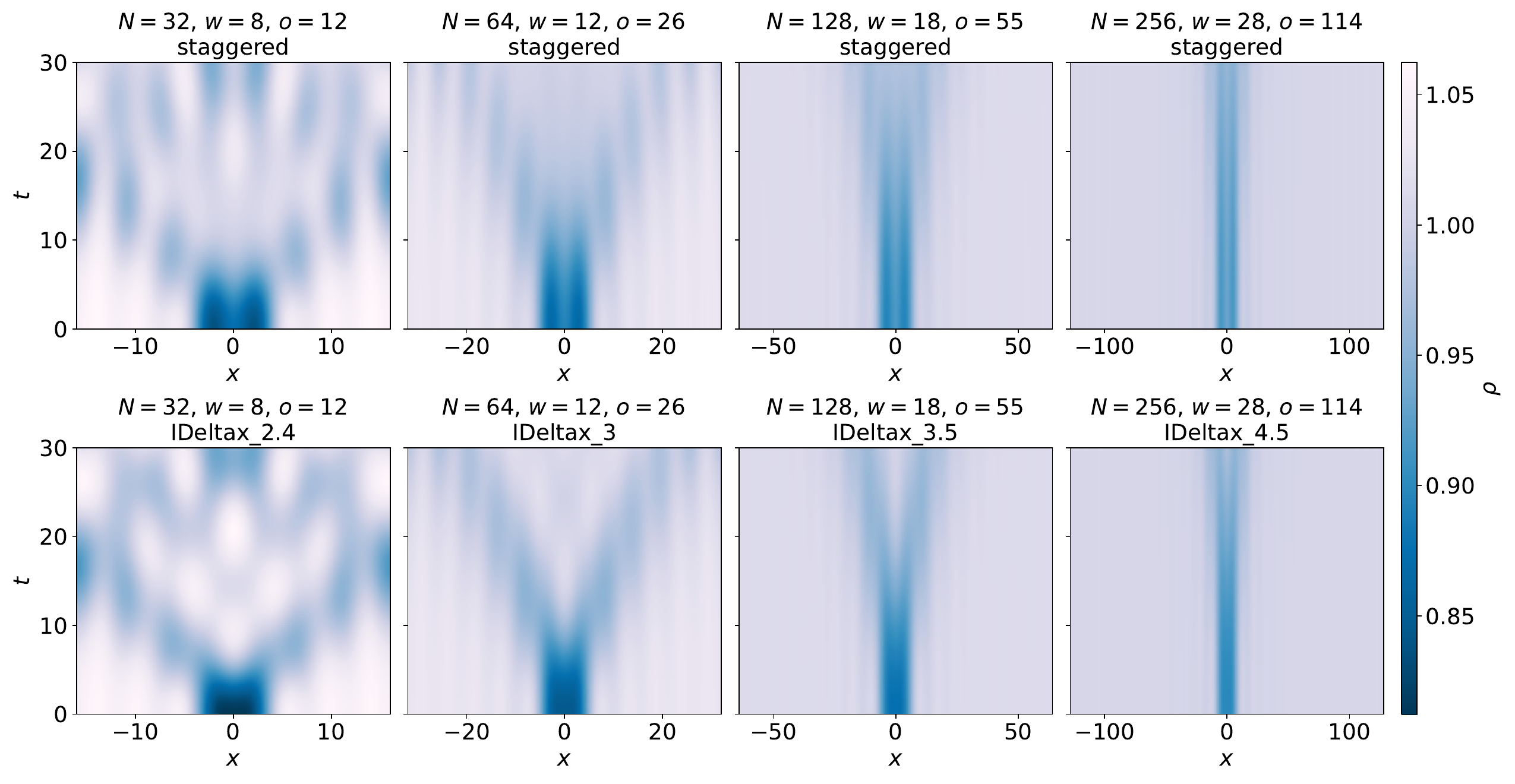}
    \caption{Colormaps for the density profiles of $n=2$-hole staggered (top) and variational (bottom) wavepackets at $c=1024$.}
\end{figure}

\begin{figure}[H]
    \centering
    \includegraphics[width=\linewidth]{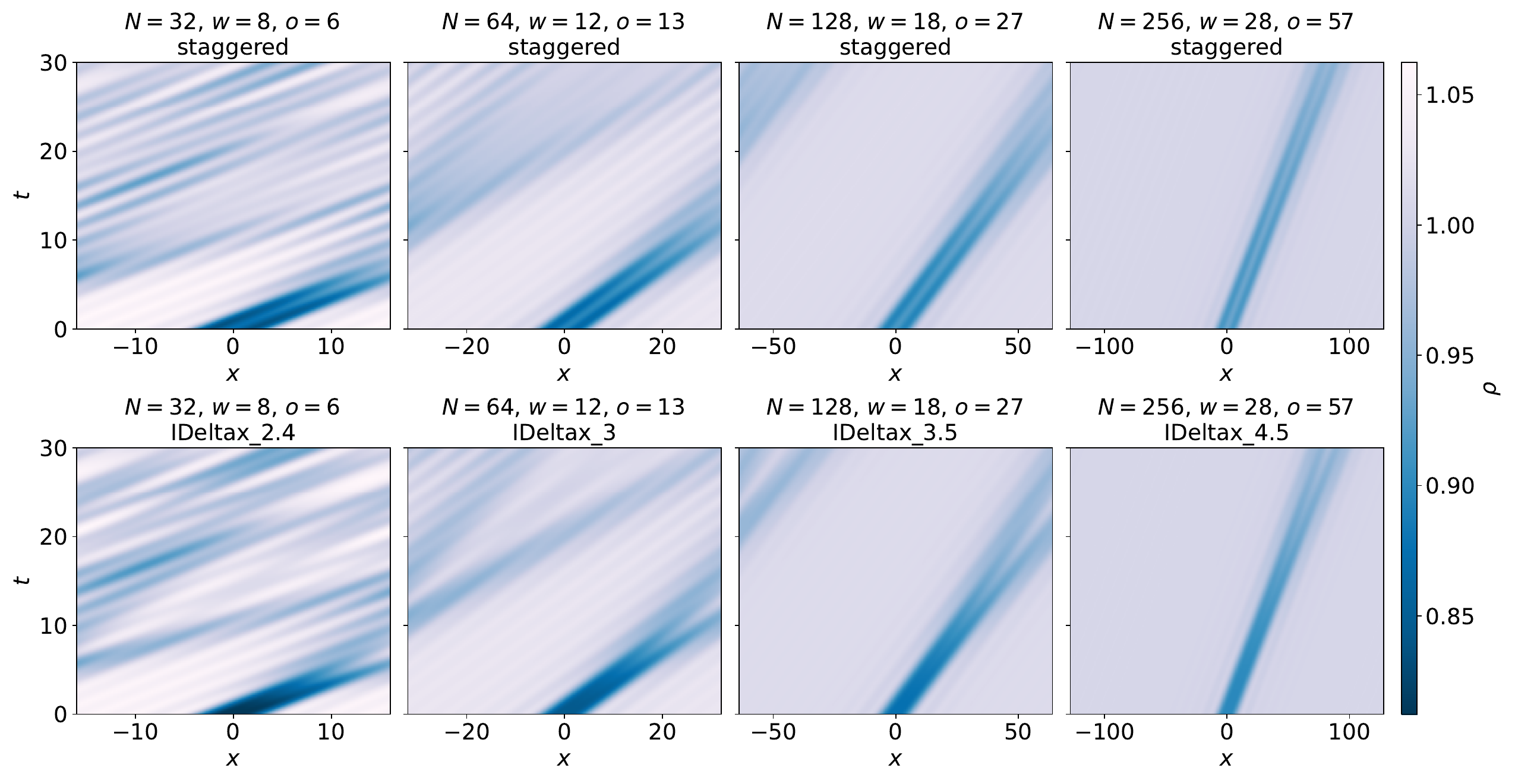}
    \caption{Colormaps for the density profiles of $n=2$-hole staggered (top) and variational (bottom) wavepackets at $c=1024$ prepared with finite group velocity.}
\end{figure}

\subsection*{$c=4$:}
\begin{figure}[H]
    \centering
    %staggered
    \subfloat[]{\includesvg[width=0.34\linewidth]{sections/figures/wavepacket_profiles/N_scaling_plots/stag_scaling_plots/c_4/tStar_0p0_c_4_n_2_N_scaling_stag.svg}}
    \subfloat[]{\includesvg[width=0.34\linewidth]{sections/figures/wavepacket_profiles/N_scaling_plots/stag_scaling_plots/c_4/tStar_7p0_c_4_n_2_N_scaling_stag.svg}}
    \subfloat[]{\includesvg[width=0.34\linewidth]{sections/figures/wavepacket_profiles/N_scaling_plots/stag_scaling_plots/c_4/tStar_14p0_c_4_n_2_N_scaling_stag.svg}}\\
    %variational
    \subfloat[]{\includesvg[width=0.34\linewidth]{sections/figures/wavepacket_profiles/N_scaling_plots/var_scaling_plots/c_4/tStar_0p0_c_4_n_2_N_scaling_var.svg}}
    \subfloat[]{\includesvg[width=0.34\linewidth]{sections/figures/wavepacket_profiles/N_scaling_plots/var_scaling_plots/c_4/tStar_7p0_c_4_n_2_N_scaling_var.svg}}
    \subfloat[]{\includesvg[width=0.34\linewidth]{sections/figures/wavepacket_profiles/N_scaling_plots/var_scaling_plots/c_4/tStar_14p0_c_4_n_2_N_scaling_var.svg}}
    \caption{Scaling plots for $n=2$-hole staggered (top) and variational (bottom) wavepackets at $c=4$.} 
\label{fig:c_4_n_2_N_scaling_stag_var}
\end{figure}

\begin{figure}[H]
    \centering
    \includegraphics[width=\linewidth]{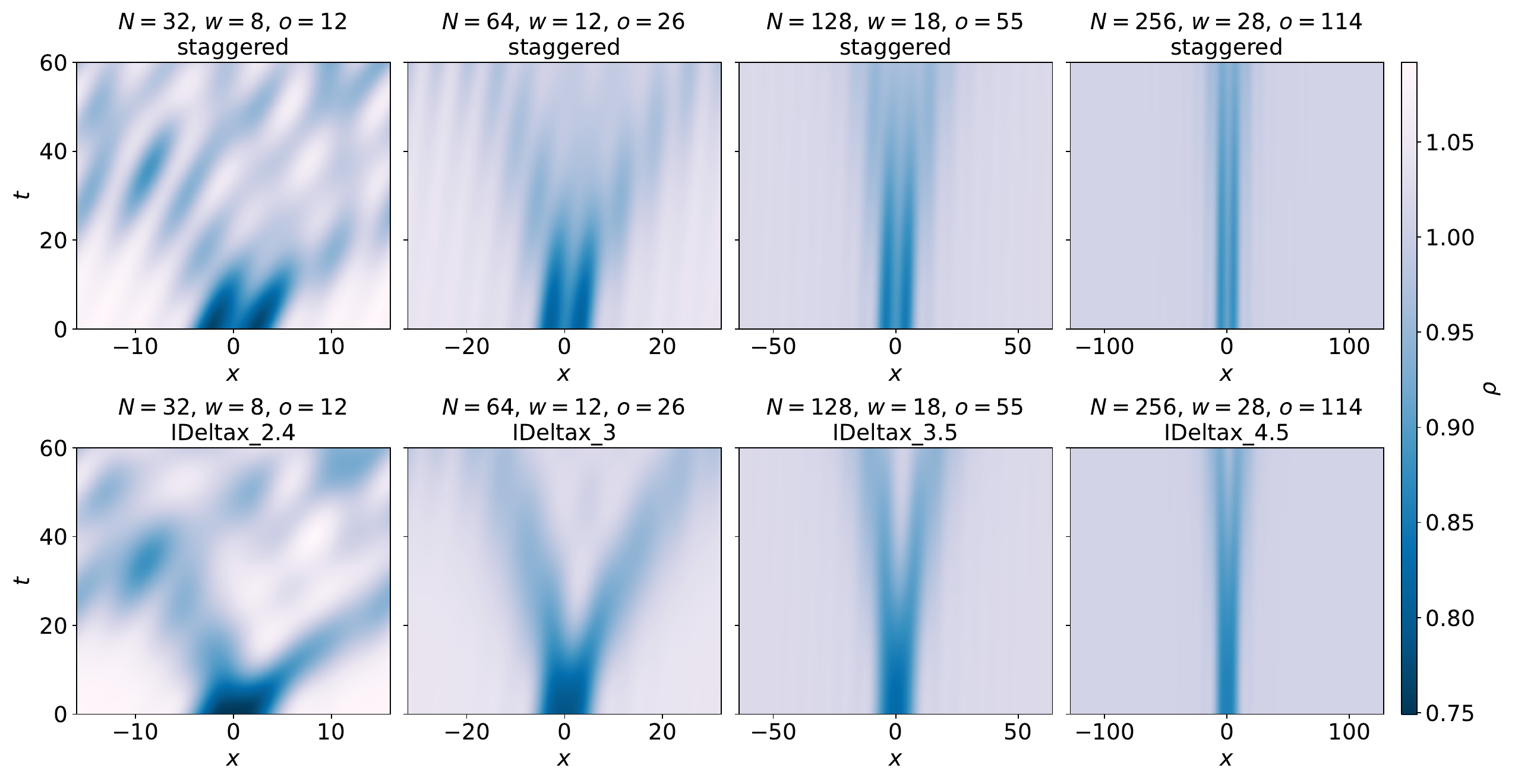}
    \caption{Colormaps for the density profiles of $n=2$-hole staggered (top) and variational (bottom) wavepackets at $c=4$.}
\end{figure}

\begin{figure}[H]
    \centering
    \includegraphics[width=\linewidth]{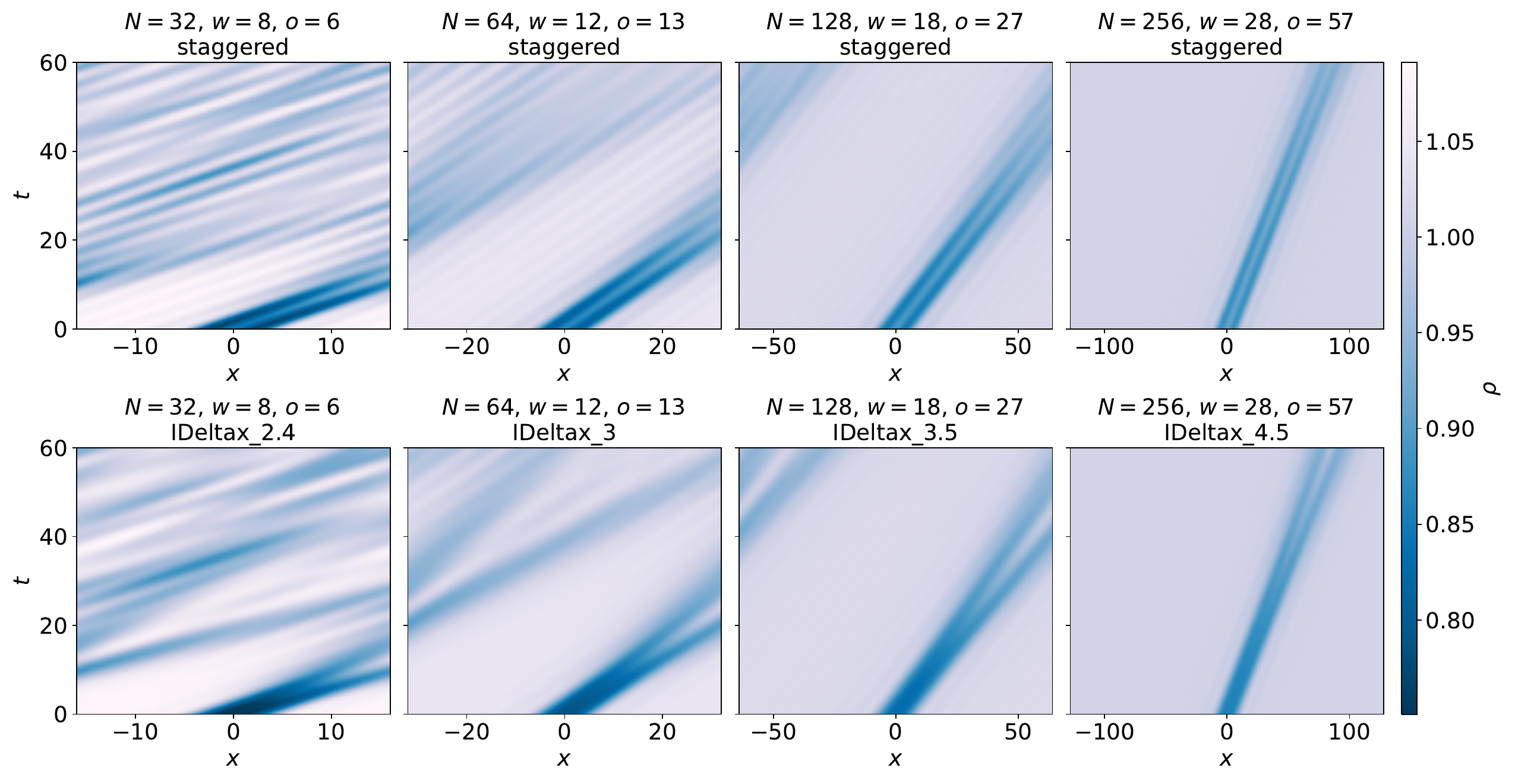}
    \caption{Colormaps for the density profiles of $n=2$-hole staggered (top) and variational (bottom) wavepackets at $c=4$ prepared with finite group velocity.}
\end{figure}

\subsection*{Staggered vs. variational wavepackets at large $N$}
\begin{figure}[H]
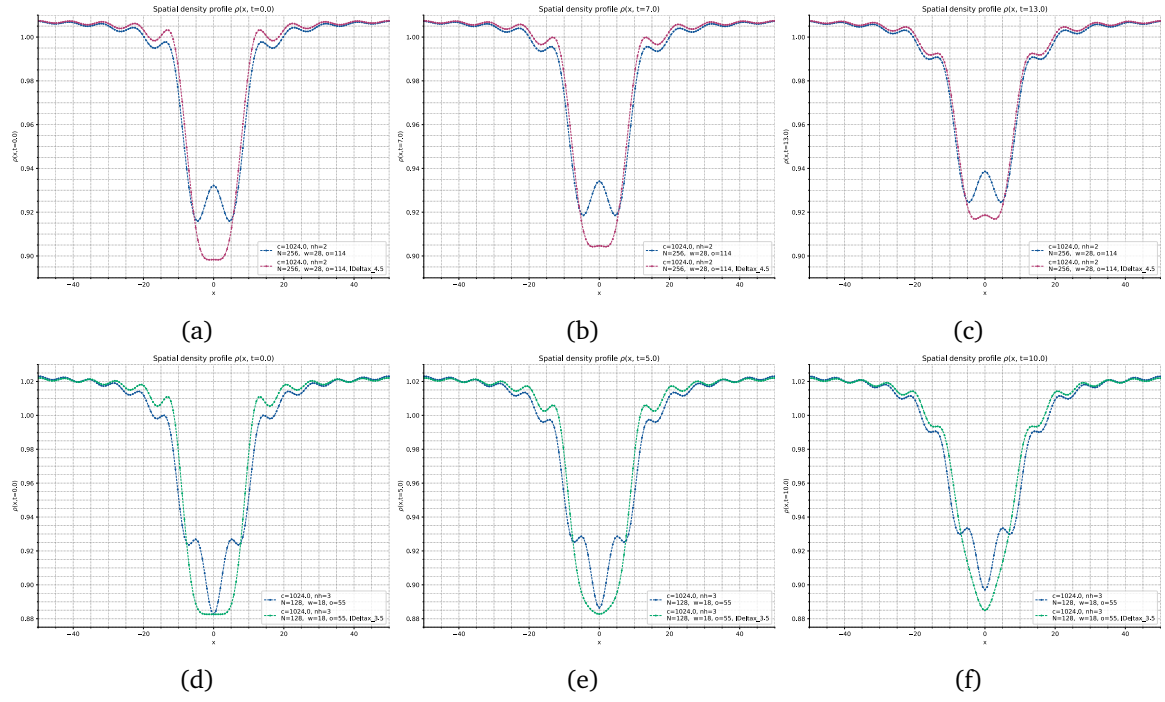

    \centering
    \subfloat[]{\includesvg[width=0.34\linewidth]{sections/figures/wavepacket_profiles/profile_comparisons/scaling_finals/c_1024/tStar_0p0_c_1024_N_256_L_256_nholes_2_width_28_offset_114_staggered_IDeltax_4.5.svg}}
    \subfloat[]{\includesvg[width=0.34\linewidth]{sections/figures/wavepacket_profiles/profile_comparisons/scaling_finals/c_1024/tStar_7p0_c_1024_N_256_L_256_nholes_2_width_28_offset_114_staggered_IDeltax_4.5.svg}}
    \subfloat[]{\includesvg[width=0.34\linewidth]{sections/figures/wavepacket_profiles/profile_comparisons/scaling_finals/c_1024/tStar_13p0_c_1024_N_256_L_256_nholes_2_width_28_offset_114_staggered_IDeltax_4.5.svg}}
    \\
    \subfloat[]{\includesvg[width=0.34\linewidth]{sections/figures/wavepacket_profiles/profile_comparisons/scaling_finals/c_1024/tStar_0p0_c_1024_N_128_L_128_nholes_3_width_18_offset_55_staggered_IDeltax_3.5.svg}}
    \subfloat[]{\includesvg[width=0.34\linewidth]{sections/figures/wavepacket_profiles/profile_comparisons/scaling_finals/c_1024/tStar_5p0_c_1024_N_128_L_128_nholes_3_width_18_offset_55_staggered_IDeltax_3.5.svg}}
    \subfloat[]{\includesvg[width=0.34\linewidth]{sections/figures/wavepacket_profiles/profile_comparisons/scaling_finals/c_1024/tStar_10p0_c_1024_N_128_L_128_nholes_3_width_18_offset_55_staggered_IDeltax_3.5.svg}}
\caption{Time evolution of $n=2$-hole (top) and  $n=3$-hole (bottom) staggered and variational wavepackets at $c=1024$.} 
\label{fig:c_1024_n_2_n_3_stag_var_final}
\end{figure}

Below, we compare the variational and staggered wavepackets at the largest computed step of the scaling protocol for the four combinations of the following parameters: $n=2$-hole (top row) and $n=3$-hole (bottom row) wavepackets at $c=1024$ (Fig. \ref{fig:c_1024_n_2_n_3_stag_var_final}) and $c=4$ (Fig. \ref{fig:c_4_n_2_n_3_stag_var_final}). While both types of wavepackets show considerable gain in lifetime with $N$ (as demonstrated in Figs. \ref{fig:c_1024_n_2_N_scaling_stag_var} and \ref{fig:c_4_n_2_N_scaling_stag_var}), the staggered wavepacket shows a considerably more stable profile under time evolution in comparison with its variational counterpart in all four categories.
\subsection*{$c=1024$:}

\subsection*{$c=4$:}
\begin{figure}[H]
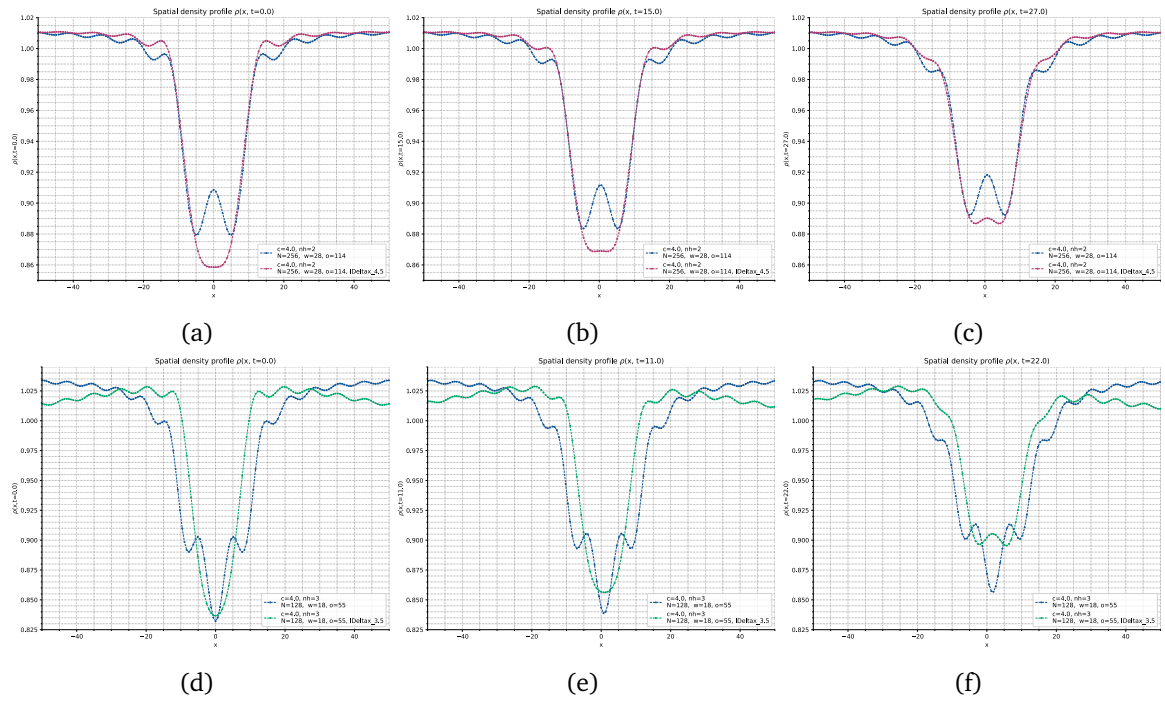

    \centering
    \subfloat[]{\includesvg[width=0.34\linewidth]{sections/figures/wavepacket_profiles/profile_comparisons/scaling_finals/c_4/tStar_0p0_c_4_N_256_L_256_nholes_2_width_28_offset_114_staggered_IDeltax_4.5.svg}}
   \subfloat[]{\includesvg[width=0.34\linewidth]{sections/figures/wavepacket_profiles/profile_comparisons/scaling_finals/c_4/tStar_15p0_c_4_N_256_L_256_nholes_2_width_28_offset_114_staggered_IDeltax_4.5.svg}}
  \subfloat[]{\includesvg[width=0.34\linewidth]{sections/figures/wavepacket_profiles/profile_comparisons/scaling_finals/c_4/tStar_27p0_c_4_N_256_L_256_nholes_2_width_28_offset_114_staggered_IDeltax_4.5.svg}}\\
    \subfloat[]{\includesvg[width=0.34\linewidth]{sections/figures/wavepacket_profiles/profile_comparisons/scaling_finals/c_4/tStar_0p0_c_4_N_128_L_128_nholes_3_width_18_offset_55_staggered_IDeltax_3.5.svg}}
    \subfloat[]{\includesvg[width=0.34\linewidth]{sections/figures/wavepacket_profiles/profile_comparisons/scaling_finals/c_4/tStar_11p0_c_4_N_128_L_128_nholes_3_width_18_offset_55_staggered_IDeltax_3.5.svg}}
    \subfloat[]{\includesvg[width=0.34\linewidth]{sections/figures/wavepacket_profiles/profile_comparisons/scaling_finals/c_4/tStar_22p0_c_4_N_128_L_128_nholes_3_width_18_offset_55_staggered_IDeltax_3.5.svg}}
\caption{Time evolution of $n=2$-hole (top) and  $n=3$-hole (bottom) staggered and variational wavepackets at $c=4$.} 
\label{fig:c_4_n_2_n_3_stag_var_final}
\end{figure}
\section{Wavepacket lifetime}\label{sec:lifetime}

\noindent In Secs. \ref{sec:intro}, \ref{sec:wavepackets_finite_c}, and \ref{sec:selective_width_summations}, we outlined the relation between the lifetime of a wavepacket and parameters such as interaction strength and summation width. This reasoning was based purely on energetic grounds. Whilst effectively capturing the scaling behavior of lifetime, these arguments do not take into account the coefficients used to construct the wavepacket, which ultimately affect its dynamics. To this end, we will seek to define a measure which effectively quantifies the lifetime of a localized depletion. Given the varied dynamics of wavepackets---including (a)symmetric and (non)rigid evaporation in some cases---it is a non-trivial task to define a universal, all-encompassing measure that consistently captures a reliable estimate of lifetime across a wide range of parameters.

We begin with a simple measure tracking the central point of the depletion. Measuring the density $\rho(t)$ at this point, we may describe its deviation $\Delta \rho$ from the initial density as an even-powered (and thus time-inversion symmetric) power series in time (Eq.~\ref{eq:time_series}), yielding the (sub)leading behavior of the central-point evaporation. This in turn provides us with a concrete numerical quantifier for lifetime: the time at which the central point has evaporated by a given percentage (relative to its initial depth) can serve as such an estimate, provided that we remain within the radius of convergence of the series.

We must however ask ourselves whether such a measure, taking into account only the central point of the wavepacket, suffices to accurately and consistently capture the time evolution of the depletion as a whole. To address this question, we define a spatially-aware measure of lifetime, $R(t)$, which takes into account the wavepacket within a spatial window $(x_{-},x_{+})$ such that $\rho(x,0) \leq \rho_0$ at all points within the interval; i.e., the entire region of depletion. Considering the wavepacket at $t=0$, we may then compute its overlap with the time-evolved depletion (Eq.~\ref{eq:R_t}) to measure the amount by which it has evaporated at time $t$.

Below, we briefly introduce the theory behind each method. We then present our numerical lifetime estimates (Sec.~\ref{sec:time_tables}), and $R(t)$ lifetime plots in Sec. \ref{sec:R_t_plots}. As shown in these plots, $R(t)$ exhibits quadratic-like evaporation of the wavepacket in time. This observation is confirmed by the time-series approach, which yields leading-order quadratic evaporation of the central point in nearly all cases. As we show in Sec. \ref{sec:time_tables}, the quadratic-order lifetime estimates (obtained at 10\% evaporation of the central point) are in close agreement with those obtained via the spatially-aware $R(t)$, confirming that both serve as a utile method to obtain concrete estimates of lifetime.

\subsection{Measures for lifetime}

\subsubsection{$\Delta \rho(t)$: Central-point power series in time}

To obtain an expression for the evaporation of the central point of the wavepacket, let us consider the difference between the original profile and its time-evolved, spatially back-shifted version:
\begin{equation}
\Delta \rho (x, t) \equiv \rho(x + v_g t, t) - \rho(x, 0),
\end{equation}
in which $v_g$ is a to-be-specified group velocity.
Using the eigenstate decomposition
\begin{equation}
|\Psi (t) \rangle = \sum_\alpha \psi_\alpha e^{-i \omega_\alpha t} | \alpha \rangle
\end{equation}
gives
\begin{equation}\label{eq:Delta_rho_x_t}
\Delta \rho(x, t) = 2 \Re \sum_{\alpha < \alpha'} \psi^*_\alpha \psi_{\alpha'} \rho_{\alpha \alpha'} [ e^{i(\omega_{\alpha \alpha'} - v_g k_{\alpha \alpha'}) t} - 1] e^{-i k_{\alpha \alpha'} x}.
\end{equation}
For simplicity, we will position the initial center of the quasisoliton at $x=0$.
Let us define the parameter
\begin{equation}
\langle \hat{\omega} \rangle = \sum_{\alpha < \alpha'} \psi^*_\alpha \psi_{\alpha'} \rho_{\alpha \alpha'} \omega_{\alpha \alpha'}
\end{equation}
and similarly for the quantities $\langle \hat k^2 \rangle$ and $\langle \hat \omega \hat k \rangle$ and their higher powers.
The group velocity is now defined by setting the second-order mixed derivative of $\Delta \rho$ evaluated at $(x,t)=(0,0)$ to zero:
\begin{equation}
\left.\frac{d^2}{dx dt} \Delta \rho(x,t)\right|_{(0,0)} = 0 \hspace{5mm} \rightarrow \hspace{5mm}
v_g = \langle \hat{\omega}\hat{k} \rangle / \langle \hat{k}^2 \rangle.
\end{equation}
We may now expand $\Delta \rho(t) \equiv \Delta \rho(x=0, t)$ into a power series in $t$. Taking the real part of the summation as denoted in Eq. \eqref{eq:Delta_rho_x_t} amounts to an even-powered series in t:
\begin{equation}\label{eq:time_series}
\Delta \rho (t) = \sum_{n=1}^\infty c_{2n} t^{2n}, \hspace{5mm}
c_{2n} = \frac{(-1)^n}{{(2n)!}} 2 \langle (\hat{\omega} - v_g \hat{k})^{2n} \rangle.
\end{equation}
The time at which the $n$th term of the series reaches a fraction $0<p<1$ of the average density is then simply given by $t=(p/c_n)^{1/n}$. We present comparison tables of such values for several categories of wavepacket parameters in Sec. \ref{sec:time_tables}.

\subsubsection{$R(t)$: Time-dependent depletion integral}\label{sec:R_t}

We now introduce a lifetime measure that takes into account the full region of depletion. We thus aim to obtain estimates that are robust to sufficiently small asymmetry and non-rigidity in the wavepacket's time evolution.
We identify the ``center-of-mass'' of the depletion as
\begin{equation}
    x_\text{cm}(t) = \frac{\int_0^L dx\,x\,[\rho_\text{max}(t)-\rho(x,t)]}{\int_0^L dx\,[\rho_\text{max}(t)-\rho(x,t)]},
\end{equation}
with periodicity of the spatial domain properly accounted for. We define the depletion region as the interval $[x_-, x_+]$, understood modulo $L$, where $\rho(x_\pm, 0)=1$, and quantify the fractional mass displacement at time $t$ as 
\begin{equation}
     R(t) = \frac{ \int_{x_-}^{x_+} dx\,|\rho(x+x_{cm}(t),t)-\rho(x,0)|}{\int_{x_-}^{x_+} dx\,[1-\rho(x,0)]}.
\end{equation}\label{eq:R_t}
The lifetime can then be defined as the smallest $t$ for which $R(t)$ reaches a prescribed threshold $R_\text{threshold}$, e.g. 10\%. As measured in this way, the lifetime takes into account changes in the spatial profile of the wavepacket with time, so long as such changes (such as asymmetric or non-rigid evaporation) are sufficiently small.\footnote{It can be argued that wavepackets showing large asymmetric or non-rigid evaporation do not constitute suitable quasi-soliton candidates in the first place.}

\subsection{Effect of wavepacket parameters on lifetime}

\subsubsection{Comparison tables of lifetime estimates}\label{sec:time_tables}

We are interested in the time at which the central point of the wavepacket has evaporated by an amount $\Delta \rho << \rho_0$. We thus look for a point in time when the central point of the initial profile has evaporated by a fraction $0<p<<1$ of the average density $\rho_0$. Let us denote the time at which the dominant term in the series expansion (\ref{eq:time_series}) reaches this fraction as $t_{c_n}^{*}$. This is then simply given by $t_{c_n}^{*}$ = $(p/c_n)^{1/n}$.

Below, we give these times for the first two terms of the series, $c_2 t^2$ and $c_4 t^4$. So long as we consider times well below the smaller of these, we can rest assured of the convergence of the series. As evident below, in most cases $c_2 t^2$ yields the smaller timescale by approximately a factor of 2, demonstrating a predominantly quadratic time evolution, which we also observe as we extend the lifetime measure beyond the central point to include the shape of the depletion, as shown in Sec. \ref{sec:R_t_plots}.

We note that since the time series focuses on only the central point of the wavepacket, one cannot presume \emph{a priori} that estimates thereby obtained would fairly represent the overall evaporation across the entire profile, especially in cases where the depletion exhibits nonrigid evaporation. However, we find that the time series estimates show remarkable agreement with those obtained via the spatially-aware $R(t)$ method. We further note that these values accurately reflect the effect of various parameters (such as interaction strength, number of holes, summation width, etc.) on lifetime, which we also observe in the time evolution of the spatial profile, as shown in the tables below. We thus conclude that, despite taking into account only a single spatial point, the time series estimates capture the quiddity of the wavepacket's time evolution in almost all cases. Note: all time estimates have been rounded to the nearest integer, unless indicated otherwise.

\subsubsection*{Coefficients: staggered vs. variational}

\begin{center}
\begin{minipage}{0.49\textwidth}
\begin{center}
{\footnotesize(i) $c=4$, $N=L=256$, $n=2$, $w$=28, $o$=114}\\
%%%%%%%%%%%%%%%%%%%%%%%%%%%%%%%%%%%%%%%%%%%%%%%%%%%%%%%%%%%
\vspace{1mm}
\begin{tabular}{||c|c|c|c||}
\hline
{\footnotesize Coefficients} & $t_{R_{10\%}}$ & $t_{c_2}^{*}$ & $t_{c_4}^{*}$ \\
 \hline
 {\footnotesize Staggered }        & 22  & 25 & 49 \\
 \hline
{\footnotesize $\Delta x = 4.5$}  &  20 & 17 & 39  \\
\hline
\end{tabular}\\
%%%%%%%%%%%%%%%%%%%%%%%%%%%%%%%%%%%%%%%%%%%%%%%%%%%%%%%%%%%
\vspace{5mm}
%%%%%%%%%%%%%%%%%%%%%%%%%%%%%%%%%%%%%%%%%%%%%%%%%%%%%%%%%%%
{\footnotesize (iii) $c=1024$, $N=L=100$, $n=2$, $w$=10, $o$=45}\\
\vspace{1mm}
\begin{tabular}{||c|c|c|c||}
\hline
{\footnotesize Coefficients} &  { $t_{R_{10\%}}$} & ${t_{c_2}^{*}}$ &   {$t_{c_4}^{*}$} \\ 
 \hline
 {\footnotesize Staggered }       & 15  & 14  & 29 \\
 \hline
 {\footnotesize $\Delta x = 1.5$} & 12 &  10  & 23\\
\hline
\end{tabular}
%%%%%%%%%%%%%%%%%%%%%%%%%%%%%%%%%%%%%%%%%%%%%%%%%%%%%%%%%%%
\end{center}
\end{minipage}
\begin{minipage}{0.49\textwidth}
\begin{center}
%%%%%%%%%%%%%%%%%%%%%%%%%%%%%%%%%%%%%%%%%%%%%%%%%%%%%%%%%%%
{\footnotesize (ii) $c=1024$, $N=L=256$, $n=2$, $w$=28, $o$=114}\\
\vspace{1mm}
\begin{tabular}{||c|c|c|c||}
\hline
{\footnotesize Coefficients} & $t_{R_{10\%}}$ & $t_{c_2}^{*}$ & $t_{c_4}^{*}$ \\
 \hline
{\footnotesize Staggered }      & 11  & 13 & 27\\
 \hline
{\footnotesize $\Delta x = 4.5$} & 10 & 9  & 20\\
\hline
\end{tabular}\\
%%%%%%%%%%%%%%%%%%%%%%%%%%%%%%%%%%%%%%%%%%%%%%%%%%%%%%%%%%%
\vspace{5mm}
%%%%%%%%%%%%%%%%%%%%%%%%%%%%%%%%%%%%%%%%%%%%%%%%%%%%%%%%%%%
{\footnotesize (iv) $c=1024$, $N=L=128$, $n=3$, $w$=18, $o$=55}\\
\vspace{1mm}
\begin{tabular}{||c|c|c|c||}
\hline
{\footnotesize Coefficients} & $t_{R_{10\%}}$ & $t_{c_2}^{*}$ & $t_{c_4}^{*}$ \\
 \hline
 {\footnotesize Staggered }          & 8 & 9 & 19\\
 \hline
 {\footnotesize $\Delta x = 3.5$}    & 8 & 417 & 15 \\
\hline
\end{tabular}
%%%%%%%%%%%%%%%%%%%%%%%%%%%%%%%%%%%%%%%%%%%%%%%%%%%%%%%%%%%
\end{center}
\end{minipage}
\end{center}
%%%%%%%%%%%%%%%%%%%%%%%%%%%%%%%%%%%%%%%%%%%%
\textit{(i) Time evolution of corresponding wavepackets shown in Fig. \ref{fig:c_4_n_2_n_3_stag_var_final} (a-c); lifetime curves in Fig. \ref{c_1024_N_128_n_2_3_w_18_o_55_stag_var}(a).}
Both methods yield an evaporation of $\sim 10\%$ at $t\approx20$, in line with direct observations. The prediction again points to slightly longer-lived staggered wavepackets, in line with our observations and $R(t)$ estimates.\\\\
%%%%%%%%%%%%%%%%%%%%%%%%%%%%%%%%%%%%%%%%%%%%
\textit{(ii) Corresponding to wavepackets shown in Figs. \ref{fig:c_1024_n_2_n_3_stag_var_final}(a-c) and \ref{c_1024_N_128_n_2_3_w_18_o_55_stag_var}(b).}
We note here that both methods demonstrate a $\sim 2$-fold increase in lifetime from $c=1024$ to $c=4$ for both the staggered and the variational wavepackets, in line with our expectation of longer lifetimes for decreasing $c$ as discussed in Sec. \ref{sec:wavepackets_finite_c}.\\\\
%%%%%%%%%%%%%%%%%%%%%%%%%%%%%%%%%%%%%%%%%%%%
\textit{(iii) Time evolution and lifetime curves of corresponding wavepackets shown in Fig. \ref{fig:c_1024_N_100_n_2_w_10_o_45_stag_var}.}
We directly observe $10\%$ evaporation of the central point of the staggered wavepacket by $t=15$, in line with results from our $R(t)$ method and that of the time-series. We note here a slight advantage in the lifetime of the staggered wavepacket, which we also observe directly and via the $R(t)$ method.\\\\
%%%%%%%%%%%%%%%%%%%%%%%%%%%%%%%%%%%%%%%%%%%%
\textit{(iv) Corresponding to wavepackets shown in Figs. \ref{fig:c_1024_n_2_n_3_stag_var_final}(d-f) and \ref{c_1024_N_128_n_2_3_w_18_o_55_stag_var}(c).}
In this case, the central point of the variational wavepacket demonstrates a predominantly quartic behavior, as evidenced by the order-of-magnitude ratio between the $t_{c_2}^*$ and $t_{c_4}^*$ terms. This behavior is clearly visible in Figs. 
\ref{fig:c_1024_n_2_n_3_stag_var_final}(d-f), which demonstrate a flat depletion with the central point remaining nearly fixed as nearby points evaporate. This constitutes a case of clearly non-rigid evaporation, for which the spatially-aware $R(t)$ method yields more reliable estimates.
%%%%%%%%%%%%%%%%%%%%%%%%%%%%%%%%%%%%%%%%%%%%
\subsubsection*{Interaction parameter $c$}
\begin{minipage}{0.49\textwidth}
\begin{table}[H]
\centering{{\footnotesize(i) $N=L=100$, $n=1$, $w=10$, $o=45$, staggered}}\vspace{1mm} 
\label{}
\begin{tabularx}{0.9\textwidth} { 
  || >{\centering\arraybackslash}X   
  | >{\centering\arraybackslash}X 
  | >{\centering\arraybackslash}X 
  | >{\centering\arraybackslash}X ||}
 \hline
$c$ \vspace{3pt} & $t_{R_{10\%}}$ & $t_{c_2}^{*}$ & $t_{c_4}^{*}$ \\
 \hline
 0.1        & > 80 & 144 & 326\\
 \hline
 1          & 44 & 43 & 98 \\
 \hline
 10         & 16 & 15 & 34  \\
 \hline
 100        & 11 & 11 & 25  \\ 
 \hline
 1000       & 11 & 11  & 24 \\
  \hline
\end{tabularx}
\end{table}
\end{minipage}
\begin{minipage}{0.49\textwidth}
\begin{table}[H]
\centering{{\footnotesize (ii) $N=L=100$, $n=2$, $w$=10, $o$=45, staggered}}\vspace{1mm} 
\label{}
\begin{tabularx}{0.9\textwidth} { 
  || >{\centering\arraybackslash}X   
  | >{\centering\arraybackslash}X 
  | >{\centering\arraybackslash}X 
  | >{\centering\arraybackslash}X ||}
 \hline
 $c$ \vspace{3pt}  & $t_{R_{10\%}}$ & $t_{c_2}^{*}$ & $t_{c_4}^{*}$ \\
 \hline
 0.1        & > 80  & 57  & 143 \\
 \hline
 1          &  56    & 43  & 88 \\
 \hline
 10         &  20   & 19  & 40  \\
 \hline
 1000       & 14    & 14  & 29 \\
  \hline
\end{tabularx}
\vspace{5mm}
\end{table}

\end{minipage}\vspace{5mm}

\begin{figure}[H]
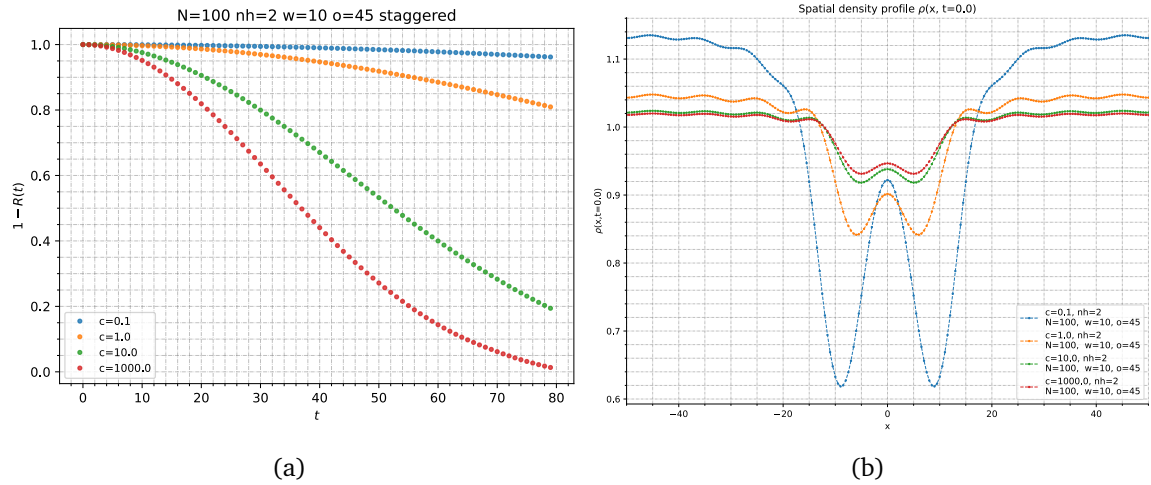

    \centering
    \subfloat[]{\includesvg[width=0.52\linewidth]{sections/figures/lifetime_plots/lifetime_comparisons/c_scaling_N_100_nholes_2_w_10_o_45_v2_2.svg}}
    \subfloat[]{\includesvg[width=0.5\linewidth]{sections/figures/wavepacket_profiles/profile_comparisons/t0_c_scaling_n_2_N_100_w_10_o_45.svg}}
    \caption{Lifetime curves (a) and initial profiles (b) of $n=2$-hole wavepackets at multiple values of $c$, $N=L=100$, $w=10$, $o=45$.}
    \label{fig:c_scaling_n_2_N_100_w_10_o_45}
\end{figure}

\noindent \textit{(i) Corresponding to wavepackets shown in Fig. \ref{fig:c_scaling_N_100_n_1_w_10_o_45}.}\\
Here, we note that the two methods show remarkable agreement, and capture the improvement in lifetime with decreasing $c$ in the expected manner, as shown in the dispersion curvature plots in Sec. \ref{sec:wavepackets_finite_c}. \\\\
%%%%%%%%%%%%%%%%%%%%%%%%%%%%%%%%%%%%%%%
\noindent \textit{(ii) Corresponding  $c=1$ wavepacket shown in Fig. \ref{fig:c_1_n_2_o_45_stag}; lifetime curves in Fig. \ref{fig:c_scaling_n_2_N_100_w_10_o_45}.}\\\\
Similar to the single-hole case, we observe a gain in lifetime with decreasing $c$. We note, however, that the agreement between the two estimates diminishes at small $c$, especially as compared to the $n=1$ case.

\subsubsection*{Number of holes $n$}
\begin{figure}[H]
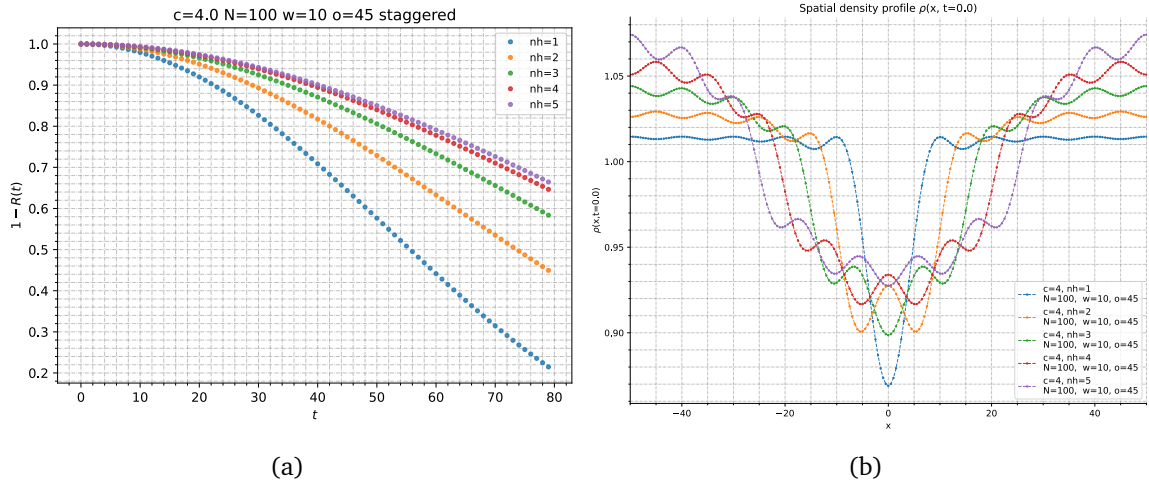

    \subfloat[]{\includesvg[width=0.52\linewidth]{sections/figures/lifetime_plots/lifetime_comparisons/c_4_N_100_n_1_2_3_4_5_w_10_o_45_v2_2.svg}}
   \subfloat[]{\includesvg[width=0.5\linewidth]{sections/figures/wavepacket_profiles/profile_comparisons/t0_c_4_n_1_2_3_4_5_N_100_w_10_o_45.svg}}
    \caption{Lifetime curves (a) and initial profiles (b) of multi-hole wavepackets at $c=4$, $N=L=100$, $w=10$, $o=45$.}
    \label{fig:c_4_N_100_n_1_2_3_4_5_w_10_o_45}
\end{figure}

\begin{table}[H]
\centering
\caption{$c=4$ $N=L=100$, $w=10$, $o=45$, staggered}
\label{}
\begin{tabularx}{0.45\textwidth} { 
  || >{\centering\arraybackslash}X   
  | >{\centering\arraybackslash}X 
  | >{\centering\arraybackslash}X 
  | >{\centering\arraybackslash}X ||}
 \hline
 $n$ \vspace{3pt}  & $t_{R_{10\%}}$ & $t_{c_2}^{*}$ & $t_{c_4}^{*}$ \\
 \hline
 1       &  22  & 21  & 49 \\
 \hline
 2        & 29 &  26  & 54  \\
 \hline
 3        & 34  & 38  & 103   \\
 \hline
 4        & 40  &  37  & 96  \\
 \hline
 5        & 40  &  63  & 124   \\
\hline
\end{tabularx}
\end{table}
\textit{Corresponding lifetime curves in Fig. \ref{fig:c_4_N_100_n_1_2_3_4_5_w_10_o_45}}.\\\\
Once again, both estimates are in line with our observation that increasing the number of holes improves the lifetime of the wavepacket, as also depicted in lifetime curves in Fig. \ref{fig:c_4_N_100_n_1_2_3_4_5_w_10_o_45}.

\subsubsection*{Summation offset $o$}

\begin{table}[H]
{\raggedright \caption{$N=L=100$, $w=10$, staggered}\label{table:t_series_o}}
{\hspace{26mm} \centering (i) $n=1$ \hspace{55mm} (ii) $n=2$}\vspace{1mm}\\
\begin{tabularx}{0.45\textwidth}{
  || >{\centering\arraybackslash}X
  | >{\centering\arraybackslash}X
  | >{\centering\arraybackslash}X
  | >{\centering\arraybackslash}X
  | >{\centering\arraybackslash}X ||}
 \hline
 $c$ & $o$ & $t_{R_{10\%}}$ & $t_{c_2}^{*}$ & $t_{c_4}^{*}$ \\
 \hline
 1024 & 0  & 11 & 10 & 24 \\
 \cline{2-5}
      & 45 & 11 & 10 & 24 \\
 \hline
 1    & 0  & 31 & 29 & 67 \\
 \cline{2-5}
      & 45 & 45 & 43 & 98 \\
 \hline
\end{tabularx}
\hspace{2mm}
\begin{tabularx}{0.45\textwidth}{
  || >{\centering\arraybackslash}X
  | >{\centering\arraybackslash}X
  | >{\centering\arraybackslash}X
  | >{\centering\arraybackslash}X
  | >{\centering\arraybackslash}X ||}
 \hline
 $c$ & $o$ & $t_{R_{10\%}}$ & $t_{c_2}^{*}$ & $t_{c_4}^{*}$ \\
 \hline
 1024 & 0  & 14 & 14 & 29 \\
 \cline{2-5}
      & 45 & 14 & 14 & 29 \\
 \hline
 1    & 0  & 39 & 31 & 64 \\
 \cline{2-5}
      & 45 & 56 & 43 & 88 \\
 \hline
\end{tabularx}
\end{table}

\begin{figure}[H]
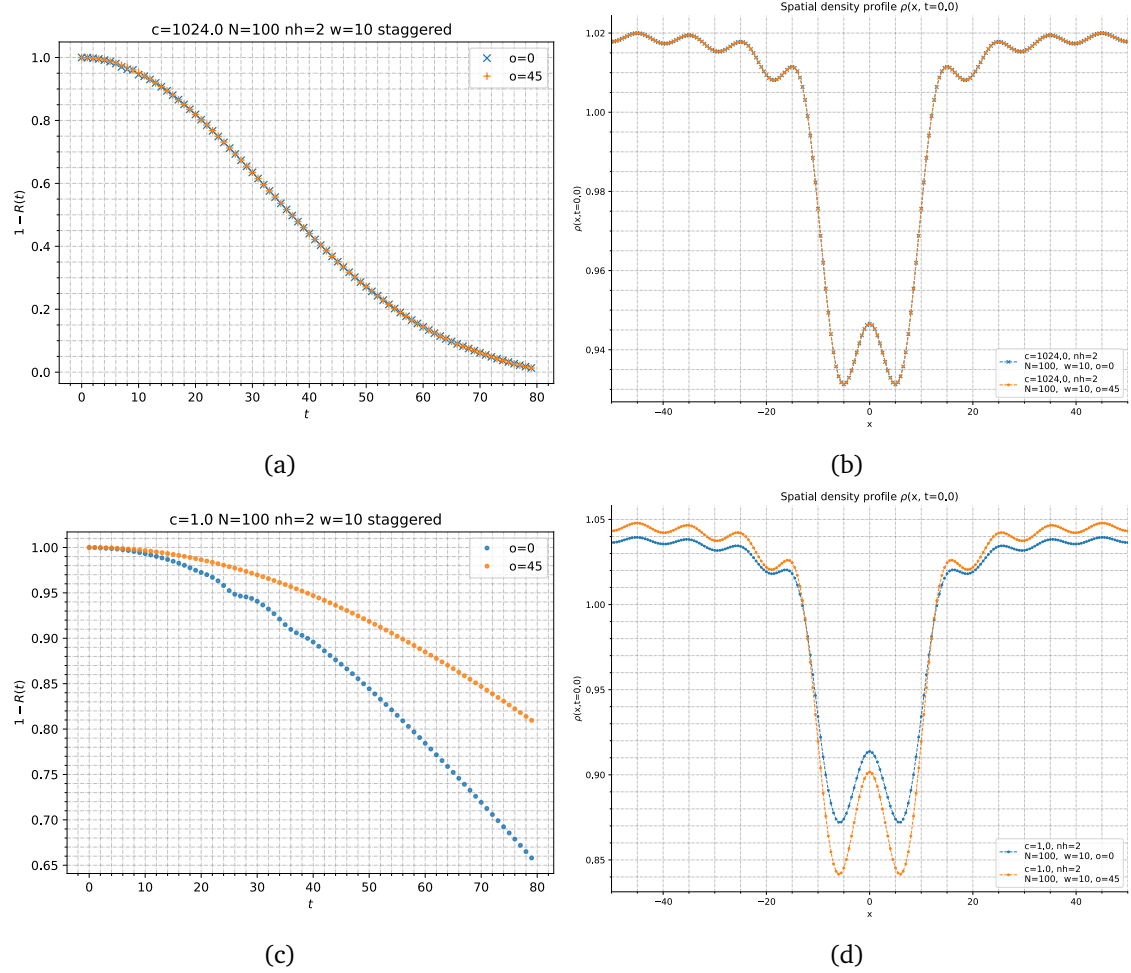

   \subfloat[]{\includesvg[width=0.5\linewidth]{sections/figures/lifetime_plots/lifetime_comparisons/R_t_vs_o/c_1024_N_100_n_2_w_10_o_0_45_stag_v2_2.svg}}
   \subfloat[]{\includesvg[width=0.5\linewidth]{sections/figures/lifetime_plots/lifetime_comparisons/R_t_vs_o/tStar_0p0_c_1024_N_100_L_100_nholes_2_width_10_offset_0_45_staggered.svg}}\\
   \subfloat[]{\includesvg[width=0.5\linewidth]{sections/figures/lifetime_plots/lifetime_comparisons/R_t_vs_o/c_1_N_100_n_2_w_10_o_0_45_stag_v2_2.svg}}
    \subfloat[]{\includesvg[width=0.5\linewidth]{sections/figures/lifetime_plots/lifetime_comparisons/R_t_vs_o/tStar_0p0_c_1_N_100_L_100_nholes_2_width_10_offset_0_45_staggered.svg}}
    \caption{$R(t)$: Lifetime curves (left) and initial profiles (right) of $n$=2-hole staggered wavepackets at $c=\{1,~1024\}, ~N=L=100$, width = 10, offset = $\{0,~45\}$.}
    \label{fig:c_1_1024_N_100_n_2_w_10_o_0_45}
\end{figure}

%%%%%%%%%%%%%%%%%%%%%%%%%%%%%%%%%%%%%%%%%%%%
\textit{Corresponding to lifetime curves shown in Figs. \ref{fig:c_1_1024_N_100_n_1_w_10_o_0_45} and \ref{fig:c_1_1024_N_100_n_2_w_10_o_0_45}}.\\\\
Precisely as expected---grounded on both dispersion-based arguments presented in Sec. \ref{sec:selective_width_summations} and lifetime curves in Figs. \ref{fig:c_1_1024_N_100_n_1_w_10_o_0_45} and \ref{fig:c_1_1024_N_100_n_2_w_10_o_0_45}---introducing an offset at large $c$ (where the curvature of the dispersion approaches a constant value) has no effect on the lifetime of the wavepacket, but displays a marked advantage at finite $c$ with placement of the summation interval in the middle of the Fermi sea ($o=45$). Again, the agreement between the two methods subsides at smaller $c$. We note that both the time-series and $R(t)$ method predict a $\sim 1.4$-fold gain in lifetime at $c=1$ with placement of the interval in the middle of the Fermi sea. This is in line with the energetics-based arguments presented in Sec. \ref{sec:selective_width_summations}, where the inverse curvature of the type II dispersion at $c=1$ is shown to be $\sim1.5$ times larger at the center of the Fermi sea than near the edge (Figs. \ref{fig:typeII_disp} and \ref{fig:typeII_inv_curv}). 

\subsubsection*{Summation width $w$}
\begin{table}[H]
\centering
\caption{$c=4$, $N=L=300$, $n=1$, staggered}
\label{}
\begin{tabularx}{0.65\textwidth} { 
  || >{\centering\arraybackslash}X   
  | >{\centering\arraybackslash}X 
  | >{\centering\arraybackslash}X 
  | >{\centering\arraybackslash}X ||}
 \hline
 $w$ \vspace{3pt}   & $t_{R_{10\%}}$ & $t_{c_2}^{*}$ & $t_{c_4}^{*}$ \\
 \hline
 {\small $300 ~~(o=0)$}       &  <1   & 0.25  &  0.52 \\
 \hline
  {\small $150 ~~(o=75)$}     &   1   & 0.9   &  1.9  \\
 \hline
  {\small $60 ~~(o=120)$}     &   5   &   5   &  12   \\
 \hline
  {\small $30 ~~(o=135)$}      &   23   &  21  &  48   \\
 \hline
  {\small $16 ~~(o=142)$}      &   78   &  75   &  168  \\
\hline

\end{tabularx}
\end{table}
\textit{Corresponding to lifetime curves shown in Fig. \ref{fig:c_4_n_1_w_scaling}}.\\\\
In line with the scaling arguments presented in Sec. \ref{sec:scaling_arg}, both methods yield a drastic improvement with each step of reducing the summation interval, yielding similar gain factors at each step. We observe this scaling in the lifetime curves presented in Fig. \ref{fig:c_4_n_1_w_scaling}.

\subsubsection*{Scaling with system size $N$, summation width $w$}
\begin{table}[H]
\centering
\caption{$c=1024$, $n=2$, variational}
\label{}
\begin{tabularx}{0.9\textwidth} { 
  || >{\centering\arraybackslash}X   
  | >{\centering\arraybackslash}X 
  | >{\centering\arraybackslash}X 
  | >{\centering\arraybackslash}X ||}
 \hline
 $N, w$ \vspace{3pt}   & $t_{R_{10\%}}$ & $t_{c_2}^{*}$ & $t_{c_4}^{*}$ \\
 \hline
$N=32, w=8$ {\small($o=12$, $\Delta x=2.4$)}            & 1.8   &  1.6  &  3.8   \\
 \hline
 $N=64, w=12$ {\small($o=26, ~\Delta x=3$)}            & 3.4   &  2.9  &  6.7   \\
 \hline
 $N=128, w=18$ {\small($o=55, \Delta x=3.5$)}          &   6   & 5.2   &  12   \\
 \hline
 $N=256, w=28$ {\small($o=114, \Delta x=4.5$)}  &  10   & 8.8   &  20   \\
\hline
\end{tabularx}
\end{table}
Note that $N=L$ for all entries. Corresponding to wavepackets shown in Fig. \ref{fig:c_1024_n_2_N_scaling_stag_var}(d-f) and lifetime curves in Fig. \ref{fig:c_1024_n_2_N_scaling_var_R_t}.\\\\
Here, we note that the time series underestimates $R(t)$ results by $\sim 10-15\%$, but displays the same gain factor (1.7- to 1.9-fold increase) at each step as does $R(t)$. The scaling behavior is also shown in Fig. \ref{fig:c_1024_n_2_N_scaling_var_R_t_rescaled}. 
\subsubsection{Plots of depletion integrals vs. time}\label{sec:R_t_plots}

\begin{figure}
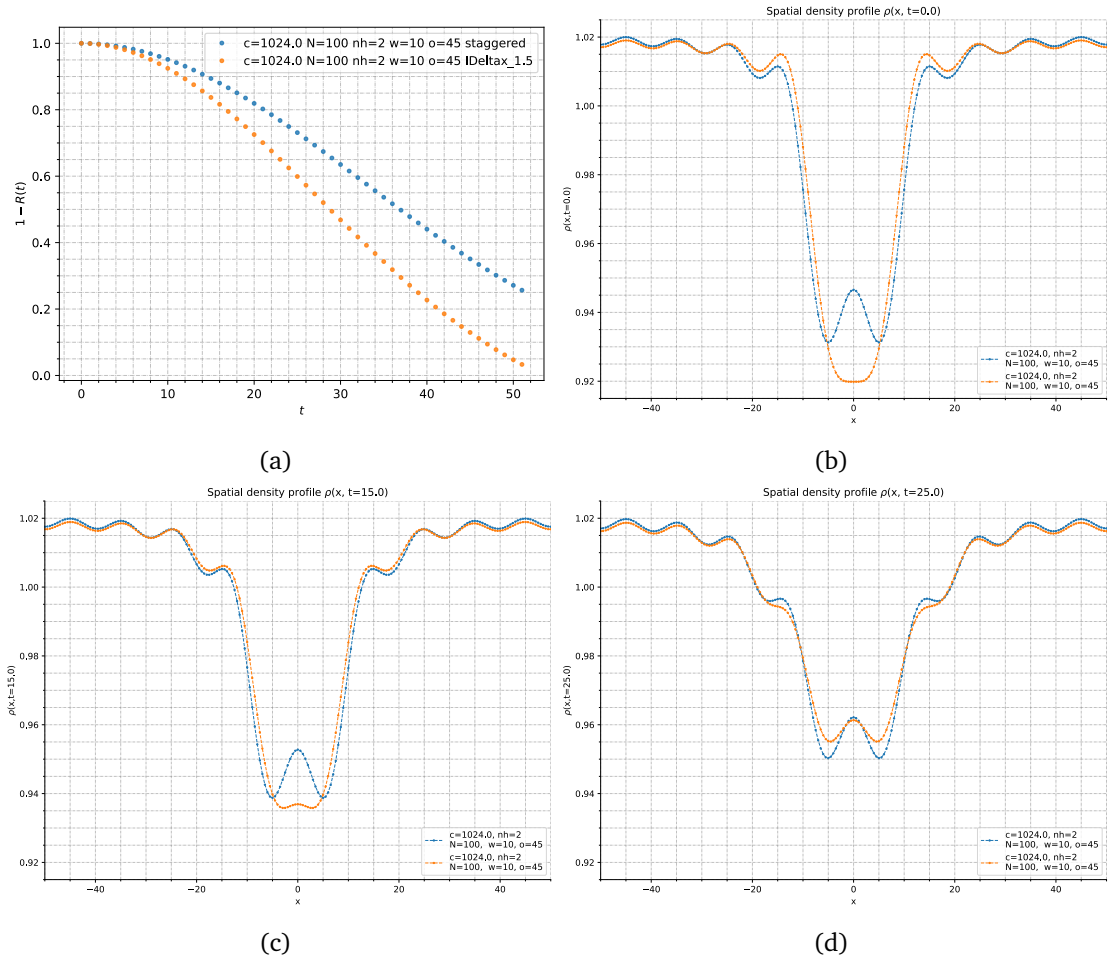

\centering
    \subfloat[]{\includesvg[width=0.49\linewidth]{sections/figures/lifetime_plots/lifetime_comparisons/c_1024_N_100_L_100_nholes_2_width_10_offset_45_stag_IDeltax.svg}}
    \subfloat[]{\includesvg[width=0.49\linewidth]{sections/figures/wavepacket_profiles/profile_comparisons/tStar_0p0_c_1024_N_100_L_100_nholes_2_width_10_offset_45_staggered_IDeltax_1p5.svg}}\\
    \subfloat[]{\includesvg[width=0.49\linewidth]{sections/figures/wavepacket_profiles/profile_comparisons/tStar_15p0_c_1024_N_100_L_100_nholes_2_width_10_offset_45_staggered_IDeltax_1p5.svg}}
    \subfloat[]{\includesvg[width=0.49\linewidth]{sections/figures/wavepacket_profiles/profile_comparisons/tStar_25p0_c_1024_N_100_L_100_nholes_2_width_10_offset_45_staggered_IDeltax_1p5.svg}}
    \caption{Depletion integral (a) and time-evolved profiles (b)-(d) of variational vs. staggered $n=2$-wavepackets with $N=L=100$, $w=10$, $o=45$ at $c=1024$.} \label{fig:c_1024_N_100_n_2_w_10_o_45_stag_var}
\end{figure}

\begin{figure}
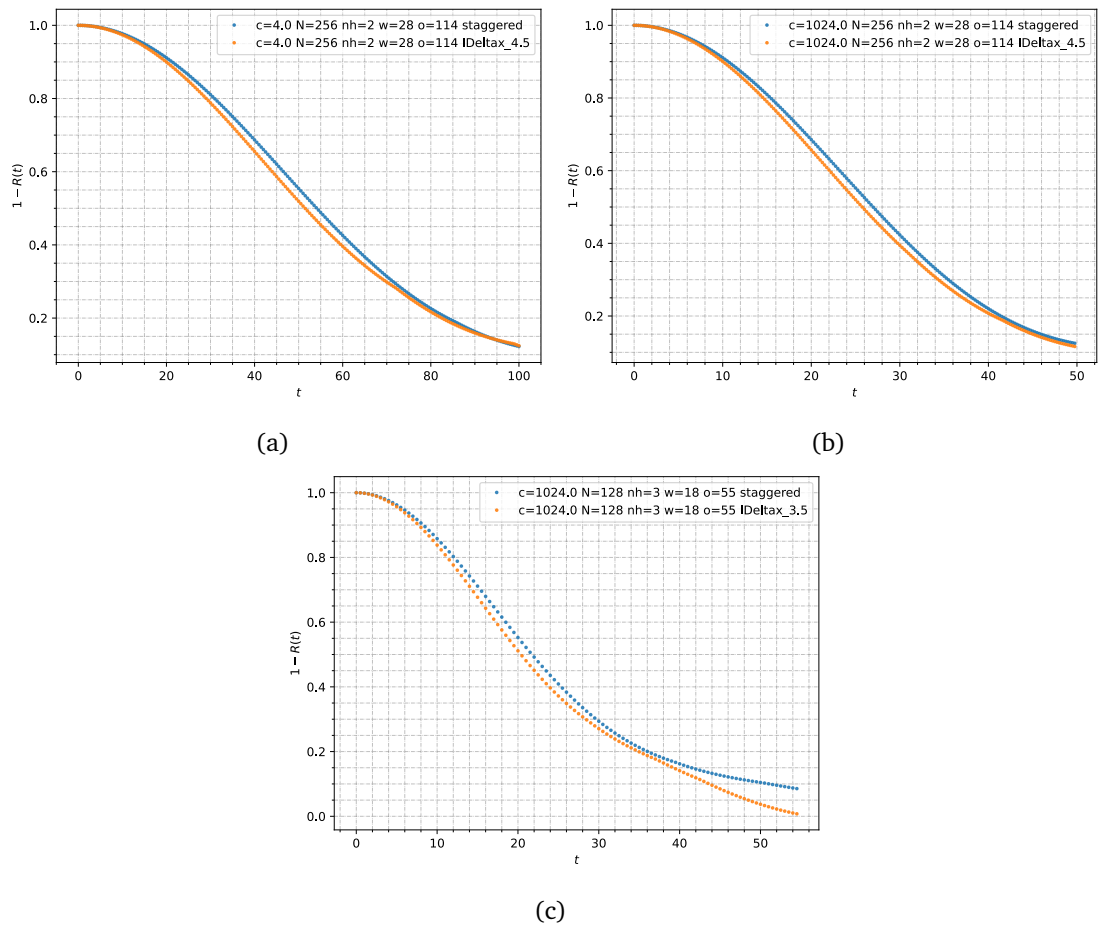

\centering
    \subfloat[]{\includesvg[width=0.49\linewidth]{sections/figures/lifetime_plots/lifetime_comparisons/c_4_N_256_L_256_nholes_2_width_28_offset_11_staggerd_IDeltax_4p5.svg}}
    \subfloat[]{\includesvg[width=0.49\linewidth]{sections/figures/lifetime_plots/lifetime_comparisons/c_1024_N_256_L_256_nholes_2_width_28_offset_11_staggerd_IDeltax_4p5.svg}}\\
    \subfloat[]{\includesvg[width=0.49\linewidth]{sections/figures/lifetime_plots/lifetime_comparisons/c_1024_N_128_L_128_nholes_3_width_18_offset_55_staggered_IDeltax_3p5_R_t.svg}}
    \caption{Lifetime curves of variational vs. staggered $n=2$-hole (a), (b) and 3-hole (c) wavepackets at $c=4, 1024$.} \label{c_1024_N_128_n_2_3_w_18_o_55_stag_var}
\end{figure}

Let us now examine the full behavior of the depletion---as defined in Eq. \ref{eq:R_t}---under time evolution. The plots below (Figs. \ref{fig:c_1024_N_100_n_2_w_10_o_45_stag_var} and \ref{c_1024_N_128_n_2_3_w_18_o_55_stag_var}) depict the same categories as presented and discussed in Sec. \ref{sec:time_tables}, with the corresponding wavepackets shown in Figs. \ref{fig:c_4_n_2_n_3_stag_var_final} and \ref{fig:c_1024_n_2_n_3_stag_var_final}.
\noindent We note that the 2-hole wavepackets at $c=4$ take nearly twice as long to evaporate than their $c=1024$ counterparts, which is also clearly visible in the time-evolved profiles shown in Figs. \ref{fig:c_4_n_2_n_3_stag_var_final} and \ref{fig:c_1024_n_2_n_3_stag_var_final}. We observe the same two (or more)-fold increase lifetime in many cases comparing wavepackets at $c \sim \mathcal{O}(1)$ and $c \sim \mathcal{O}(10^3)$.
The $R(t)$ plots in Fig. \ref{c_1024_N_128_n_2_3_w_18_o_55_stag_var} suggest that the lifetimes of the variational and staggered are nearly the same, whereas their profiles in Figs. \ref{fig:c_4_n_2_n_3_stag_var_final} and \ref{fig:c_1024_n_2_n_3_stag_var_final}, as well as the corresponding time-series estimates presented in section \ref{sec:time_tables}, indicate that the variational wavepacket evaporates faster. Here we may conclude that there is likely no single measure which, by itself, accurately captures every aspect of time evolution for an arbitrary set of parameters.

\subsubsection*{Effect of $n$ on $R(t)$}

\subsubsection*{Effect of $c$ on $R(t)$}
\begin{figure}[H]
    \centering
    \subfloat[]{\includesvg[width=0.52\linewidth]{sections/figures/lifetime_plots/lifetime_comparisons/c_scaling_N_100_n_1_w_10_o_45_v2_2.svg}}
    \subfloat[]{\includesvg[width=0.5\linewidth]{sections/figures/wavepacket_profiles/profile_comparisons/t0_c_scaling_n_1_N_100_w_10_o_45.svg}}
    \caption{}
    \caption{Lifetime curves (a) and initial profiles (b) of one-hole wavepackets at multiple values of $c$, $N=L=100$, $w=10$, $o=45$.}
    \label{fig:c_scaling_N_100_n_1_w_10_o_45}
\end{figure}

\subsubsection*{Effect of $o$ on $R(t)$}
\textbf{n=1}
\begin{figure}[H]
   \subfloat[]{\includesvg[width=0.5\linewidth]{sections/figures/lifetime_plots/lifetime_comparisons/R_t_vs_o/c_1024_N_100_n_1_w_10_o_0_45_stag.svg}}
   \subfloat[]{\includesvg[width=0.5\linewidth]{sections/figures/lifetime_plots/lifetime_comparisons/R_t_vs_o/tStar_0p0_c_1024_N_100_L_100_nholes_1_width_10_offset_0_45_staggered.svg}}\\
   \subfloat[]{\includesvg[width=0.5\linewidth]{sections/figures/lifetime_plots/lifetime_comparisons/R_t_vs_o/c_1_N_100_n_1_w_10_o_0_45_stag.svg}}
    \subfloat[]{\includesvg[width=0.5\linewidth]{sections/figures/lifetime_plots/lifetime_comparisons/R_t_vs_o/tStar_0p0_c_1_N_100_L_100_nholes_1_width_10_offset_0_45_staggered.svg}}
    \caption{$R(t)$: Lifetime curves (left) and initial profiles (right) of $n$=1-hole staggered wavepackets at $\{c=1,~1024\}, ~N=L=100$, width = 10, offset = $\{0,~45\}$.}
    \label{fig:c_1_1024_N_100_n_1_w_10_o_0_45}
\end{figure}

\textbf{n=2}

\subsubsection*{$R(t)$ vs. $w$; Verification of lifetime scaling with $w^{-2}$}

\begin{figure}[H]
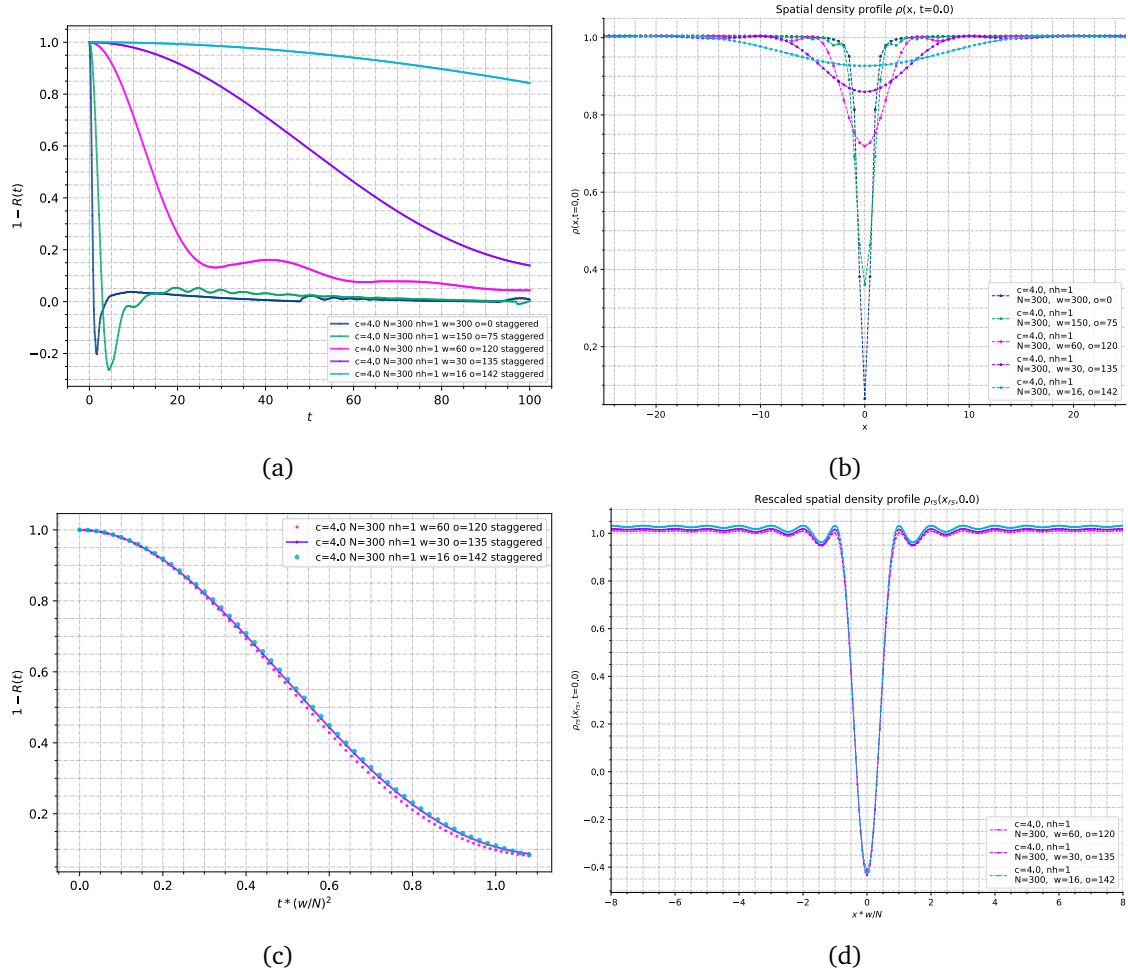

    \subfloat[]{\includesvg[width=0.5\linewidth]{sections/figures/lifetime_plots/lifetime_comparisons/R_t_vs_w/c_4_N_300_n_1_staggered_width_scaling_R_t.svg}\label{fig:c_4_n_1_w_scaling_R_t}}
   \subfloat[width=0.2\linewidth][]{\includesvg[width=0.5\linewidth]{sections/figures/lifetime_plots/lifetime_comparisons/R_t_vs_w/tStar_0p0_c_4_N_300_L_300_n_1_w_300_150_60_30_16_staggered.svg}}\\
    \subfloat[width=0.1\textwidth][]{\includesvg[width=0.5\linewidth]{sections/figures/lifetime_plots/lifetime_comparisons/R_t_vs_w/c_4_N_300_n_1_staggered_width_scaling_full.svg}\label{fig:c_4_n_1_w_scaling_rescaled_R_t}}
    \subfloat[width=0.1\textwidth][]{\includesvg[width=0.5\linewidth]{sections/figures/lifetime_plots/lifetime_comparisons/R_t_vs_w/tStar_0p0_c_4_N_300_n_1_w_60_30_16_staggered_rescaled.svg}\label{fig:c_4_n_1_w_scaling_rescaled_profiles}}
    \caption{Effect of summation width $w$ on $n=1$-hole staggered wavepacket lifetime and initial profile: Lifetime curves (a) and corresponding initial profiles (b) for single-hole wavepackets at $c=4, ~N=L=300$, $w=\{300,~150,~60,~30,~16\}$. Scaling of lifetime with $w^{-2}$ (c) and initial profiles rescaled by $w/N$ (d) for $w=\{60, ~30, ~16\}$.}
   \label{fig:c_4_n_1_w_scaling}
\end{figure}
In Fig. \ref{fig:c_4_n_1_w_scaling_rescaled_profiles}, we rescale the initial profiles in the following manner:
\begin{center}
$x_{rs} = x\cdot \frac{w}{N}$\\
$\rho_{rs} = \frac{(\rho - \rho_{avg})}{w/N} +\rho_{avg}$, ~~
$\rho_{avg} := 1 + 1/N$.
\end{center}
For the lifetime plots in Fig. \ref{fig:c_4_n_1_w_scaling_rescaled_R_t} the scaling relation is $t_{rs}=t \cdot \frac{w^2}{N^2}$, using a quadratic relation between energies and momenta. The lifetime curves in Fig. \ref{fig:c_4_n_1_w_scaling_R_t} demonstrate a significant improvement of lifetime with decreasing summation width: comparing for instance wavepackets with $w=60$ and $w=16$, we observe an order-of-magnitude improvement in lifetime, highlighting the drastic gain in stability. Let us make a brief remark on the first two curves ($w$=\{300,~150\}) in Fig. \ref{fig:c_4_n_1_w_scaling_R_t}: given their exceedingly quick evaporation (as compared with smaller summation-width cases), it is natural to expect that the displaced mass quickly exceeds that of the initial depletion, thus leading to negative values as a consequence. Thus, lifetime curves should be taken into account only for times \textit{prior} to this threshold: the more slowly (and less drastically) the wavepacket changes shape, the longer we can take $R(t)$ to be a reliable reflection of its behavior.

In  Fig. \ref{fig:c_4_n_1_w_scaling_rescaled_R_t}, we clearly observe the expected scaling behavior of lifetime with $w^2$, as we outlined in Sec. \ref{sec:intro} using energetics-based arguments. More significantly, the adherence of the lifetime curves to this scaling persists to very long times: the plot includes points up to a final non-rescaled times of $t_f \approx \{27, ~110, ~385\}$ for wavepackets with with $w = \{60,~ 30,~ 16\}$, respectively. We thus conclude that despite deviations from quadraticity of the dispersion at finite-$c$, the $w^2$ scaling accurately captures the improvement in stability of single-hole staggered wavepackets.

\subsubsection*{$R(t)$ vs. $N$; Effect of $N$-scaling on lifetime}
\subsubsection*{Staggered scaling with $N$}

\begin{figure}[H]
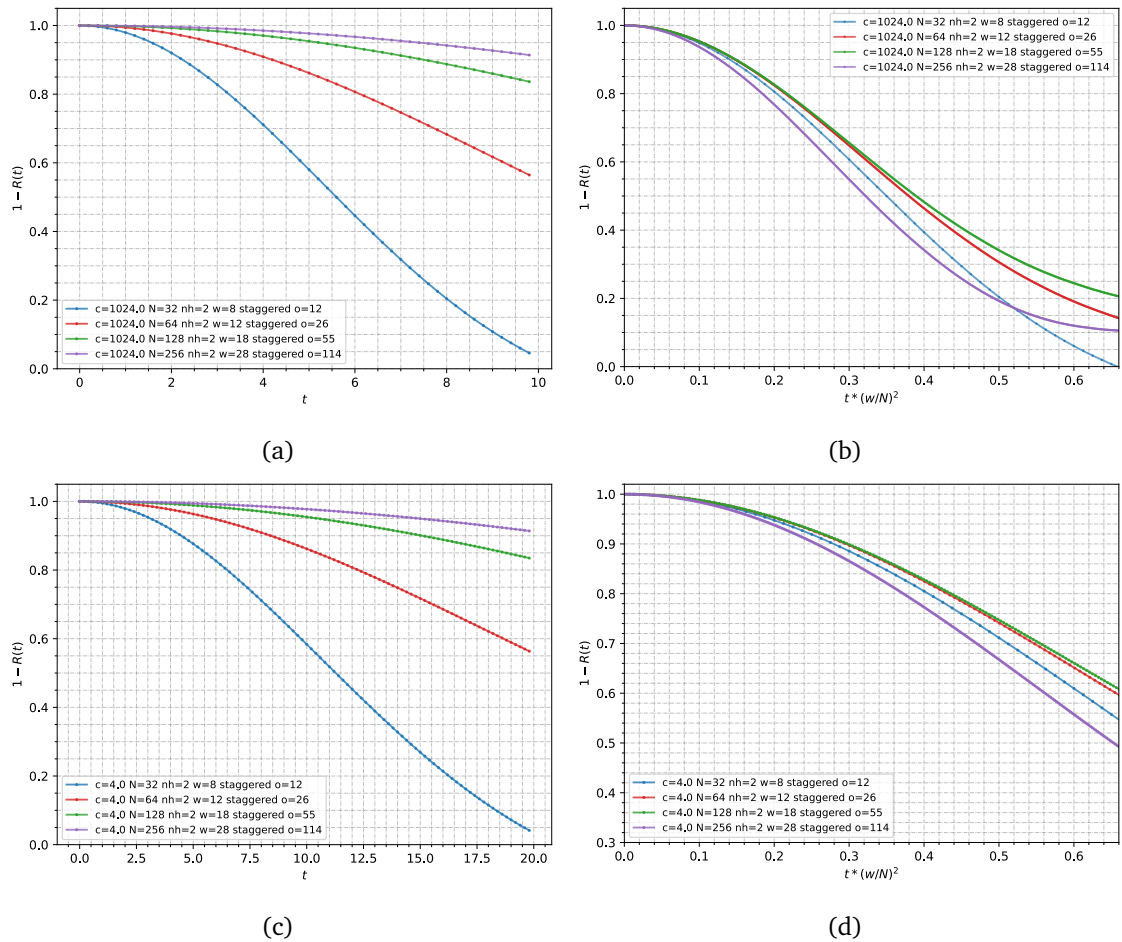

    \subfloat[]{\includesvg[width=0.5\linewidth]{sections/figures/lifetime_plots/lifetime_comparisons/R_t_vs_N_scaling/c_1024_n_2_N_scaling_stag_t_10.svg}\label{fig:c_1024_n_2_N_scaling_stag_R_t}}
    \subfloat[]{\includesvg[width=0.5\linewidth]{sections/figures/lifetime_plots/lifetime_comparisons/R_t_vs_N_scaling/c_1024_n_2_N_scaling_stag_rescaled.svg}\label{fig:c_1024_n_2_N_scaling_stag_R_t_rescaled}}\\
    \subfloat[]{\includesvg[width=0.5\linewidth]{sections/figures/lifetime_plots/lifetime_comparisons/R_t_vs_N_scaling/c_4_n_2_N_scaling_stag_t_20.svg}\label{fig:c_4_n_2_N_scaling_stag_R_t}}
    \subfloat[]{\includesvg[width=0.5\linewidth]{sections/figures/lifetime_plots/lifetime_comparisons/R_t_vs_N_scaling/c_4_n_2_N_scaling_stag_rescaled.svg}\label{fig:c_4_n_2_N_scaling_stag_R_t_rescaled}}
    \caption{Scaling of lifetime curves for $n=2$-hole staggered wavepackets at $c=1024$ (a) and $c=4$ (c). Plots in the right column show the curves rescaled by $(w/N)^2$.}
\end{figure}
\begin{figure}[H]
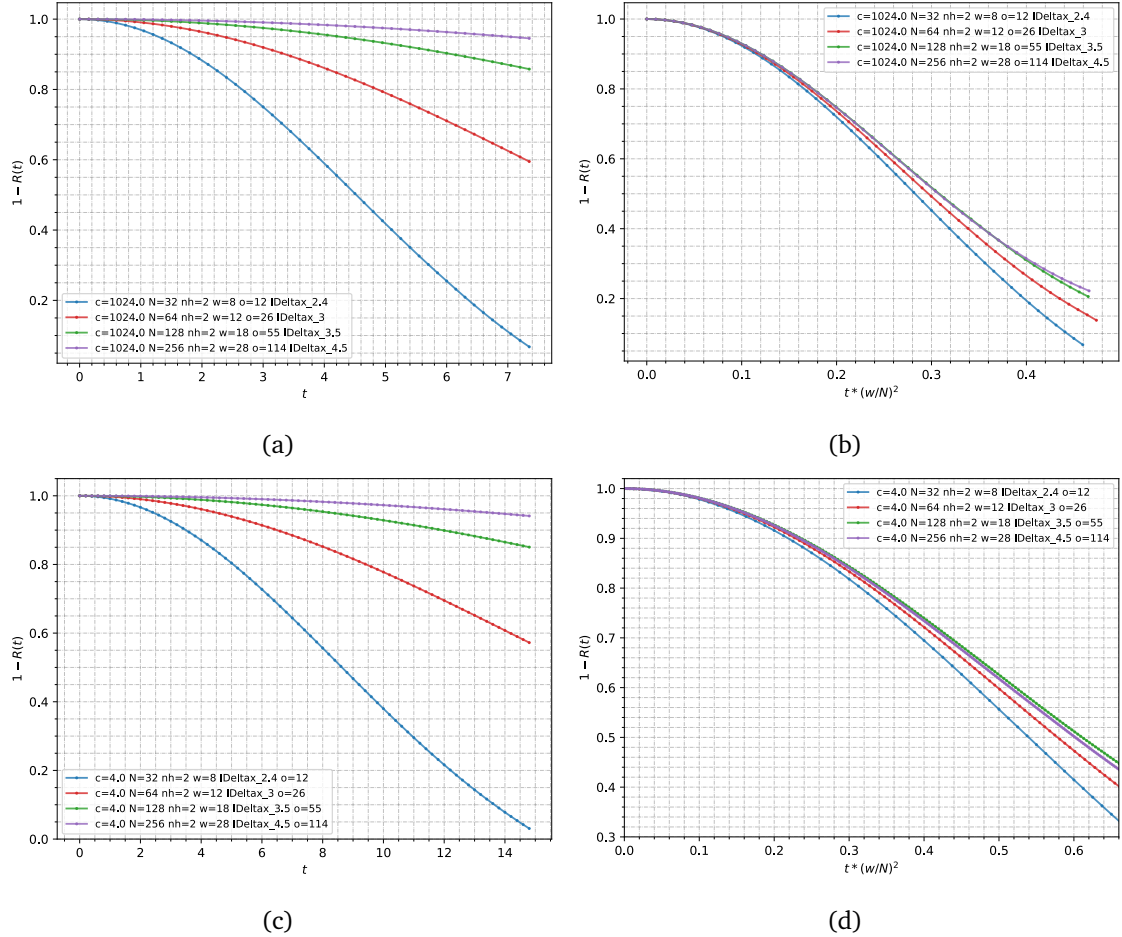

    \subfloat[]{\includesvg[width=0.5\linewidth]{sections/figures/lifetime_plots/lifetime_comparisons/R_t_vs_N_scaling/scaling_c_1024_n_2_IDeltax.svg}\label{fig:c_1024_n_2_N_scaling_var_R_t}}
    \subfloat[]{\includesvg[width=0.5\linewidth]{sections/figures/lifetime_plots/lifetime_comparisons/R_t_vs_N_scaling/scaling_c_1024_n_2_IDeltax_rescale_full.svg}\label{fig:c_1024_n_2_N_scaling_var_R_t_rescaled}} \\
    \subfloat[]{\includesvg[width=0.5\linewidth]{sections/figures/lifetime_plots/lifetime_comparisons/R_t_vs_N_scaling/c_4_n_2_N_scaling_var_t_15.svg}\label{fig:c_4_n_2_N_scaling_var_R_t}}
   \subfloat[]{\includesvg[width=0.5\linewidth]{sections/figures/lifetime_plots/lifetime_comparisons/R_t_vs_N_scaling/c_4_n_2_N_scaling_var_rescaled.svg}\label{fig:c_4_n_2_N_scaling_var_R_t_rescaled}} 
    \caption{Scaling of lifetime curves for $n=2$-hole variational wavepackets at $c=1024$ (a) and $c=4$ (c). Plots in the right column show a rescaling by by $(w/N)^2$.}
\end{figure}

Here, we note that the rescaled lifetime curves in the case of the $n=2$ staggered wavepacket adhere less strongly to the $w^2$-scaling than those of either the single-hole staggered (Fig. \ref{fig:c_4_n_1_w_scaling_rescaled_R_t}) or the $n=2$ variational wavepacket (Figs. \ref{fig:c_4_n_2_N_scaling_var_R_t_rescaled} and \ref{fig:c_1024_n_2_N_scaling_var_R_t_rescaled}), for which all parameters (other than coefficients used to construct the wavepacket) are the same as those presented here. We note that one common feature among the cases that display clear $w^2$-scaling behavior is the single depletion (regardless of $n$) of the initial profile, which is no longer the case as we move to $n=2$-hole staggered wavepackets (for wavepacket profiles, see Figs. \ref{fig:c_1024_n_2_N_scaling_stag_var} and \ref{fig:c_4_n_2_N_scaling_stag_var}). Whether this deviation from the predicted $w^2$ scaling is a function of initial profile---or that of the specific lifetime measure used---requires further investigation. We conclude by noting that
regardless of the specific scaling behavior, the $n=2$-staggered wavepackets continue to enjoy a significant improvement to lifetime under the scaling protocol.

\subsubsection*{Variational scaling with $N$}
%%%%%%%%%%%%%%%%
%%%%%%%%%%%%%%%%%
The lifetime curves here demonstrate scaling with summation width at sufficiently early times. The time evolution of the corresponding density profiles are displayed Fig. \ref{fig:c_1024_n_2_N_scaling_stag_var} and Fig. \ref{fig:c_4_n_2_N_scaling_stag_var}.  The significance of the scaling herein displayed further lies in the fact that despite containing variational amplitudes on $n=2$-hole states---and thus more complex dynamics than the single-hole, uniform-amplitude wavepackets---these curves still largely exhibit the predicted $w^2$ scaling.
\section{Conclusion} \label{sec:conclusion}
Throughout this work, we have built localized, long-lived wavepackets using superpositions of type-II eigenstates of the 1D Bose gas at arbitrary interaction strength. We have shown that a simple procedure involving \textit{selective summations}---reduction of momentum width in the set of eigenstates used to build the wavepacket---greatly improves its lifetime. We have shown that the  lifetime $\tau$ and momentum interval $\Delta k$  follow the scaling relation  $\tau \sim \big((\Delta k)^{2} \cdot \omega_0''\big)^{-1}$  (where $\omega_0''$ is the curvature of the dispersion at the center of the interval $\Delta k$), and quantitatively verified the scaling relation for the $n=1$-hole uniform-amplitude and $n=2$-hole variational wavepackets at $c \sim \mathcal{O}(1)$ and  $c \sim \mathcal{O}(10^3)$.  To this end, we have introduced two independently defined measures of lifetime---one using a time-series expansion of the central point in the wavepacket, and the other taking into account the entire region of depletion---and shown quantitatively the effect of various parameters on lifetime, including eigenstate coefficients, interaction strength $c$, summation offset $o$, and scaling with system size $N$. We have further demonstrated quantitatively the improvement in lifetime of multiple-hole ($n\geq2$) wavepackets as compared to their single-hole counterparts. Using selective-summation multiple-hole wavepackets, we have built long-lived wavepackets at intermediate-$c$, well outside of both the semi-classical regime and the $c \rightarrow \infty$ limit mappable to free fermions.
As for initial profiles, we have shown that using a variational approach to maximize the depletion in a chosen spatial window $\Delta x$, we may build localized, single-depletion profiles using multiple-hole eigenstates. While the uniform-amplitude formulation generally yields longer-lived wavepackets, both variational and uniform amplitude wavepackets gain considerable lifetime under scaling with system size, illuminating an open road towards building classical-like solitons in a purely quantum system.

Looking ahead, we note a plethora of possibilities to trek forward: most prominently, we note that our wavepackets have thus far been solivagant, traveling through the system without meeting any other localized excitations. Given that robustness under collision is a key feature of a solitonic excitation, examining such collision dynamics is a natural next step; we note however that the integrable 1D Bose gas should yield such robustness as a mere consequence of the presence of factorized scattering. As for choice of coefficients, there exists a myriad of choices---and a plentitude of variational protocols alone---well worth pursuing. Lastly, the approaches here studied are not limited to the 1D Bose gas: one may aim to build such solitonic excitations using similar strategies in other quantum integrable systems. We leave these, and the many other open routes, to future adventures.

\paragraph{Code availability}
The code used to produce the data in this paper can be found in the following repositories: \href{https://git.jscaux.org/jscaux/LiebLiniger-Quasisolitons-TypeII}{git.jscaux.org/jscaux/LiebLiniger-Quasisolitons-TypeII} (for quasisoliton-related executables and scripts) and \href{https://git.jscaux.org/jscaux/abacus-LL-min}{git.jscaux.org/jscaux/abacus-LL-min} (for the ABACUS (new version) integrability-handling backend).

\paragraph{Author contributions}
All authors contributed to every aspect of research execution and manuscript writing.

\paragraph{Funding information}
AS and JSC acknowledge support from the Institute of Physics of the University of Amsterdam. CV is partially supported by the Project ``Frontiere Quantistiche'' (Dipartimenti di Eccellenza) of the Italian Ministry of University and Research and by ``Iniziativa Specifica Quantum'' of INFN.

\begin{appendix}
\numberwithin{equation}{section}
\section{Derivation of Density Matrix Elements}\label{sec:app_dme}
In this appendix, we present a derivation of the matrix elements of the density operator $\hat \rho(x) \equiv \hat \Psi^\dag(x) \hat \Psi(x)$ between two Bethe states $|\{\mu\}_N\rangle$ and $|\{\lambda\}_N\rangle$ in the Tonks-Girardeau regime, by performing a $c \rightarrow \infty$ expansion on all relevant quantities. Such matrix elements are clearly related to those of the integrated density $\hat Q_1(x) = \int_0^x \hat \rho(x')\,dx'$, representing the number of particles in the interval $[0, x]$. We thus consider the form factor
\begin{equation}\label{eq:FF}
    F_N(\{\mu\}, \{ \lambda \}) = \frac{\langle \{\mu\}_N | \hat Q_1(x) | \{\lambda\}_N \rangle}{\lVert\{\mu\}_N \rVert\,\lVert\{\lambda\}_N\rVert},
\end{equation}
where the state norms are given by the Gaudin-Korepin formula \cite{Gaudin1981, Korepin1982, Gaudin_2014}
\begin{equation}\label{eq:norm}
    \lVert\{\lambda\}_N\rVert^2 = c^N\prod_{j>k}\frac{\lambda_{jk}^2 + c^2}{\lambda_{jk}^2}\text{det}_N\left[\delta_{jk}\left(L+\sum_l K_{jl}\right)-K_{jk}\right], 
\end{equation}
with
\begin{equation}
    K_{jk} = \frac{2c}{\lambda_{jk}^2+c^2}
\end{equation}
the Cauchy kernel and $\lambda_{jk} \equiv \lambda_j-\lambda_k$. The matrix elements of $\hat Q_1(x)$ obtained from the Algebraic Bethe Ansatz are \cite{Izergin, Slavnov1989, Slavnov}
\begin{equation}
    \langle \{\mu\}_N | \hat Q_1(x) | \{\lambda\}_N \rangle = \left\{\exp\left[ix\sum_m (\lambda_m - \mu_m)\right]-1\right\}\partial_\alpha\sigma_N^\alpha(\{\mu\},\{\lambda\})_{\alpha = 0}, 
\end{equation}
where
\begin{equation}
  \partial_\alpha\sigma_N^\alpha(\{\mu\},\{\lambda\})_{\alpha = 0}  = \prod_j (V_j^+  -V_j^-)\prod_{j,k}\Bigg{(}\frac{\lambda_{jk} + ic}{\mu_j-\lambda_k}\Bigg{)}\frac{\text{det}_N[\delta_{jk}+U_{jk}(\lambda_p)]}{V_p^+-V_p^-}.
\end{equation}
Here $\lambda_p$ is an arbitrary element of the set $\{\lambda\}_N$, and the factors $V_j^\pm$ and $U_{jk}$ are defined by
\begin{equation}\label{eq:Vj}
    V_j^{\pm}= \prod_m \frac{\mu_m - \lambda_j \pm ic}{\lambda_m - \lambda_j \pm ic},
\end{equation}
\begin{equation}
    U_{jk}(\lambda_p)= \frac{i(\mu_j - \lambda_j)}{V_j^+ - V_j^-}\prod_{m\neq j}\left(\frac{\mu_m - \lambda_j}{\lambda_m - \lambda_j}\right)\left(K_{jk}-K_{pk}\right).
\end{equation}
In particular, the normalized matrix elements of $\hat\rho(x)$ are
\begin{align}\label{eq:DME}
    \rho_{\mu\lambda}(x) \equiv \frac{\langle \{\mu\}_N | \hat\rho(x) | \{\lambda\}_N \rangle}{\lVert \{\mu\}_N\rVert \lVert \{\lambda\}_N \rVert} &= \frac{\partial}{\partial x}F_N(\{\mu\},\{\lambda\}) \nonumber\\
    &= e^{i(P_\lambda-P_\mu)x} \frac{i(P_\lambda -P_\mu)\partial_\alpha\sigma_N^\alpha(\{\mu\},\{\lambda\})_{\alpha = 0}}{\lVert\{\mu\}_N \rVert\,\lVert\{\lambda\}_N\rVert}\nonumber\\
    &\equiv e^{i(P_\lambda-P_\mu)x} \rho_{\mu\lambda}(0),
\end{align}
where $P_\lambda = \sum_j \lambda_j = \frac{2\pi}{L}\sum_j I^\lambda_j$ and $P_\mu = \sum_j \mu_j = \frac{2\pi}{L}\sum_j I^\mu_j$ are the momenta of the two Bethe states.

We shall now expand all the relevant quantities in series of $1/c$ for large $c$.

\subsubsection*{Expansion of the rapidities}

The series expansion of the scattering phase shift \eqref{eq:phaseshift} is
\begin{equation}
    \phi(\lambda) =2 \arctan \!\left(\frac{\lambda}{c}\right) = 2\sum_{k=0}^\infty \frac{(-1)^k}{2k+1}\left(\frac{\lambda}{c}\right)^{2k+1}.
\end{equation}
Substituting this expression into the Bethe equations \eqref{eq:Bethe_eq}, we obtain 
\begin{equation}
    \lambda_j = \frac{2\pi}{L}I^\lambda_j - \frac{2}{L}\sum_{k=0}^\infty\frac{(-1)^k}{2k+1}\sum_{l=1}^N\frac{(\lambda_j-\lambda_l)^{2k+1}}{c^{2k+1}}.
\end{equation}
Expanding the rapidities as
\begin{equation}
    \lambda_j = \sum_{a=0}^\infty \frac{\lambda_j^{(a)}}{c^a},
\end{equation}
where $\lambda_j^{(a)}$ is the coefficient of the $a^\text{th}$ order of the expansion in $1/c$, with $\lambda_j^{(0)} = 2\pi I_j/L$, and retaining only the $k=a=0$ term, we find
\begin{equation}
    \lambda_j^{(1)} = -\frac{2}{L}\left(N\lambda_j^{(0)} - \sum_{l=1}^N \lambda_l^{(0)}\right),
\end{equation}
therefore
\begin{equation}\label{eq:lambda_exp}
    \lambda_j = \frac{2\pi I^\lambda_j}{L}\left(1-\frac{2N}{cL}\right) + \frac{2 P_\lambda}{cL} + \mathcal O(c^{-2}).
\end{equation}

\subsubsection*{Expansion of the Cauchy kernel}

The Cauchy kernel is
\begin{equation}
    K_{jk} = \frac{2c}{\lambda_{jk}^2+c^2} = 2\sum_{n=0}^\infty (-1)^n\frac{\lambda_{jk}^{2n}}{c^{2n+1}} = \frac{2}{c} - \frac{2 \lambda_{jk}^2}{c^3} + \mathcal O(c^{-5}).
\end{equation}
To leading order, the difference $K_{jk}-K_{pk}$ is thus
\begin{equation}\label{eq:K_exp}
    K_{jk} - K_{pk} = -\frac{2}{c^3}(\lambda_{jk}^2-\lambda_{pk}^2)+\mathcal O(c^{-5})= -\frac{2}{c^3}\lambda_{jp}(\lambda_{jk}+\lambda_{pk})+\mathcal O(c^{-5}).
\end{equation}

\subsubsection*{Expansion of $V_j^\pm$}

We have
\begin{align}
    V_j^\pm &= \prod_m \frac{\pm 1 - i(\mu_m-\lambda_j)/c}{\pm 1 - i(\lambda_m-\lambda_j)/c} = \prod_m \left[1 \mp \frac{i}{c}(\mu_m-\lambda_m)+\mathcal O(c^{-2})\right]\nonumber\\
    &= 1 \mp \frac{i}{c}\sum_m (\mu_m - \lambda_m) + \mathcal O(c^{-2}) = 1 \pm \frac{i(P_\lambda-P_\mu)}{c} + \mathcal O(c^{-2})\label{eq:V_exp}
\end{align}
and therefore
\begin{equation}
    \prod_j \left(V_j^+ - V_j^-\right) = \left[\frac{2i}{c}(P_\lambda -P_\mu)\right]^N + \mathcal O(c^{-(N+1)}).
\end{equation}

\subsubsection*{Expansion of $U_{jk}$}
Inserting the expansions \eqref{eq:lambda_exp}, \eqref{eq:K_exp}, and \eqref{eq:V_exp} into
\begin{equation}
    U_{jk}(\lambda_p)= \frac{i(\mu_j - \lambda_j)}{V_j^+ - V_j^-}\prod_{m\neq j}\left(\frac{\mu_m - \lambda_j}{\lambda_m - \lambda_j}\right)\left(K_{jk}-K_{pk}\right),
\end{equation}
we obtain
\begin{align}
    U_{jk}(\lambda_p) &= -\frac{1}{L}\left[(I_j^\mu-I_j^\lambda)\frac{\pi}{P_\lambda-P_\mu}\left(c-\frac{2N}{L}\right)-1\right]\nonumber\\
    &\times \prod_{m\neq j} \frac{I_m^\mu - I_j^\lambda -\frac{P_\lambda-P_\mu}{\pi}(c-\frac{2N}{L})^{-1}}{I_m^\lambda - I_j^\lambda} \left[\frac{2}{c^3}\lambda_{jp}(\lambda_{jk}+\lambda_{pk})\right] +\cdots = \mathcal O(c^{-2}).
\end{align}

\subsubsection*{Expansion of $\partial_\alpha\sigma_N^\alpha$}

Let us now consider
\begin{equation}
  \partial_\alpha \sigma_N^\alpha(\{\mu\},\{\lambda\})_{\alpha = 0}  = \prod_j (V_j^+  -V_j^-)\prod_{j,k}\left(\frac{\lambda_{jk} + ic}{\mu_j-\lambda_k}\right)\frac{\det_N[\delta_{jk}+U_{jk}(\lambda_p)]}{V_p^+-V_p^-}.
\end{equation}
It is useful to absorb the factor $\prod_j (V_j^+  -V_j^-)$ into the determinant, defining $\tilde U_{jk} \equiv (V_j^+ - V_j^- )U_{jk}$. The determinant then becomes $\det_N[\delta_{jk}(V_j^+  -V_j^-)+\tilde U_{jk}(\lambda_p)]$. The first term is $\mathcal O(c^{-1})$, while the second is $\mathcal O(c^{-3})$, therefore to leading order this quantity becomes
\begin{equation}
    {\rm det}_N[\delta_{jk}(V_j^+  -V_j^-)] = {\rm det}_N(\delta_{jk})\prod_j (V_j^+  -V_j^-) \simeq \left[\frac{2i}{c}(P_\lambda-P_\mu)\right]^N.
\end{equation}
We thus have 
\begin{equation}
     \partial_\alpha \sigma_N^\alpha(\{\mu\},\{\lambda\})_{\alpha = 0} \simeq \prod_{j,k}\left(\frac{\lambda_{jk}+ic}{\mu_j - \lambda_k}\right) \left[\frac{2i}{c}(P_\lambda-P_\mu)\right]^N\frac{1}{V_p^+ - V_p^-}.
\end{equation} 
Recalling that $\lambda_{jk}$ is at most $\mathcal{O}(1)$, to leading order $\prod_{j,k}(\lambda_{jk}+ic) \simeq (ic)^{N^2}$, and therefore
\begin{equation}\label{eq:d_alpha}
     \partial_\alpha \sigma_N^\alpha(\{\mu\},\{\lambda\})_{\alpha = 0} \simeq \frac{i^{N^2+N-1}c^{N^2-N+1}[2(P_\lambda-P_\mu)]^{N-1}}{\prod _{j,k}(\mu_j - \lambda_k)}.
\end{equation}

\subsubsection*{Expansion of the norm}

Inserting the leading-order expansion of the Cauchy kernel, $K_{jk} \simeq 2/c$, into the expression for the norm,
\begin{equation}
    \lVert\{\lambda\}_N\rVert^2 = c^N\prod_{j>k}\frac{\lambda_{jk}^2 + c^2}{\lambda_{jk}^2}\text{det}_N\left[\delta_{jk}\left(L+\sum_l K_{jl}\right)-K_{jk}\right], 
\end{equation}
the leading order of the determinant is
\begin{equation}
    {\rm det}_N \left[\delta_{jk}\left(L+\sum_{l=1}^N \frac{2}{c}\right)-\frac{2}{c}\right] \simeq {\rm det}_N[\delta_{jk}L] =L^N.
\end{equation}
Moreover, $\prod_{j>k} (\lambda_{jk}^2+c^2) \simeq \prod_{j>k} c^2 = c^{N^2-N}$, therefore
\begin{equation}
    \lVert\{\lambda\}_N\rVert^2 \simeq \frac{c^{N^2}L^N}{\prod_{j>k} \lambda_{jk}^2},
\end{equation}
and
\begin{equation}\label{eq:norm_exp}
    \frac{1}{\lVert \{\mu\}_N\rVert \lVert \{\lambda\}_N \rVert} \simeq \frac{\prod_{j>k} |\lambda_{jk}| |\mu_{jk}|}{c^{N^2}L^N}.
\end{equation}

\subsubsection*{}
Substituting Eqs. \eqref{eq:d_alpha} and \eqref{eq:norm_exp} into Eq. \eqref{eq:DME}, we thus obtain
\begin{equation}\label{eq:DME_exp}
\rho_{\mu\lambda}(x) \simeq e^{i(P_\lambda-P_\mu)x} \frac{i^{N(N+1)}(P_\lambda-P_\mu)^N}{L}\left(\frac{2}{cL}\right)^{N-1} \frac{\prod_{j>k}|\lambda_{jk}| |\mu_{jk}|}{\prod_{j,k} (\mu_j-\lambda_k)}.
\end{equation}
Our purpose is now to simplify, and possibly work out explicitly, the products on the right hand side. In Sec. \ref{app:1h_DME}, we consider the case in which the Bethe states differ by \emph{one} quantum number. In Sec. \ref{app:nh_DME} we will generalize to the case in which the Bethe states differ by $n$ quantum numbers.

\subsection{One-hole states}\label{app:1h_DME}

Consider the case in which the states $|\{\lambda\}_N\rangle$ and $|\{\mu\}_N\rangle$ differ by a single quantum number indexed by $j_{\rm hole}$, i.e., $I_j^\lambda = I_j^\mu$ for all $j \neq j_{\rm hole}$, while $I_{j_{\rm hole}}^\lambda \neq I_{j_{\rm hole}}^\mu$. We split the product over $j, k$ in the denominator of Eq. \eqref{eq:DME_exp} as
\begin{equation}
    \prod_{j,k}(\mu_j - \lambda_k) = \prod_{j=1}^N(\mu_j - \lambda_j) \prod_{j<k}(\mu_j-\lambda_k)\prod_{j>k}(\mu_j-\lambda_k) \equiv p_1 p_2 p_3.
\end{equation}
Recalling that [Eq. \eqref{eq:lambda_exp}]
\begin{equation}
    \mu_j-\lambda_k = \frac{2\pi}{L}(I_j^\mu - I_k^\lambda) - \frac{2N}{cL}\left[\frac{2\pi}{L}(I_j^\mu - I_k^\lambda)+\frac{P_\lambda-P_\mu}{N}\right] + \mathcal O(c^{-2}),
\end{equation}
and noting that $P_\lambda-P_\mu = \frac{2\pi}{L}(I_{j_{\rm hole}}^\lambda-I_{j_{\rm hole}}^\mu)$, the first factor, $p_1$, is to leading order
\begin{equation}
    p_1 = \prod_{j \neq j_{\rm hole}}  (\mu_j - \lambda_j) (\mu_{j_{\rm hole}} - \lambda_{j_{\rm hole}}) = -\left(-\frac{2}{cL}\right)^{N-1} (P_\lambda-P_\mu)^N +\mathcal O(c^{-1}).
\end{equation}
The second factor, $p_2$, is
\begin{align}
    p_2 &= \prod_{j<k} (\mu_j-\lambda_k) = (-1)^{N(N-1)/2}\prod_{j>k} (\lambda_j-\mu_k) \nonumber\\
    &= (-1)^{N(N-1)/2} \prod_{\substack{j>k \\ k \neq k_{\rm hole}}} (\mu_j-\mu_k) \prod_{j>k_{\rm hole}} (\lambda_j-\mu_k)+\mathcal O(c^{-1}).
\end{align}
Taking $k_{\rm hole}=N$, the second factor on the right hand side is not present, so that
\begin{equation}
    p_2 = (-1)^{N(N-1)/2} \prod_{j>k} (\mu_j-\mu_k)+\mathcal O(c^{-1}).
\end{equation}
Lastly, $p_3$ is
\begin{equation}
    p_3 = \prod_{j>k}(\mu_j-\lambda_k) = \prod_{j>k} (\lambda_j-\lambda_k)+\mathcal O(c^{-1}).
\end{equation}
It follows that
\begin{equation}
    p_1p_2p_3 = \left(\frac{2}{cL}\right)^{N-1}(P_\lambda-P_\mu)^{N} (-1)^{N(N-3)/2}\prod_{j>k} \lambda_{jk} \mu_{jk} + \mathcal O(c^{-1}).
\end{equation}
Substituting this expression back into Eq. \eqref{eq:DME_exp}, and noting that $i^{N(N+1)}(-1)^{N(3-N)/2}=1$, we then obtain
\begin{equation}\label{eq:DME_exp_int}
    \rho_{\mu\lambda}(x) \simeq e^{i(P_\lambda-P_\mu)x} \frac{{\rm sgn} \left(\prod_{j>k} \lambda_{jk} \mu_{jk}\right)}{L}.
\end{equation}

\begin{figure}[H]
    \centering \includegraphics[width=0.7\textwidth]{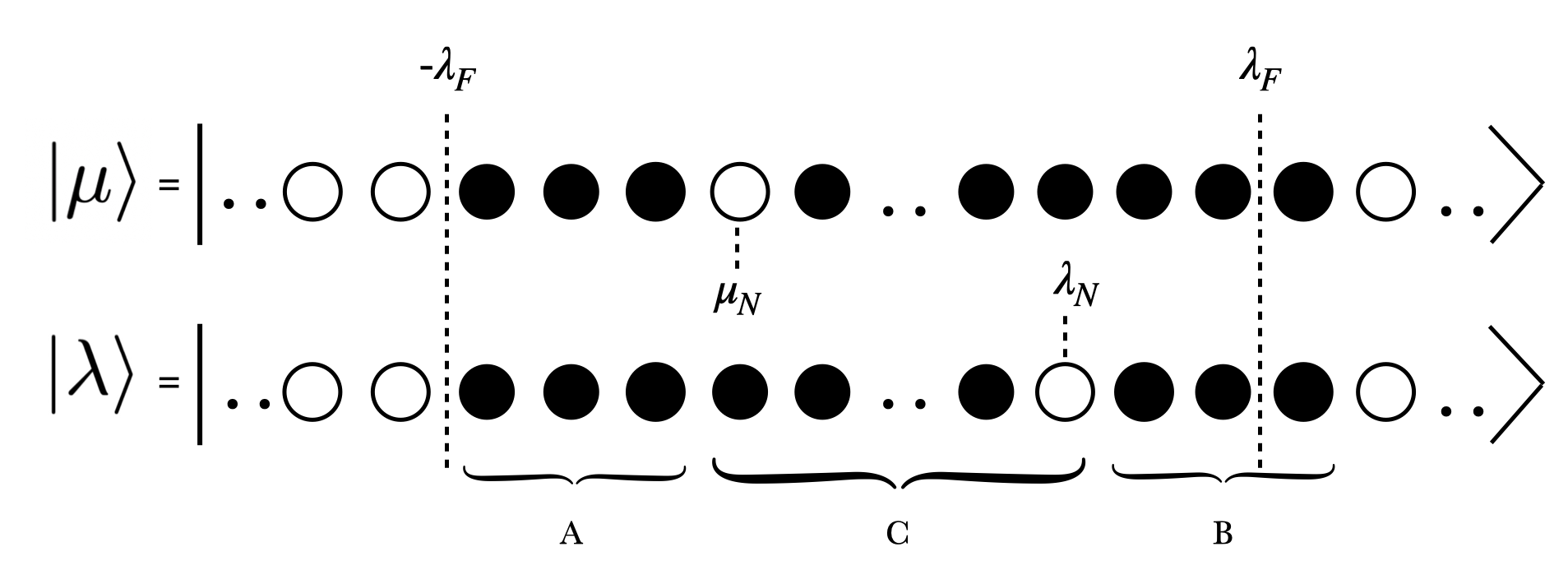}
    \caption{Rapidities in single-hole states $\ket{\mu}$ and $\ket{\lambda}$}
    \label{fig:TG_DME_chart}
\end{figure}

We would like to simplify this expression to the extent possible. Let us define
\begin{equation}
    \xi \equiv \prod_{j>k}\text{sgn}[(\lambda_j - \lambda_k)(\mu_j - \mu_k)].
\end{equation}
We may write this as
\begin{equation}
    \xi = \prod_{k=1}^{N-1} \underbrace{\text{sgn}[(\lambda_N - \lambda_k)(\mu_N-\mu_k)]}_{j=N \text{ (hole factor)}}\prod_{j=2}^{N-1}\prod_{k<j}\underbrace{\text{sgn}[(\lambda_j-\lambda_k)(\mu_j-\mu_k)]}_{\text{no hole index}}.
\end{equation}
Choosing again to label index of the hole as $j_\text{hole}=N$ guarantees that the second product does not include the index of the hole. This ensures that rapidities increase monotonically with the index; therefore, $\text{sgn}[(\lambda_j-\lambda_k)(\mu_j-\mu_k)] = 1$ for these factors. We thus have 
\begin{equation}
\xi = \prod_{\substack{k=1}}^{N-1} \text{sgn}[(\lambda_N - \lambda_k)(\mu_N-\mu_k)] = \prod_{k=1}^{N-1}\text{sgn}[(I_N^\lambda - I_k^\lambda)(I_N^\mu - I_k^\mu)].
\end{equation}
Note that 
\begin{equation}
    I_N^\lambda - I_k^\lambda = \left(\frac{N+1}{2}-h\right) - \left(-\frac{N+1}{2}+k\right) =(N+1-h)-k  
\end{equation}
and similarly, 
\begin{equation}
    I_N^\mu - I_k^\mu = (N+1-h')-k. 
\end{equation}
Let us define $(N+1-h) \equiv a $ and $(N+1-h') \equiv b$. We then have
\begin{equation}
    \xi = \prod_{k=1}^{N-1} \text{sgn}[(a - k)(b - k)].
\end{equation}
Note that $a,b >0$ for $h, h' \in [1,N]$. To determine $\text{sgn}[(a - k)(b - k)]$ we must therefore take into account the relative position between the hole and the rapidity with index $k$. For this, we make use of Fig. \ref{fig:TG_DME_chart}. Note that for $ k \in A$ and $k \in B$,  $\text{sgn}[(a - k)(b - k)]=1$ as both factors $(a - k)$ and $(b - k)$ take on the same sign. For $ k \in C$, $\text{sgn}[(a - k)(b - k)]= -1$; $\xi$ is therefore entirely determined by the number of elements in $C$; in other words, the number of particles between the holes in states $|\{\mu\}_N\rangle$ and $|\{\lambda\}_N\rangle$:
\begin{equation}
    \xi = \text{sgn}\left(\prod_{k \in C} (a-k)(b-k) \right) = \prod_{k \in C}(-1) = (-1)^{|C|},
\end{equation}
where $|C| = |\tilde I_{h'}^{\mu} - \tilde I_h^{\lambda}| - 1$, with $\tilde I_h^\lambda$ and $\tilde I_{h'}^\mu$ the hole quantum numbers in the two states.
For convenience, we write $\xi$ in exponential form and in terms of the hole parameters $h$ and $h'$, noting that $\tilde I^{h'}_{\mu} - \tilde I^h_{\lambda} \in \mathbb{Z}$:
\begin{equation}
    \xi = (-1)^{|\tilde I^{h'}_{\mu} - \tilde I^h_{\lambda}| - 1} = -e^{i\pi(h' - h)}.
\end{equation}
We conclude that
\begin{equation}
    \rho_{\mu\lambda}(x) \simeq e^{i(P_\lambda-P_\mu)x} \left[\frac{N}{L}\delta_{hh'}-\frac{e^{i\pi(h'-h)}}{L}(1-\delta_{hh'})\right].
\end{equation}

\subsection{Generalization to $n$-hole states}\label{app:nh_DME}

The above reasoning can be directly generalized to the case in which the two Bethe states differ by $n$ quantum numbers. In this case, state $\ket{\{\lambda\}_N}$ can be obtained from state $\ket{\{\mu\}_N}$ by punching $n$ holes, with $n$ an integer between 1 and $N$, and adding an equal number of particles. In other words, the set of quantum numbers $\{I^\lambda\}$ differs from the set $\{I^\mu\}$ by $n$ elements. We parameterize the quantum numbers $\{I^\mu\}$ in increasing order, so that $I^\mu_1 < I^\mu_2 < \dots < I^\mu_N$. To minimize the number of $I^\lambda_j$'s that are different from $I^\mu_j$'s for the same value of $j$, we parameterize $\{I^\lambda\}$ so as to keep unchanged the indices of the quantum numbers that remain occupied, while we parameterize in increasing order the new quantum numbers. For example,
\begin{figure}[H]
    \centering
    \includegraphics[width=\textwidth]{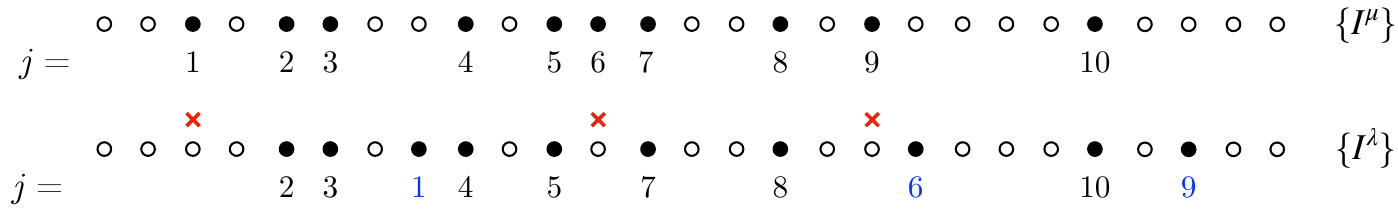}
\end{figure}
We will denote by $\mathscr H$ the set of indices $j$ corresponding to the new quantum numbers. The number of elements in $\mathscr H$ is $n$. In the example above, $n=3$ and $\mathscr H = \{1,6,9\}$. With these conventions, we obtain \cite{Cesare}
\begin{align}\label{mateltg}
    \rho_{\mu\lambda}(x) &= e^{i(P_\lambda-P_\mu)x}\frac{[-(P_\lambda-P_\mu)]^n}{L}\left(\frac{2}{c L}\right)^{n-1}(-1)^\ell\nonumber\\
    &\times\text{sgn}\biggl({\textstyle \prod_{\substack{j>k\\k \notin \mathscr H}} \mu_{jk} \prod_{\substack{j>k\\j \notin \mathscr H}} \lambda_{jk}}\biggr)\frac{\prod_{\substack{j>k\\k \in \mathscr H}} |\mu_{jk}| \prod_{\substack{j>k\\j \in \mathscr H}}|\lambda_{jk}|}{\prod_{\substack{j,k \\ j \in \mathscr H}} (\mu_j - \lambda_k)},
\end{align}
where $\ell$ is the number of elements in the product $\prod_{j>k,\,k\in \mathscr H}(\lambda_j-\mu_k)$. 

In the simplest case $n=1$, where the states $\ket{\{\lambda\}_N}$ and $\ket{\{\mu\}_N}$ differ by only one quantum number, we can always take this quantum number to be the one of index $N$, i.e. $\mathscr H = \{N\}$, and the other quantum numbers parameterized in increasing order. Then the products on the right-hand side of Eq. \eqref{mateltg} simplify to $(\mu_N-\lambda_N)^{-1} = [-(P_\lambda-P_\mu)]^{-1}$, so that we get back to Eq. \eqref{eq:DME_exp_int}.

In the case $n>1$, by taking $\mathscr H = \{N, N-1, \dots, N-n+1\}$, we get
\begin{align}
    \rho_{\mu\lambda}(x) &= e^{i(P_\lambda-P_\mu)x}\frac{[-(P_\lambda-P_\mu)]^n}{L}\left(\frac{2}{cL}\right)^{n-1}(-1)^{n(n-1)/2}\nonumber\\
    &\times \text{sgn}\biggl({\textstyle \prod_{\substack{j>k\\k \notin \mathscr H}} \mu_{jk} \lambda_{jk}}\biggr)\frac{\prod_{j>k} |\mu_{jk}| |\lambda_{jk}|}{\prod_{j,k}(\mu_j-\lambda_k)}\biggr|_{j,k \in \mathscr H}.
\end{align}
Notice that the factors $\mu_j-\lambda_k$ in the denominator are always non-zero, therefore the matrix elements are well-behaved. This equation implies that in the Tonks-Girardeau regime the density matrix elements between Bethe states differing by more than one quantum number are zero.
\section{Correspondence between numerics and analytics in TG}
\subsection{Numeric vs. analytic density matrix elements}
\begin{figure}[H]
    \centering
    \begin{subfigure}{0.49\linewidth}
        \includegraphics[width=\textwidth]{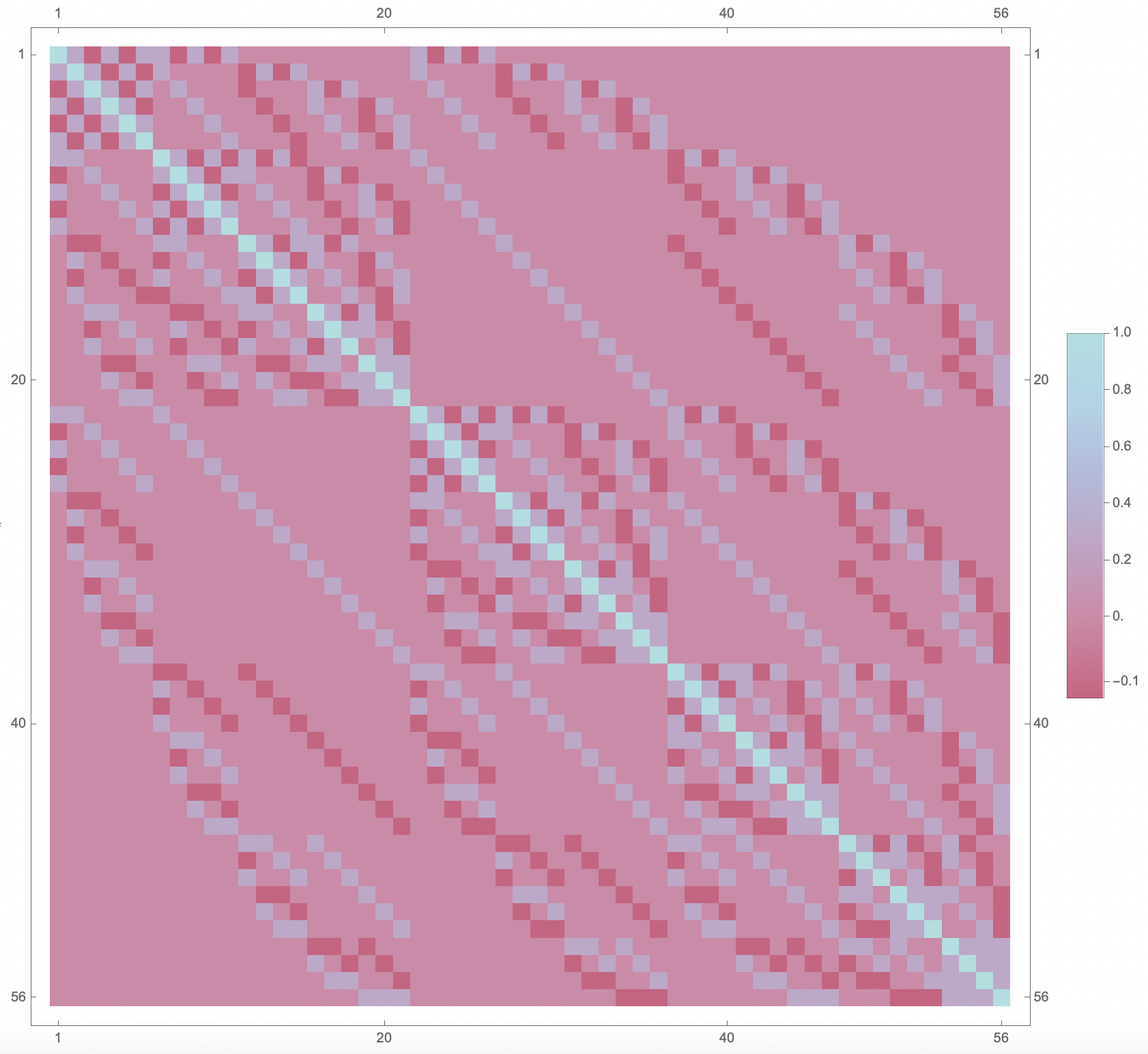}
        \subcaption[0.1\linewidth]{Analytically obtained density matrix elements for $n=$3 holes in Tonks-Girardeau regime with $N=8$.}
     \end{subfigure}
         \begin{subfigure}{0.49\linewidth}
         \includegraphics[width=\textwidth]{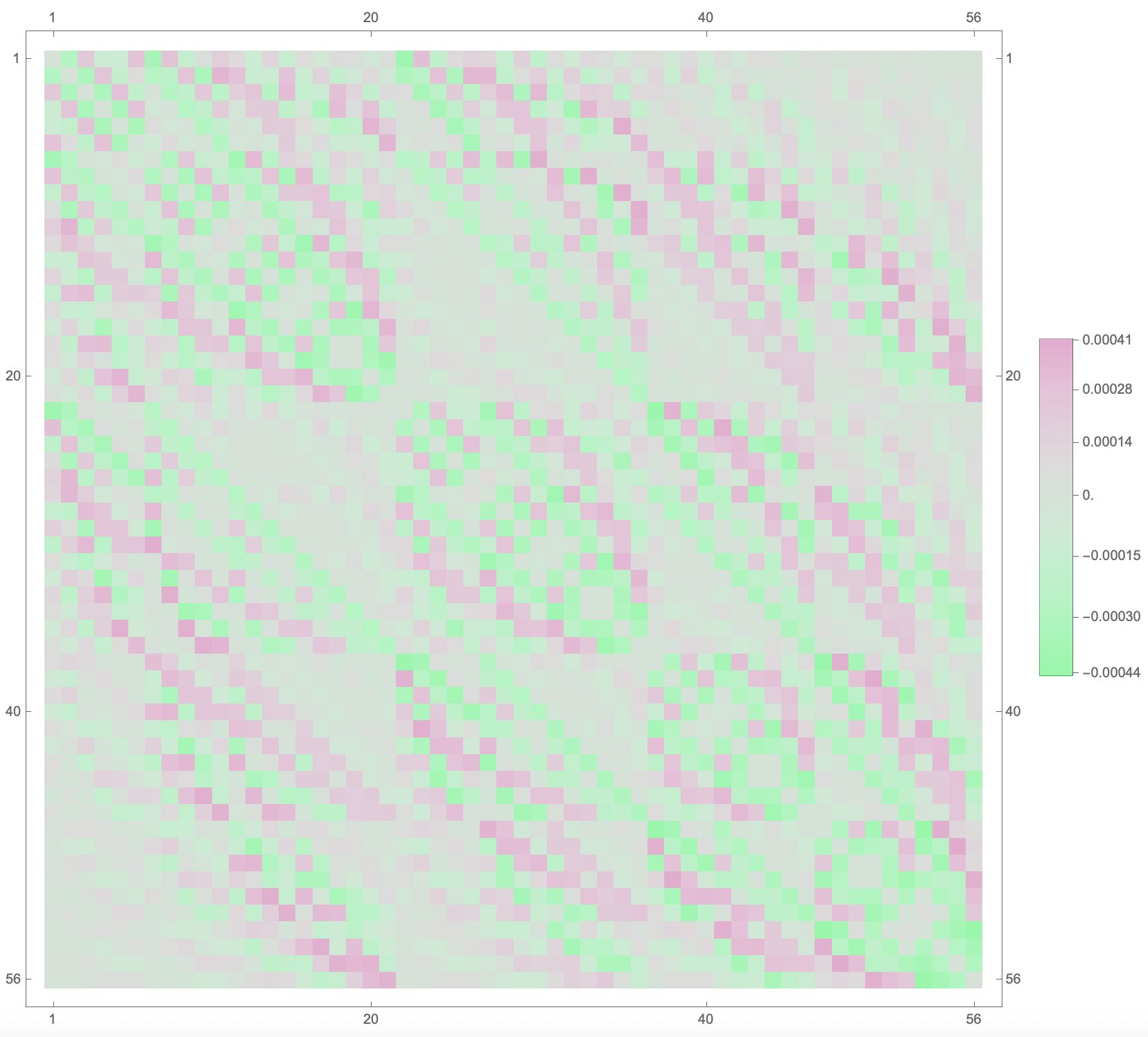}
        \subcaption[0.1\linewidth]{Difference between analytical and numerical DMEs for $n=3$ in TG. $N=8$; $c=1024$ in numerics. As shown, values agree to order $10^{-4}.$}
    \end{subfigure}
\end{figure}

\subsection{Numeric vs. analytic $n = 1, 2, 3, 4$ wavepackets}\label{app:exact_vs_num}

\begin{figure}[H]
    \centering
    \includegraphics[width=0.9\linewidth]{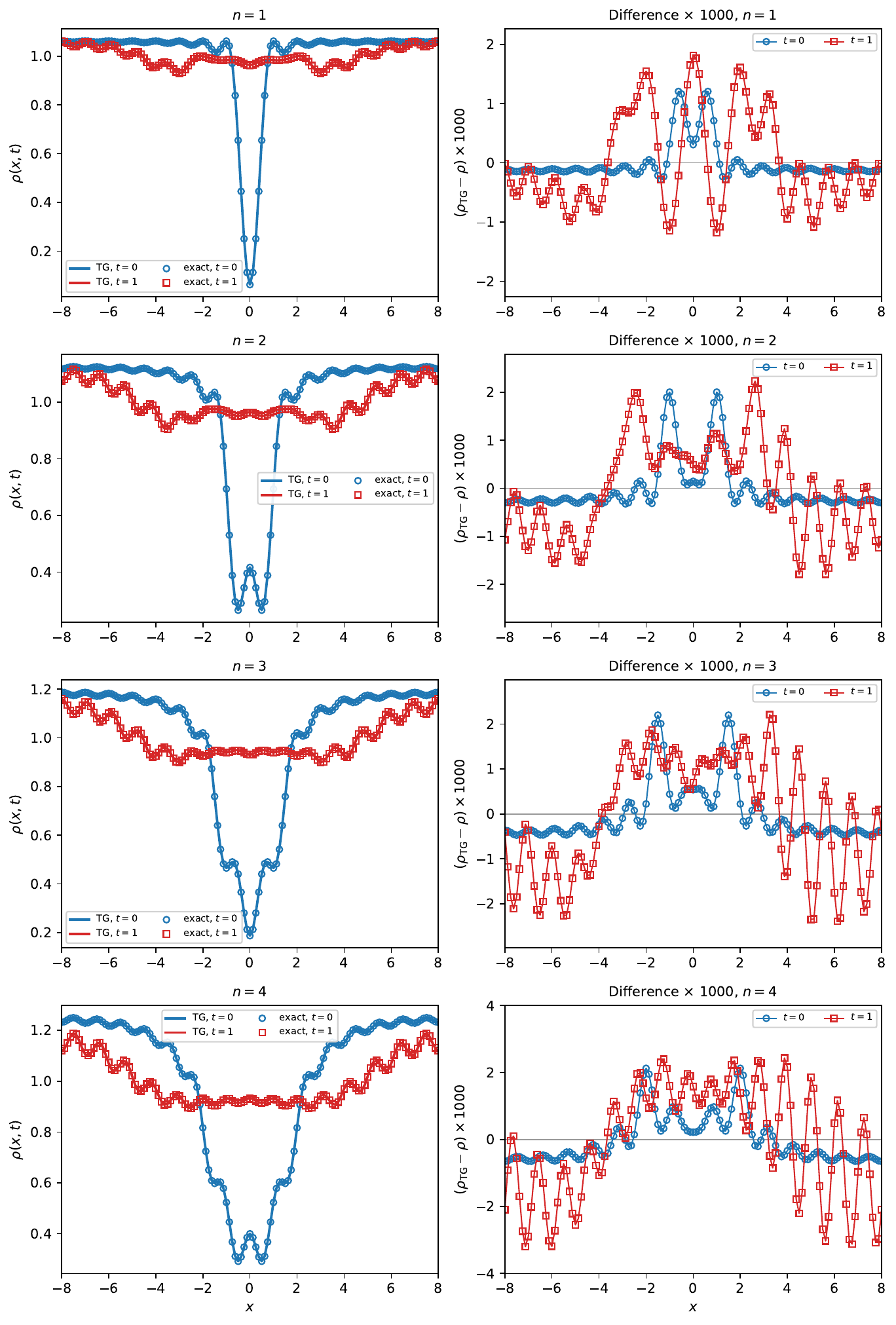}
    \caption{Comparison of full-width analytical TG density profiles and numerical ones, obtained for $c=1024$, $N=L=16$, and $n=1,\dots, 4$ holes.}
    \label{}
\end{figure}

\begin{figure}[H]
    \centering
    \includegraphics[width=0.9\linewidth]{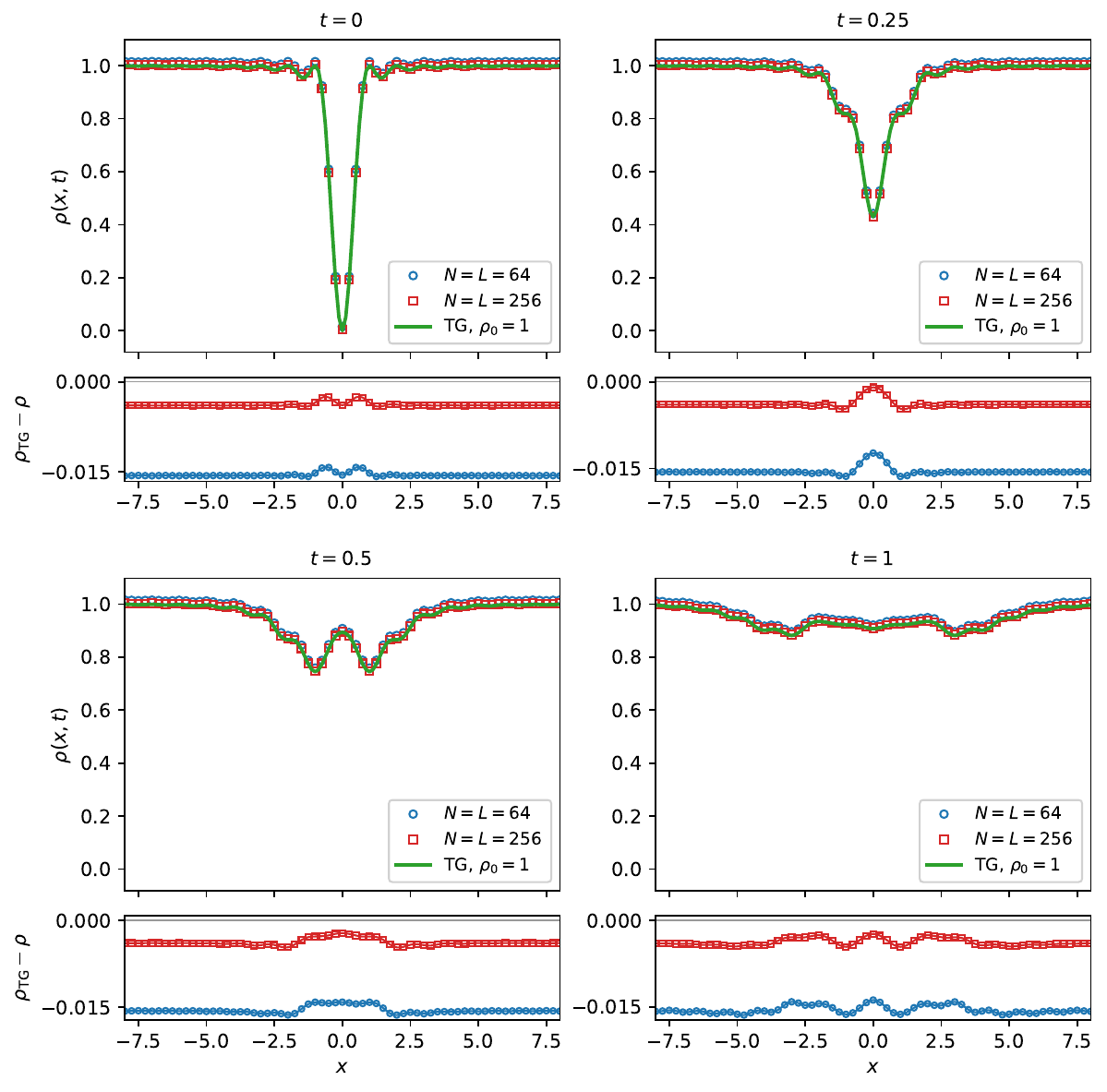}
    \caption{Comparison of the full-width 1-hole analytical TG density profile in the thermodynamic limit and numerical ones, obtained for $c=1024$ and two different system sizes, $N=L=64$ $N=L=256$.}
    \label{fig:app:TG_n_1_analytics_numerics}
\end{figure}
\subsection{Points of fixed density}\label{app:fixed_pts}

Consider the $n=1$ uniform-amplitude wavepacket in the Tonks-Girardeau regime. We observe a fixed point at $x=0$ for a wavepacket with parameters $x_0 = 0$ and $N=L$ even (Fig. \ref{fig:fixed_density}). We aim to show that $\rho(x=0,t) =\frac{N+1}{L} \hspace{5pt} \forall \hspace{5pt} t$.

\emph{Proof}: We can express the particle density analytically as follows:
\begin{equation}
    \rho(x,t)=\frac{N}{L} - \frac{2}{NL}\sum_{h<h'} \cos \left\{\frac{2\pi}{L}(h'-h)\left(x-x_0 +\frac{L}{2}-\frac{2\pi}{L}(N+1-h'-h)t\right)\right\}
\end{equation}
for weights $C_{h,h'}=e^{i\frac{2\pi}{L}(h-h')x_0}$. Now using $\eta_{-} \equiv \frac{2\pi}{L}(h-h')$ and  $\eta_{+} \equiv \frac{2\pi}{L}(h+h'-(1+N))$, we have
\begin{equation}
    \rho(x,t)=\frac{N}{L} - \frac{2}{NL}\sum_{\eta_{-}} \sum_{\eta_{+}} \cos\big{\{}\eta_{-}(x-x_0 +L/2)-\eta_{-}\eta_{+}t\big{\}}, 
\end{equation} 
where $\frac{2 \pi}{L}(1-N) \leq \eta_{-} \leq -\frac{2\pi}{L}$ and $\frac{2 \pi}{L}(2-N) \leq \eta_{+} \leq \frac{2\pi}{L}(N-2)$.

Choosing $x_0=0$, we have
\begin{equation}
    \rho(x=0,t)=\frac{N}{L} - \frac{2}{NL}\sum_{\eta_{-}, \hspace{3pt} \eta_{+}} \cos\big{\{}\eta_{-}(L/2-\eta_{+}t)\big{\}}.
\end{equation}
\textbf{1.} Consider terms with $\eta_{+} = 0:$
\begin{equation}
    \sum_{\eta_{+}=0, \hspace{1pt}\eta_{-}} \cos\big{\{}\eta_{-}(L/2-\eta_{+}t)\big{\}}
    = \sum_{\frac{2\pi}{L}(1-N)}^{-\frac{2\pi}{L}} \cos\big{\{}\eta_{-}(L/2)\big{\}} 
    = \sum_{h-h'=1-N}^{-1} \cos\big{\{}\pi(h-h')\big{\}},
\end{equation}
where the summation is over $(1-N) \leq h-h' \leq -1$ and therefore contains $N-1$ total terms. However, since $\eta_{+}$ is fixed to zero, the index of summation must progress in steps of two in order to keep $h'+h$ constant.

\begin{figure}[t]
\centering
    \subfloat[]{\includesvg[width=0.45\linewidth]{sections/figures/fixed_points/c_102400_N_10_L_10_nholes_1_width_4_offset_3_staggered_Nx_300_Nt_300_tmax_4.svg}\label{fig:fixedPts_n1}}
   \subfloat[]{\includesvg[width=0.45\linewidth]{sections/figures/fixed_points/c_102400_N_10_L_10_nholes_2_width_4_offset_3_staggered_Nx_300_Nt_300_tmax_4.svg}\label{fig:fixedPts_n2}}
   \newline
   \centering
    \subfloat[]{\includesvg[width=0.45\linewidth]{sections/figures/fixed_points/c_102400_N_10_L_10_nholes_3_width_4_offset_3_staggered_Nx_300_Nt_300_tmax_4.svg}\label{fig:fixedPts_n3}}
    \subfloat[]{\includesvg[width=0.45\linewidth]{sections/figures/fixed_points/c_102400_N_10_n_1_2_3_w_4_o_3_stag_tmax_4.svg}\label{fig:fixedPts_c_102400}}
    \caption{Fig. \ref{fig:fixedPts_n1}, \ref{fig:fixedPts_n2}, \ref{fig:fixedPts_n3}: Time evolution of the 1-, 2-, and 3-hole  staggered wavepackets at c=102400. Fig. \ref{fig:fixedPts_c_102400}: Corresponding points of fixed density obtained numerically, plotted over a single recurrence time of $t_{rec}\approx4$.}
    \label{fig:fixed_density}
\end{figure}

Therefore, there are only $N/2$ terms in the sum for $\eta_{+} =0$. Furthermore, since $N$ is even, all quantities $h-h'$ are odd, and thus:
\begin{equation}
    \sum_{\eta_{+}=0, \hspace{1pt}\eta_{-}} \cos\big{\{}\pi(h-h')\big{\}} = -\frac{N}{2}.
\end{equation}
\textbf{2.} Now consider all remaining terms, which have $\eta_{+} \neq 0$:
 \begin{equation}
 \begin{split}
    \sum_{\eta_{+}\neq0, \hspace{1pt}\eta_{-}} \cos\big{\{}\eta_{-}(L/2)-\eta_{-}\eta_{+}t)\big{\}} =&
    \sum_{\eta_{+}\neq0, \hspace{1pt}\eta_{-}} \cos\big{\{}\eta_{-}(L/2)\big{\}} \cdot\cos\big{\{}\eta_{-}\eta_{+}t\big{\}} + \underbrace{\sin\big{\{}\eta_{-}(L/2)\big{\}}\cdots}_{0 \hspace{2pt} \forall \hspace{2pt} \eta_{-}} 
    \\=& \sum_{\eta_{+}\neq0, \hspace{1pt}\eta_{-}} \underbrace{\cos\big{\{}\pi(h-h')\big{\}}}_{\pm1}\cdot \cos\big{\{}\eta_{+}\eta_{-}t\big{\}}.
    \end{split}
\end{equation}
Due to symmetry (for $N$ even), these terms cancel, leaving only the contribution from $\eta_{+}=0$:
\begin{equation}
    \rho(x=0,t)=\frac{N}{L} + \frac{2}{NL}\Big{(}\frac{N}{2}\Big{)} = \frac{N+1}{L},
\end{equation}
which is the observed density at the fixed point. $\Box$

Presumably, a similar symmetry holds in other cases where a fixed point is observed: for $n>1$ at different values of $x$, well outside of the Tonks-Girardeau regime (down to at least $c=10$), and for large system sizes.
\end{appendix}

\bibliography{References.bib}

\end{document}